\documentclass[preprint,prd,nofootinbib,tightenlines,amssymb]{revtex4}
\usepackage[dvipdfmx]{graphicx}
\usepackage{bm}
\usepackage{amsmath}
\usepackage{hhline}
\usepackage{color}
\usepackage{xcolor}
\usepackage{slashed}
\usepackage{relsize}
\usepackage[T1]{fontenc}
\usepackage{upquote}
\usepackage{changes} 

\newcommand{\Eq}{&=&}

\newcommand{\white}[1]{{\color[rgb]{1,1,1} #1}}
\newcommand{\black}[1]{{\color[rgb]{0,0,0} #1}}
\newcommand{\bs}[1]{{\boldsymbol{#1}}}
\newcommand{\hs}[1]{{\hspace{#1}}}
\newcommand{\vs}[1]{{\vspace{#1}}}
\newcommand{\tf}[1]{{\textsf{#1}^{}}}
\newcommand{\tx}[1]{{\text{#1}^{}}}
\newcommand{\nn}{\nonumber\\}
\newcommand{\rx}{\raisebox{1pt}{}}
\newcommand{\sx}[2]{{\scalebox{#1}{#2}}}
\newcommand{\dd}{{\mathrm{d}}}
\newcommand{\XXXNN}{X\hs{-0.03cm}X\hs{-0.03cm}X \to \bar{N}\hs{-0.03cm}\bar{N}}
\newcommand{\XXNXN}{X\hs{-0.03cm}X\hs{-0.03cm}N \to \bar{X}\hs{-0.03cm}\bar{N}}
\newcommand{\XNNXX}{X\hs{-0.03cm}N\hs{-0.03cm}N \to \bar{X}\hs{-0.03cm}\bar{X}}
\newcommand{\NNXX}{N\hs{-0.03cm}\bar{N} \to X\hs{-0.03cm}\bar{X}}
\newcommand{\XXNN}{X\hs{-0.03cm}\bar{X} \to N\hs{-0.03cm}\bar{N}}
\newcommand{\NNXXX}{\bar{N}\hs{-0.03cm}\bar{N} \to X\hs{-0.03cm}X\hs{-0.03cm}X}
\newcommand{\XXNNNN}{X\hs{-0.03cm}\bar{X} \to N\hs{-0.03cm}\bar{N}\hs{-0.03cm}N\hs{-0.03cm}\bar{N}}
\newcommand{\xfo}{x^{}_\tf{f.o.}}
\newcommand{\xfi}{x^{}_\tf{f.i.}}
\newcommand{\lambdaXS}{\lambda_{X\hs{-0.03cm}S}}
\newcommand{\SSXX}{S\hs{-0.01cm}\bar{S} \to X\hs{-0.03cm}\bar{X}}
\newcommand{\XX}{X\hs{-0.03cm}\bar{X}}
\newcommand{\NN}{N\hs{-0.03cm}\bar{N}}
\newcommand{\Zf}{\mathbb{Z}^{}_4}

\begin{document}
\baselineskip=14.5pt \parskip=2.5pt

\vspace*{3em}

\preprint{KIAS-22002}

\title{\large Scalar and Fermion Two-component SIMP Dark Matter \\with an Accidental $\,\mathbb{Z}^{}_\mathbf{4}$ Symmetry}

\author{
Shu-Yu\,\,Ho,\footnote[1]{phyhunter@kias.re.kr}
Pyungwon\,\,Ko,\footnote[2]{pko@kias.re.kr}
and Chih-Ting\,\,Lu\footnote[3]{timluyu@kias.re.kr}}

\affiliation{
Korea Institute for Advanced Study, Seoul 02455, Republic of Korea
\vspace{3ex}}

\begin{abstract}
In this paper, we construct for the first time a two-component strongly interacting massive particles (SIMP) dark matter (DM) model, where a complex scalar and a vector-like fermion play the role of the SIMP DM candidates.\,\,These two particles are stable due to an accidental $\Zf$ symmetry after the breaking of a U$(1)^{}_\tf{D}$ gauge symmetry.\,\,By introducing one extra complex scalar as a mediator between the SIMP particles, this model can have $3 \to 2$ processes that determine the DM relic density.\,\,On the other hand, the SIMP DM particles can maintain kinetic equilibrium with the thermal bath until the DM freeze-out temperature via the U$(1)^{}_\tf{D}$ gauge couplings.\,\,Most importantly, we find an unavoidable two-loop induced $2 \to 2$ process tightly connecting to the $3 \to 2$ process that would redistribute the SIMP DM number densities after the chemical freeze-out of DM.\,\,Moreover, this redistribution would significantly modify the predictions of the self-interacting cross section of DM compared with other SIMP models.\,\,It is crucial to include the two-loop induced $2 \to 2$ annihilations to obtain the correct DM phenomenology.
\end{abstract}

\maketitle

\section{Introduction}\label{sec:1}

"What is dark matter (DM)?" and "Where does DM come from?" are two very questions that drive countless particle physicists and cosmologists to work day and night to solve these problems.\,\,As of now, the only thing we know for sure is that it contributes about 26\,\% energy density in the present universe.\,\,The remaining energy density is dominantly attributed to dark energy which is another mystery physicists aim to understand.

The first question concerns the particle nature of DM, such as mass, spin, and fundamental interactions.\,\,Firstly, the mass of DM can spread over a very broad range from $10^{-15}\,\tx{GeV}$ to $10^{15}\,\tx{GeV}$\,\cite{Baer:2014eja}.\,\,Secondly, it could be comprised of a scalar boson, a vector boson, a Dirac fermion, a Majorana fermion, or a Rarita-Schwinger fermion.\,\,Thirdly, it may possess interactions to the ordinary matter other than the gravitational interaction.\,\,The second question asks about the production mechanism of DM.\,\,As we know, it can be produced thermally or non-thermally in the early universe.\,\,Lastly, there is a possibility that the universe contains more than one kind of DM just like the visible world exist many stable particles such as the electron, proton, and neutrinos.\,\,Indeed, there are many efforts along this direction~\cite{Hochberg:2014kqa,Katz:2020ywn,Choi:2021yps,Baek:2013dwa,Aoki:2016glu,
Daido:2019tbm,Herms:2019mnu,Yaguna:2021rds,DiazSaez:2021pmg}.

The most popular thermally-produced DM is weakly interacting massive particles (WIMP) \cite{Lee:1977ua}, where the annihilation cross sections of DM pairs into the standard model (SM) particles determine the DM relic abundance.\,\,Nonetheless, the null result of direct detection experiments has pushed the WIMP scenario into a corner, which motivates physicists to come up with new prospects for DM.\,\,The so-called secluded WIMP scenarios are still viable since they are not to be strongly constrained by direct detection experiments~\cite{Pospelov:2007mp,Pospelov:2008jd}.

Strongly interacting massive particles (SIMP) \cite{Hochberg:2014dra} is an alternative thermal DM scenario that has brought people attention due to its exotic dynamic, where the DM relic abundance is set by the self-annihilation cross sections of DM number-changing processes.\,\,In particular, the SIMP with a large self-interacting cross section can relax some inconsistencies between the N-body simulations and astrophysical observations at small-scale structures (\,$\lesssim$\,1 Mpc) of the universe.\,\,For instance, the collisionless cold DM predicts a cuspy density profile in the center of dwarf galaxy halos.\,\,However, what we observe is a relatively flat distribution \cite{Tulin:2017ara}.\,\,This is known as the core-vs-cusp problem.\,\,Besides, the collisionless cold DM also predicts dozens of large sub-halos with speeds $v > 25$ km/s in the Milky Way and M31, but no such halos have been discovered \cite{Brooks:2012vi}.\,\,This is commonly named the too-big-to-fail problem.

\begin{figure}[t!]
\centering
\hs{0.5cm}
\includegraphics[width=0.55\textwidth]{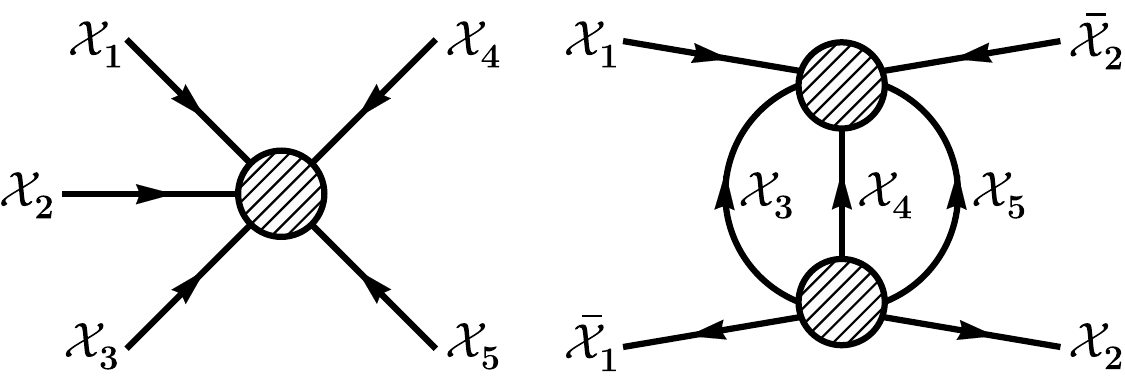}
\vs{-0.3cm}
\caption{The Feynman diagrams of the $3 \to 2$ and the two-loop induced 
$2 \to 2$ processes in the $r$SIMP scenario, where ${\cal X}^{}_i$ denotes the SIMP 
particle and the arrow represents the dark charge flow.}
\label{fig:multiSIMP}
\end{figure}

With the above considerations, we study in Ref.\,\cite{Ho:2021ojb} the multi-component SIMP scenario by using the effective operator method.\,\,As in the single-component SIMP scenario, the DM relic abundance is determined by the reaction rate of the $3 \to 2$ process as shown in the left graph of Fig.\,\ref{fig:multiSIMP}.\,\,Surprisingly, we notice that in this scenario there is an irreducible two-loop induced $2 \to 2$ number-conserving process\footnote{Here the number-conserving means the total DM number is conserved.\,\,However, the individual DM density would change due to the $2 \to 2$ processes.} (see the right graph of Fig.\,\ref{fig:multiSIMP}) that would reshuffle the DM number densities after the chemical freeze-out of DM.\,\,We then dub this scenario as reshuffled SIMP ($r$SIMP) DM.\,\,Note that in the single-component SIMP scenario, since the external legs of such a two-loop diagram are the same particles, there is no redistribution of DM number densities due to this diagram.\,\,Intuitively, one may think that this $2 \to 2$ process is suppressed by the two-loop factor.\,\,However, for a $3 \to 2$ process to take place, it has to capture one extra DM particle whose number yield is Boltzmann-suppressed.\,\,It turns out that the reaction rate of the $2 \to 2$ process dominates over that of the $3 \to 2$ process.\,\,Furthermore, we find that the masses of   DM particles must be nearly degenerate to weaken the reshuffled effect.\,\,Otherwise, the $2 \to 2$ process would actively enforce the heavy SIMP particle annihilating into the light one, with essentially no remaining of heavy SIMP DM.\footnote{In our perspective, each DM component should have a sizable amount in multi-component DM scenarios.}

In order to make our analysis of the $r$SIMP scenario more robust and reliable, we build up a UV complete model in this paper instead of the effective theory.\,\,We consider a two-component SIMP DM model (hereafter we call it $r$SIMP model), where the DM is comprised of a complex scalar and a vector-like fermion.\footnote{The two-component SIMP model with complex scalar and vector-like fermion is constructed in this paper for the first time.\,\,In Ref.\,\cite{Choi:2021yps}, such a possibility based on U$(1)^{}_\tf{D} \to \mathbb{Z}^{}_2 \times \mathbb{Z}^{}_3$ 
was mentioned in the "footnote 2", but without explicit construction.}\,\,In this model, the DM particles have an accidental $\,\Zf$ charge after a U$(1)^{}_\tf{D}$ symmetry breaking.\footnote{Note that this discrete symmetry does not inherit from a gauge symmetry by the Krauss-Wilczek manner \cite{Krauss:1988zc}.}\,\,If this U$(1)^{}_\tf{D}$ symmetry is promoted to a gauge symmetry, then a vector-portal interaction naturally arises between the SIMP DM and SM particles.\,\,This interaction is necessary for the SIMP scenario to prevent the heat up of DM due to the $3 \to 2$ process before the chemical freeze-out of DM.\,\,This is known as the SIMP conditions~\cite{Hochberg:2015vrg,Hochberg:2018rjs}.

Following this setup, we then explicitly compute the annihilation cross sections of the $3 \to 2$ and $2 \to 2$ processes and solve the coupled Boltzmann equations numerically to get the correct number densities of DM.\,\,We find that the reshuffled phenomena still occur in the UV complete model.\,\,Thus, our previous effective operator analysis of the $r$SIMP scenario is valid.\,\,Also, the form of the $2 \to 2$ annihilation cross section derived by the effective operator is consistent with the one in this UV complete model if we treat the cut-off scale as the mediator mass in the two-loop diagram.\,\,Again, we emphasize that the $2 \to 2$ process in the multi-component SIMP scenario is generic and cannot be ignored in DM phenomenology, especially in estimating the DM relic abundance.\,\,Adding number-conserving $2 \to 2$ processes to number-changing $3 \to 2$ processes in multi-component SIMP models will not only change the fractions of DM particles but also the total DM number densities.\,\,It can dramatically modify model parameters that accommodate the correct relic density compared with only involving $3 \to 2$ processes.

In most of the SIMP models, the DM is assumed to be a complex scalar in order to have the DM number changing $3 \rightarrow 2$ processes be allowed.\,\,And typically one has to choose large enough quartic or cubic couplings of the scalar DM to satisfy the DM relic density and the vacuum stability.\,\,With such couplings, the prediction of DM self-interacting cross section may be too big to be compatible with the astrophysical observations from the Bullet and Abell 3827 clusters \cite{Markevitch:2003at,Clowe:2003tk,Massey:2015dkw,Kahlhoefer:2015vua}.\,\,However, in the two-component SIMP model with complex scalar and vector-like fermion DM, this tension can be eased thanks to the reshuffled effect.\,\,For example, if the complex scalar is heavier than the vector-like fermion, the DM self-interacting cross section can be reduced since the portion of the complex scalar annihilates into the vector-like fermion due to the $2 \to 2$ process.\,\,Plus, the self-interaction of the vector-like fermion corresponds to a four-fermion interaction which is suppressed by the mass scale of the mediator at low energy.\,\,This is one of the interesting features of this model.

The structure of this paper is as follows.\,\,In the next section, we introduce the $r$SIMP model and
give a description of the relevant interactions and masses for the new particles.\,\,In Sec.\,\ref{sec:3}, we write down the formulas for the annihilation cross sections of the $3 \to 2$ and $2 \to 2$ processes.\,\,In Sec.\,\ref{sec:4}, we take into account various theoretical and
experimental constraints on this model.\,\,In Sec.\,\ref{sec:5}, we evaluate the relic abundance of the $r$SIMP DM and explain the reshuffled mechanism.\,\,In Sec.\,\ref{sec:6}, we discuss the SIMP conditions.\,\,In Sec.\,\ref{sec:7}, we show the predictions of DM self-interacting cross section in this model.\,\,Finally, we briefly mention some outlook of this model and conclude our study in Sec.\ref{sec:8}.\,\,In the appendices, we demonstrate the computations of annihilation cross sections of the $3 \to 2$ and $2 \to 2$ processes in the $r$SIMP model.

%%%%%%%%%%%%%%%%%%%%%%%%%%%%%%%%%%%%%%%%%%%%%%%%%

\section{$\bs{r}$SIMP model}\label{sec:2}

To demonstrate the redistribution of DM mass densities in the $r$SIMP scenario, we consider one vector-like fermion, $N$, and three complex singlet scalars, $X, S$, and $\Phi$ in addition to the SM particles.\,\,These new particles have dark charges under a gauged U$(1)^{}_\tf{D}$ symmetry, and all SM particles are neutral under this U$(1)^{}_\tf{D}$ symmetry.\,\,We summarize the particle contents and their charge assignments in Tab.\,\ref{tab:1}.\,\,In our setup, the $X$ and $N$ are SIMP DM candidates, and $S$ is an unstable mediator connecting these two particles.\,\,In particular, the $\Phi$ particle triggers the breaking of the U$(1)^{}_\tf{D}$ symmetry.\,\,After the 
U$(1)^{}_\tf{D}$ symmetry breaking, these new particles can possess an accidental ${}^{}\Zf$ symmetry, which stabilizes the $X$ and $N$ and make them DM.

\begin{table}[b!]
\begin{center}
\def\arraystretch{1.3}
\begin{tabular}{|c||c||c|c|c|c|}
\hline
$\vphantom{|_|^|}$                            
& ~$H$~  & ~$N$~ & ~$X$~ & ~$S$~ & ~$\Phi$~ 
\\\hline\hline 
~\,SU$(2)\vphantom{|_|^|}$~         
& ~$\mathbf{2}$~ & ~$\mathbf{1}$~ & ~$\mathbf{1}$~ & ~$\mathbf{1}$~ & ~$\mathbf{1}$~ 
\\\hline
~\,U$(1)^{}_\tf{Y}\vphantom{|_|^|}$~      
& ~$-{}^{}1/2$~ & ~$0$~ & ~$0$~ & ~$0$~ & ~$0$~ 
\\\hline
~\,U$(1)^{}_\tf{D}\vphantom{|_|^|}$~      
& ~$0$~ & ~$-{}^{}1/8$~ & ~$+{}^{}1/12$~ & ~$+{}^{}1/4$~ & ~$-{}^{}1/2$~ 
\\\hline
~$\Zf$~      
& ~$+{}^{}1$~ & ~$\pm{}^{}i$~ & ~$-1$~ & ~$-1$~ & ~$+{}^{}1$~ 
\\\hline  
\end{tabular}
\caption{Charge assignments of the fermion and scalars in the $r$SIMP model, where 
$H$ is the SM Higgs doublet and $i =\sqrt{-1}$.}
\vs{-1.0cm}
\label{tab:1}
\end{center}
\end{table}

The Lagrangian density for the scalar fields in this model is given by
\begin{eqnarray}
{\cal L}_\tf{scalar}
\,=\,
\big({\cal D}^\rho H\big)^{\hs{-0.05cm}\dag} {\cal D}_\rho {}^{} H +
\big({\cal D}^\rho X\big)^{\hs{-0.05cm}\dag} {\cal D}_\rho {}^{} X +
\big({\cal D}^\rho S\big)^{\hs{-0.05cm}\dag} {\cal D}_\rho {}^{} S +
\big({\cal D}^\rho \Phi\big)^{\hs{-0.05cm}\dag} {\cal D}_\rho {}^{} \Phi \,-
{\cal V}(H, X, S, \Phi)
~,
\end{eqnarray}
where ${}^{}{\cal D}_\rho = \partial_\rho + (i/2)^{} g^{}_\tf{W} \tau^a W^a_\rho + i g^{}_\tf{Y} {\cal Q}^{}_\tf{Y} B^{}_\rho + i g^{}_\tf{D} {\cal Q}^{}_\tf{D} C^{}_\rho{}^{}$ denotes the covariant derivative with $g^{}_\tf{W}\,(W^a_\rho)$, $g^{}_\tf{Y}\,(B^{}_\rho{}^{})$, and $g^{}_\tf{D}\,(C^{}_\rho)$ being the SU$(2)$, U$(1)^{}_\tf{Y}$, and U$(1)^{}_\tf{D}$ gauge couplings (fields), respectively\,; $\tau^a\,\big(a = 1, 2, 3)$ the Pauli matrices, and ${\cal Q}^{}_\tf{Y}\,({\cal Q}^{}_\tf{D})$ the hypercharge (dark charge) operator.\,\,The scalar potential \,${\cal V} = {\cal V}(H, X, S, \Phi)$\, is given by
\begin{eqnarray}\label{potential}
{\cal V} 
\Eq
\mu_h^2 {}^{} H^\dag \hs{-0.05cm} H +
\mu_X^2 X^\ast \hs{-0.05cm} X + 
\mu_S^2 {}^{} S^\ast \hs{-0.05cm} S +  
\mu_\phi^2 {}^{} \Phi^\ast \hs{-0.02cm} \Phi 
\nn[0.1cm]
&&
+\,
\lambda^{}_h \big(H^\dag \hs{-0.05cm} H {}^{}\big)\rx{\hs{-0.03cm}^2} +
\lambda^{}_X \big(X^\ast \hs{-0.05cm} X\big)\rx{\hs{-0.03cm}^2} +
\lambda^{}_S \big(S^\ast \hs{-0.05cm} S {}^{} \big)\rx{\hs{-0.03cm}^2} +
\lambda^{}_\phi \big(\Phi^\ast \hs{-0.02cm} \Phi\big)\rx{\hs{-0.03cm}^2} 
\nn[0.1cm]
&&
+\,
\lambda_{h X} \big(H^\dag \hs{-0.05cm} H{}^{}\big) \big(X^\ast \hs{-0.05cm} X\big) +
\lambda_{h S} \big(H^\dag \hs{-0.05cm} H{}^{}\big) \big(S^\ast \hs{-0.05cm} S {}^{} \big) +
\lambda_{h \phi} \big(H^\dag \hs{-0.05cm} H {}^{}\big) 
\big(\Phi^\ast \hs{-0.02cm} \Phi\big)
\nn[0.15cm]
&&
+\,
\lambda_{X \hs{-0.03cm} S} \big(X^\ast \hs{-0.05cm} X\big) \big(S^\ast \hs{-0.05cm} S {}^{} \big) + 
\lambda_{X \hs{-0.03cm} \phi} \big(X^\ast \hs{-0.05cm} X\big) \big(\Phi^\ast \hs{-0.01cm} \Phi\big) +
\lambda_{S \phi} \big(S^\ast \hs{-0.05cm} S {}^{} \big) \big(\Phi^\ast \hs{-0.02cm} \Phi\big) 
\nn[0.1cm]
&&
+\,
\sx{1.2}{\big(}
\lambda^{}_3 {}^{} X^3 \hs{-0.03cm} S^\ast +
\tfrac{1}{\sqrt2} {}^{}  \kappa {}^{} \upsilon^{}_\phi {}^{} S^2 \Phi +
\text{h.c.}^{}
\sx{1.2}{\big)}
~,
\end{eqnarray}
where $\upsilon^{}_\phi$ is the vacuum expectation value (VEV) of $\Phi$.\,\,The Hermiticity of the scalar potential ${\cal V}$ implies that $\mu_{h,X,S,\phi}^2$ and $\lambda_{h,X,S,\phi, h X, h S, h \phi, X \hs{-0.03cm} S, X \hs{-0.03cm} \phi, S \phi}$ must be real.\,\,For simplicity, we will choose 
$\,\lambda^{}_3\,$ and \,$\kappa$\, to be real and positive because one can redefine the scalar 
fields $X$ and $\Phi$ to absorb the phases of $\lambda^{}_3$ and $\kappa$.

Based on our setup, we require that the VEVs of the scalar fields in this model satisfy the following conditions\,\,:
\begin{eqnarray}
\langle H {}^{} \rangle 
\,=\,
\frac{1}{\sqrt2}
\begin{pmatrix}
0 \\ \,\upsilon^{}_h \,
\end{pmatrix}
~,\quad
\langle \Phi \rangle 
\,=\,
\frac{1}{\sqrt2} {}^{} \upsilon^{}_\phi
~,\quad
\langle X \rangle \,=\, \langle S \rangle \,=\, 0 
~,
\end{eqnarray}
where $\upsilon^{}_h\,\simeq\,246.22\,\,\rm{GeV}$ is the VEV of $H$.\,\,On the other hand, the $\kappa {}^{} \upsilon^{}_\phi$ terms in the potential cause the mass splitting of the real and imaginary part of the $S$ field.\,\,Thus, after spontaneously symmetry breaking, we can expand the  scalar fields around the VEVs as
\begin{eqnarray}
H \,=\, 
\frac{1}{\sqrt2}
\begin{pmatrix}
0 \\ \upsilon^{}_h + h'
\end{pmatrix}
~,\quad 
\Phi 
\,=\, 
\frac{1}{\sqrt2}\big(\upsilon^{}_\phi + \phi' {}^{}{}^{}\big) 
~,\quad 
S 
\,=\, 
\frac{1}{\sqrt2}\big(S^{}_\tx{R} + i S^{}_\tx{I}\big) 
~.
\end{eqnarray}
With these parametrizations, the minimum conditions for the scalar potential would give
\begin{eqnarray}\label{VEV}
\frac{\dd {\cal V}}{\dd \phi'}\bigg|_\tx{VEV} 
=
\upsilon^{}_\phi 
\Big({}^{}
\mu_\phi^2 + 
\lambda^{}_\phi \upsilon_\phi^2 + 
\tfrac{1}{2} \lambda^{}_{h \phi} \upsilon_h^2
\Big)
\,=\,
0
~,
\quad
\frac{\dd {\cal V}}{\dd h'}\bigg|_\tx{VEV} 
=
\upsilon^{}_h
\Big({}^{}
\mu_h^2 + 
\lambda^{}_h \upsilon_h^2 + 
\tfrac{1}{2} \lambda^{}_{h \phi} \upsilon_\phi^2
\Big)
\,=\,
0
~.
\end{eqnarray}
Solving these two equations, one can express the VEVs in terms of the quadratic and quartic couplings in the scalar potential as
\begin{eqnarray}
\upsilon^{}_\phi
\,=\,
\sqrt{
\frac{4 {}^{} \lambda^{}_h {}^{} \mu_\phi^2 - 2 {}^{} \lambda^{}_{h \phi} {}^{} \mu_h^2}
{\lambda_{h \phi}^2 - 4 {}^{} \lambda^{}_ h \lambda^{}_\phi}
}
~,\quad
\upsilon^{}_h
\,=\,
\sqrt{
\frac
{4 {}^{} \lambda^{}_\phi {}^{} \mu_h^2 - 2 {}^{} \lambda^{}_{h \phi} {}^{} \mu_\phi^2}
{\lambda_{h \phi}^2 - 4 {}^{} \lambda^{}_ h \lambda^{}_\phi}
}
~.
\end{eqnarray}
Besides, the masses of $X, S^{}_\tx{R}$, and $S^{}_\tx{I}$ are given by
\begin{eqnarray}
m_X^2
\,=\,
\mu_X^2 
+ 
\tfrac{1}{2} 
\sx{1.1}{\big(} 
\lambda^{}_{h X} {}^{} \upsilon_h^2 +
\lambda^{}_{X \hs{-0.03cm} \phi} \upsilon_\phi^2 {}^{} 
\sx{1.1}{\big)} 
~,\quad
m_{S_\tx{R},{}^{}S_\tx{I}}^2
\,=\,
\mu_S^2
+
\tfrac{1}{2}
\sx{1.1}{\big(} 
\lambda^{}_{h S}  {}^{} \upsilon_h^2 +
\lambda^{}_{S \phi} \upsilon_\phi^2 {}^{} 
\sx{1.1}{\big)}
\pm
\kappa {}^{} \upsilon_\phi^2
~.
\end{eqnarray}
Also, the $\lambda^{}_{h \phi}$ term in the scalar potential induces a mass mixing between the $h'$ and $\phi'$.\,\,In the basis $\big({}^{}{}^{}h' \,\,\, \phi'{}^{}{}^{}\big)\rx{\hs{-0.05cm}^\tf{T}}$, the corresponding mass mixing matrix is written as
\begin{eqnarray}\label{mixing}
M^2_{h \phi}
\,=\,
\begin{pmatrix}
2 {}^{} \lambda^{}_h \upsilon_h^2 & \lambda^{}_{h \phi} \upsilon^{}_h \upsilon^{}_\phi \\[0.1cm]
\,\lambda^{}_{h \phi} \upsilon^{}_h \upsilon^{}_\phi & 2 {}^{} \lambda^{}_\phi \upsilon_\phi^2
\end{pmatrix}
~.
\end{eqnarray}
Here we have used the relations in Eq.\,\eqref{VEV} to simplify the form of $M^2_{h \phi}$.\,\,Upon diagonalizing $M^2_{h \phi}$, we obtain the mass eigenstates $h$ and $\phi$ and their respective masses $m^{}_h$ and $m^{}_\phi$ given by
\begin{eqnarray}
\begin{pmatrix}
\, h' \, \\ \, \phi' \,
\end{pmatrix}
\,=\,
\begin{pmatrix}
\,\cos \alpha && -\sin\alpha \\
\,\sin\alpha && \cos \alpha
\end{pmatrix}
\begin{pmatrix}
\, h \, \\ \, \phi \,
\end{pmatrix}
\,\equiv\,
{\cal O}^{}_\alpha
\begin{pmatrix}
\, h \, \\ \, \phi \,
\end{pmatrix}
~,\quad
{\cal O}^\tf{T}_\alpha M^2_{h \phi} {\cal O}^{}_\alpha
\,=\, 
\tx{diag} \big( m_h^2 {}^{},{}^{} m_\phi^2 {}^{} \big)
~,
\end{eqnarray}
\vs{-0.3cm}
\begin{eqnarray}
m_{h,\phi}^2 
\,=\, 
\lambda^{}_h \upsilon_h^2 +
\lambda^{}_\phi \upsilon_\phi^2
\pm
\sqrt
{\big( \lambda^{}_h \upsilon_h^2 - \lambda^{}_\phi \upsilon_\phi^2 {}^{} \big)
\raisebox{0.5pt}{$\hs{-0.05cm}^2$} +
\big( \lambda^{}_{h \phi} \upsilon^{}_h \upsilon^{}_\phi {}^{} \big)
\raisebox{0.5pt}{$\hs{-0.05cm}^2$}}
~,\quad
\tan (2 {}^{} \alpha)
\,=\,
\frac
{\lambda^{}_{h \phi} \upsilon^{}_h \upsilon^{}_\phi}
{\lambda^{}_h \upsilon_h^2 - \lambda^{}_\phi \upsilon_\phi^2}
~,\quad
\end{eqnarray}
where $h$ denotes the observed Higgs boson with $m^{}_h \simeq 125.1\,\tx{GeV}$, and $\phi$ is a new neutral scalar with $m_\phi$ as a free parameter.\,\,In our study, we will assume that the mass splitting of $S^{}_\tx{R}$ and $S^{}_\tx{I}$ and the mass mixing of $h$ and $\phi$ are negligibly small for simplicity.\,\,In such cases, the masses of $S^{}_\tx{R}, S^{}_\tx{I}, h$, and $\phi$ are reduced to
\begin{eqnarray}
m_{S_\tx{R}}^2 \simeq\, m_{S_\tx{I}}^2
\equiv\,
m_S^2 
\,=\,
\mu_S^2
+
\tfrac{1}{2}
\sx{1.1}{\big(} 
\lambda^{}_{h S}  {}^{} \upsilon_h^2 + \lambda^{}_{S \phi} \upsilon_\phi^2 {}^{} 
\sx{1.1}{\big)}
~,\quad
m_h^2 
\,=\, 2 {}^{} \lambda^{}_h \upsilon_h^2
~,\quad
m_\phi^2 
\,=\, 2 {}^{} \lambda^{}_\phi \upsilon_\phi^2
~. 
\label{Scalar_mass}
\end{eqnarray}

The Lagrangian density responsible for the mass and the interactions of newly added dark fermion $N$ is given by
\begin{eqnarray}\label{Yukawa}
{\cal L}^{}_N 
\,=\, 
\overline{N} \big( i \gamma^\rho {\cal D}_\rho - m^{}_N \big) N 
-
\tfrac{1}{2}
\sx{1.1}{\big(} 
\,y^{}_N \overline{N\raisebox{0.5pt}{$^\tf{c}$}} \hs{-0.03cm} N S + \tx{h.c.}
\sx{1.1}{\big)}
~,
\end{eqnarray}
where $m^{}_N$ is the Dirac mass of $N$, $y^{}_N$ is the Yukawa coupling, and superscript $\tf{c}$ refers to the charge conjugation.\,\,Again, we will take $y^{}_N$ to be real and positive by absorbing its phase into the field $N$ or $S$ without loss of generality.\,\,Note that the $S$ particle can decay into a pair of $N$ if $m^{}_S > 2{}^{}m^{}_N$ and three $X$ particles if $m^{}_S > 3{}^{}m^{}_X$.\,\,Therefore, even $S$ has a $\,\Zf$ charge, it is still not suitable to be a DM candidate if $m^{}_S > 2 m^{}_N$ or $3{}^{}m^{}_X$.

The Lagrangian density for the ${}^{}\tx{SU}(2) \otimes \tx{U}(1)^{}_\tf{Y} \otimes \tx{U}(1)^{}_\tf{D}$ gauge bosons is given by
\begin{eqnarray}
{\cal L}_\tf{gauge}
\,=\, 
-\,\tfrac{1}{4} {}^{} W^{3 \rho\sigma} W^3_{\rho \sigma}
- \tfrac{1}{4} {}^{} B^{\rho \sigma} \hs{-0.05cm} B^{}_{\rho \sigma}
- \tfrac{1}{4} {}^{} C^{\rho \sigma} \hs{-0.03cm} C^{}_{\rho \sigma}
- \tfrac{1}{2} {}^{} s_\epsilon {}^{} B^{\rho \sigma} \hs{-0.03cm} C^{}_{\rho \sigma}
- \tfrac{1}{2} {}^{} m_C^2 {}^{}{}^{} C^{\rho} C^{}_{\rho}
~,
\end{eqnarray}
where $W^3_{\rho \sigma} = \partial^{}_\rho W^3_\sigma - \partial^{}_\sigma W^3_\rho + g^{}_\tf{W} W^1_{[\rho} W^2_{\sigma]}$, $B^{}_{\rho \sigma} = \partial^{}_\rho B^{}_\sigma - \partial^{}_\sigma B^{}_\rho$, and $C^{}_{\rho \sigma} = \partial^{}_\rho C^{}_\sigma - \partial^{}_\sigma C^{}_\rho$ are the field strength tensors of the gauge bosons, $s_\epsilon \equiv \sin \epsilon$ is the kinetic mixing parameter, and $m^{}_C = \frac{1}{2} {}^{} g^{}_\tf{D} \upsilon^{}_\phi$ coming from the $|{\cal D}^\rho \Phi|^2$ term after the $\tx{U}(1)^{}_\tf{D}$ symmetry breaking.

After the breakdown of the electroweak symmetry, the kinetic and the mass mixing matrices of the gauge fields in the basis $\big(B \,\,\, W^3 \, C{}^{}{}^{}\big)\rx{\hs{-0.05cm}^\tf{T}}$, are respectively given by
\begin{eqnarray}
K^{}_G
\,=\,
\begin{pmatrix}
1 & 0 & s_\epsilon\, \\
0 & 1 & 0 \\
\,\,s_\epsilon & 0 & 1 \\
\end{pmatrix}
~,\quad
M^2_G
\,=\,
\frac{1}{4}
\begin{pmatrix}
g_\tf{Y}^2 {}^{} \upsilon_h^2 & -{}^{}g^{}_\tf{W} {}^{} g^{}_\tf{Y} {}^{} \upsilon_h^2 & 0 \\[0.1cm]
-{}^{}g^{}_\tf{W} {}^{} g^{}_\tf{Y} {}^{} \upsilon_h^2 & g_\tf{W}^2 {}^{} \upsilon_h^2 & 0 \\[0.1cm]
0 & 0 & g_\tf{D}^2 \upsilon_\phi^2\,{}^{}{}^{} \\
\end{pmatrix}
~.
\end{eqnarray}
To write the kinetic terms into the canonical form, it is known that one can diagonalize matrix $K^{}_G$ without changing the diagonal elements by utilizing a general linear transformation ${\cal T}$, and   subsequently diagonalize $M^2_G$ by an orthogonal matrix ${\cal O}^{}_{\tf{W} \xi}$ as
\begin{eqnarray}\label{TO}
{\cal T} 
\,=\,
\begin{pmatrix}
\,\,1 & 0 & -{}^{}t_\epsilon{}^{}{}^{} \\
\,\,0 & 1 & 0 \\
\,\,0  & 0 & c_\epsilon \\
\end{pmatrix}
~,\quad
{\cal O}^{}_{\tf{W} \xi}
\,=\,
\begin{pmatrix}
\,c^{}_\tf{W} & -{}^{}s^{}_\tf{W} & 0\,{}^{} \\
\,s^{}_\tf{W} & c^{}_\tf{W} & 0\,{}^{} \\
\,0 & 0 & 1\,{}^{} \\
\end{pmatrix}
\hs{-0.2cm}
\begin{pmatrix}
\,\,1& 0 & 0\,{}^{} \\
\,\,0 & c^{}_\xi & -{}^{}s^{}_\xi  \\
\,\,0  & s^{}_\xi & c^{}_\xi  \\
\end{pmatrix}
~,
\end{eqnarray}
where $t_\epsilon \equiv \tan \epsilon{}^{}{}^{}, c_\epsilon \equiv \cos \epsilon$, and $c^{}_\theta \equiv \cos \theta$ and $s^{}_\theta \equiv \sin \theta$ with $\theta = \tf{W}, \xi$.\,\,Upon diagonalizing $M^2_G{}^{}$, we get the mass eigenstates of the gauge bosons $A, Z$, and $Z'$ as
\begin{eqnarray}\label{BWC}
\begin{pmatrix}
B \\ W^3 \\ C
\end{pmatrix}
\,=\,
{\cal T} {\cal O}^{}_{\tf{W} \xi}
\begin{pmatrix}
A \\ Z \\ \,Z'
\end{pmatrix}
~,\quad
\big( {\cal T} {\cal O}^{}_{\tf{W} \xi} \big)\rx{\hs{-0.05cm}^\tf{T}} 
M^2_G {}^{}{}^{} {\cal T}{\cal O}^{}_{\tf{W} \xi}
\,=\,
\tx{diag}\big( {}^{}{}^{} 0 {}^{},{}^{} m_Z^2 {}^{},{}^{} m_{Z'}^2 \big) ~,
\end{eqnarray}
\vs{-0.5cm}
\begin{eqnarray}\label{mixG}
\tan \tf{W} 
\,=\, \frac{g^{}_\tf{Y}}{g^{}_\tf{W}}
~,\quad
\tan (2{}^{}\xi) 
\,=\,
\frac{m_{\bar{Z}}^2 {}^{}{}^{} s^{}_{2\epsilon} {}^{}{}^{} s^{}_\tf{W}}
{m_{\bar{Z}}^2 \big( c_\epsilon^2 - s_\epsilon^2 s^2_\tf{W} \big) - m_C^2}
~,\quad
m_{\bar{Z}}^2 
\,=\, \tfrac{1}{2} \big({}^{}{}^{} g_\tf{W}^2 + g_\tf{Y}^2 \big) \upsilon_h^2
~,
\end{eqnarray}
where $s_\tf{W}^2 \simeq 0.23$, and the physical gauge boson masses are given by
\begin{eqnarray}\label{mG2}
m_A^2 
\,=\, 0
~,\quad
m_{\white{\bar{\black{Z}}}}^2 
\,=\, m_{\bar{Z}}^2 \big(1 + s^{}_\tf{W} {}^{} t_\epsilon {}^{} t_\xi {}^{} \big)
~,\quad
m_{Z'}^2 
\,=\, \frac{m_C^2}{c_\epsilon^2 \big(1 + s^{}_\tf{W} {}^{} t_\epsilon {}^{} t_\xi {}^{} \big)}
\end{eqnarray}
with $t^{}_\xi \equiv \tan \xi$.\,\,Here $A$ and $Z$ are the photon and neutral massive gauge boson in the SM, respectively, and $Z'$ is a new massive gauge boson in the dark sector.

As we shall see soon, we are interested in the case where $\epsilon \ll 1$ and $m_Z^2 \gg m_{Z'}^2$, with which the second equation in Eq.\,\eqref{mixG} with Eq.\,\eqref{mG2} is reduced to
\begin{eqnarray}
t^{}_\xi  
\,\simeq\,   
s^{}_\xi  
\,\simeq\,   
\frac{m_{\bar{Z}}^2}{m_{\bar{Z}}^2-m_C^2} {}^{} s^{}_\tf{W} \epsilon
\,\simeq\,   
\frac{m_Z^2}{m_Z^2 - m_{Z'}^2} {}^{} s^{}_\tf{W} \epsilon
\,\simeq\, 
s^{}_\tf{W} \epsilon
~.
\end{eqnarray}
With this approximation and Eqs.\,\eqref{TO} and \eqref{BWC}, the covariant derivative (here we only show the dark gauge interaction) becomes
\begin{eqnarray}\label{Drho}
{\cal D}_\rho 
\,\supset\,
i \big({}^{}{}^{}
g^{}_\tf{D} {\cal Q}^{}_\tf{D} -
g^{}_e  {}^{}{}^{} c^{}_\tf{W} \epsilon {}^{} {\cal Q}^{}_e
\big) Z_\rho' 
~,
\end{eqnarray}
where ${\cal Q}^{}_e = \frac{1}{2} \tau^3 + {\cal Q}^{}_Y$ is the electromagnetic charge in unit $g^{}_e = g^{}_\tf{W} s^{}_\tf{W} \sim 0.3$.\,\,This interaction is crucial when we discuss the kinetic equilibrium between the dark sector and the SM sector.

%%%%%%%%%%%%%%%%%%%%%%%%%%%%%%%%%%%%%%%%%%%%%%%%%

\section{Annihilation cross sections in dark sector}\label{sec:3}

\begin{figure}[t!]
\hs{0.2cm}
\centering
\includegraphics[width=0.31\textwidth]{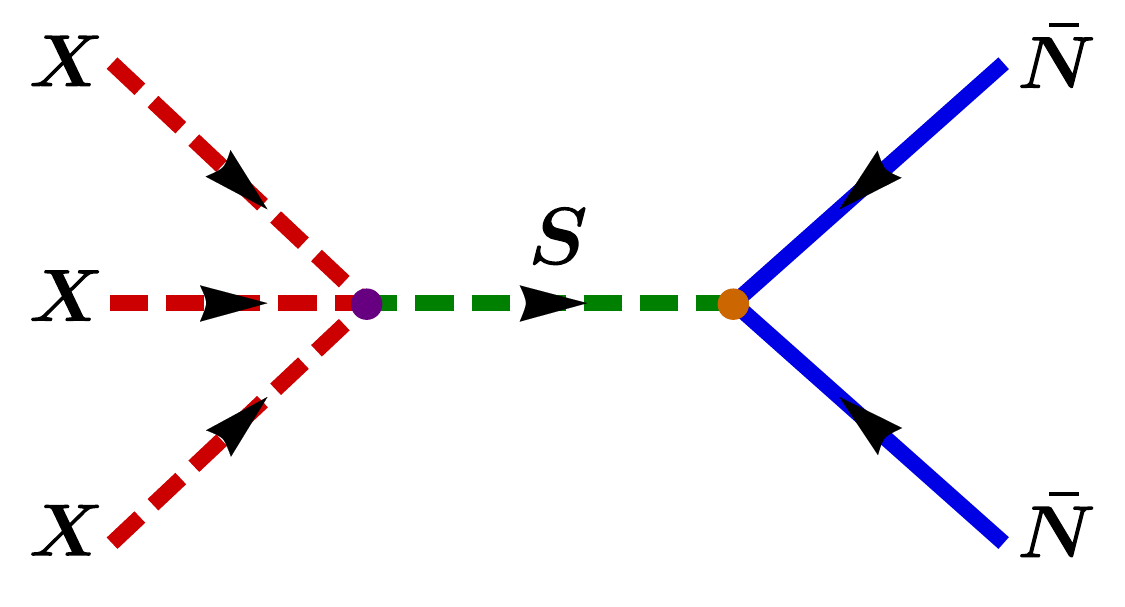}
\hs{0.2cm}
\includegraphics[width=0.31\textwidth]{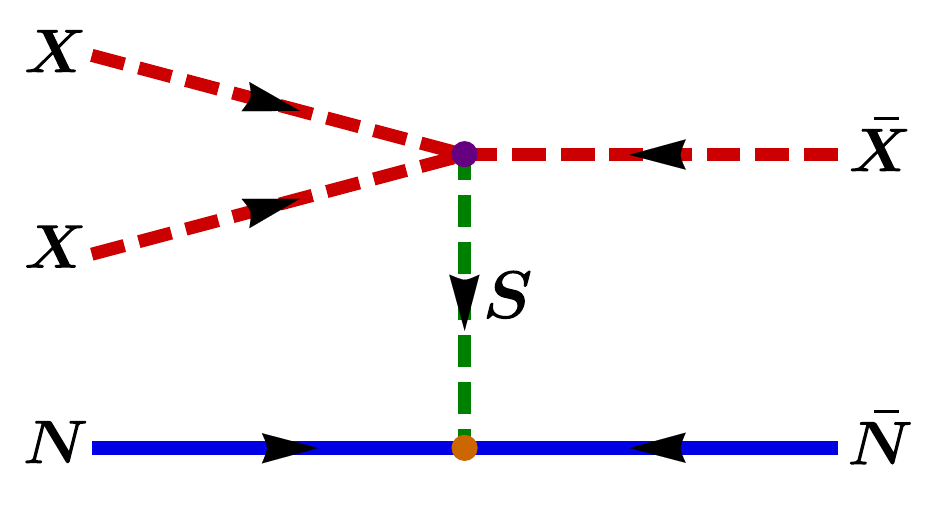}
\hs{0.2cm}
\includegraphics[width=0.31\textwidth]{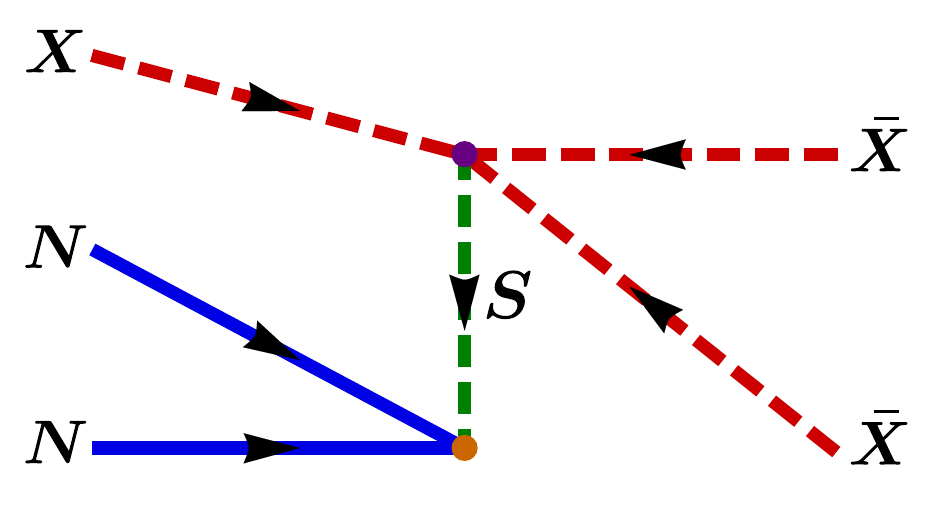}
\\[0.5cm]
\includegraphics[width=0.41\textwidth]{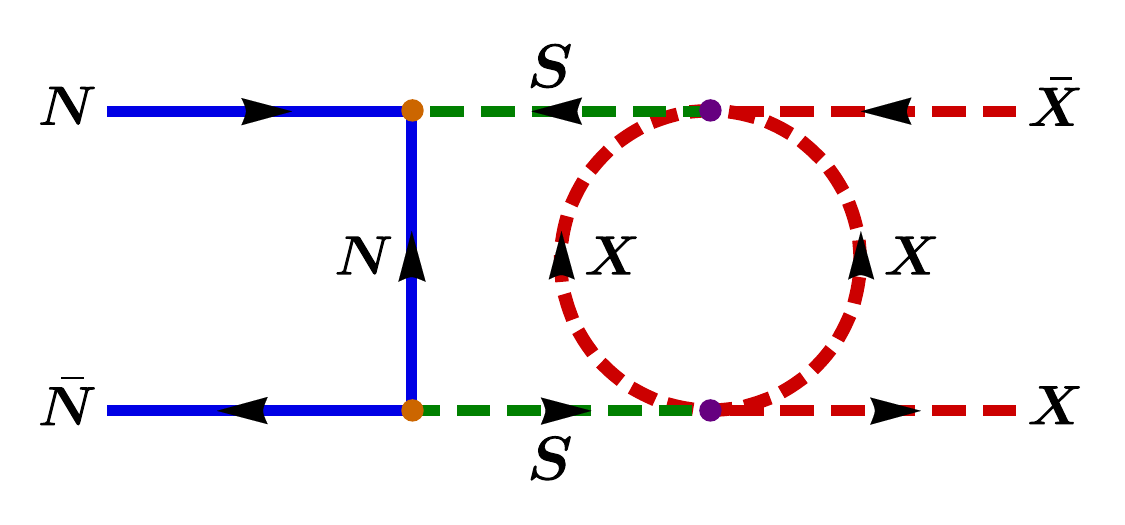}
\hs{0.3cm}
\includegraphics[width=0.41\textwidth]{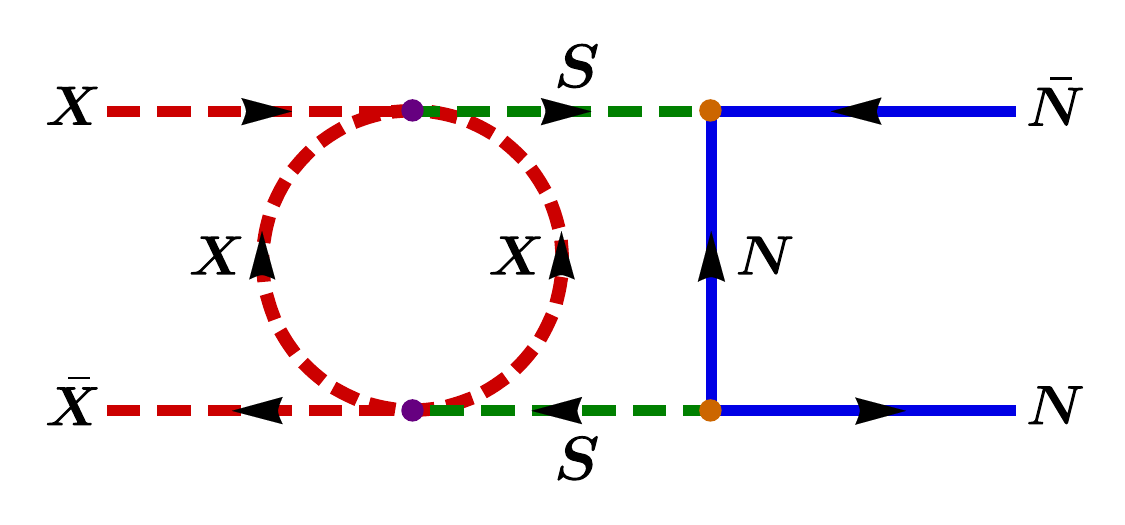}
\vs{-0.3cm}
\caption{The Feynman diagrams of the $3 \to 2$ and $2 \to 2$ processes in the $r$SIMP model, where the arrows represent the direction of dark charge flow.}
\label{fig:ann}
\end{figure}

In this section, we will present the formulas for the annihilation cross sections of $3 \to 2$ and $2 \to 2$ processes in the dark sector.\,\,The detailed derivations for these cross sections can be found in the appendices.\,\,The relevant Lagrangian of the $3 \to 2$ and $2 \to 2$ processes is given by
\begin{eqnarray}
{\cal L}^{}_\tf{ann}
\,=\,
-\,\lambda^{}_3
\sx{1.2}{\big[}
X^3 \hs{-0.03cm} S^\ast + (X^\ast)^3 \hs{-0.03cm} S {}^{}{}^{}
\sx{1.2}{\big]}
-\tfrac{1}{2}{}^{}{}^{}y^{}_N 
\sx{1.1}{\big(} \,{}^{}
\overline{N\raisebox{0.5pt}{$^\tf{c}$}} \hs{-0.03cm} N S +
\overline{N} N\raisebox{0.5pt}{$^\tf{c}$} \hs{-0.03cm} S^\ast
\sx{1.1}{\big)}
~.
\end{eqnarray}
With these interactions and the \,U$(1)^{}_\tf{D}$ charge conservation, the possible $3 \to 2$ processes are $\XXXNN, \XXNXN$, and $\XNNXX$ (here we have omitted their charge conjugation processes).\,\,For these processes to take place, the masses of $X$ and $N$ should satisfy the relation $3{}^{}m^{}_X > 2{}^{}m^{}_N > m^{}_X$, under which the $2 \to 3$ and $2 \to 4$ processes such as $\NNXXX$ and $\XXNNNN$, etc., are kinematically forbidden.\,\,On the other hand, the $2 \to 2$ processes $\NNXX$ or $\XXNN$ can be induced via the two-loop diagrams.\,\,The Feynman diagrams of these $3 \to 2$ and $2 \to 2$ processes are depicted in Fig.\,\ref{fig:ann}.

First, the non-thermally-averaged $3 \to 2$ annihilation cross sections are computed as
\begin{eqnarray}
(\sigma v^2)_{\hs{-0.03cm}\XXXNN}
\Eq
\frac{\lambda^2_3 {}^{}{}^{} y^2_N}{128 {}^{} \pi {}^{} m^5_X} 
\frac{\big({}^{}9 - 4{}^{} r_N^2\big)^{\hs{-0.05cm}3/2}}{\big({}^{}9 - r_S^2{}^{}\big)
\raisebox{1pt}{$^{\hs{-0.05cm}2}$}}
\label{XXXNN}
~,\quad
\\[0.15cm]
(\sigma v^2)_{\hs{-0.03cm}\XXNXN}
\Eq
\frac{9\sqrt{3}\,\lambda^2_3 {}^{}{}^{} y^2_N}{32{}^{}\pi{}^{}m^5_X} 
\frac{\big(1 + r^{}_N\big)\big(1 + 2{}^{}r^{}_N + 2{}^{}r_N^2 \big) \sqrt{3 + 8{}^{}r^{}_N +4{}^{}r_N^2}}
{\big(2 + r^{}_N\big)\raisebox{1pt}{$^{\hs{-0.05cm}2}$} \sx{1.0}{\big[} r_S^2\big(1 + r^{}_N\big) + 
2{}^{}r^{}_N \sx{1.0}{\big]}\raisebox{1pt}{$^{\hs{-0.05cm}2}$}}
~,
\label{XXNXN}
\end{eqnarray}
where $r^{}_{N,S} \equiv m^{}_{N,S}/m^{}_X$, and we demand that $3/2 > r^{}_N > 1/2$ and $r^{}_S > 2{}^{}r^{}_N$.\,\,Notice that the $(\sigma v^2)_{\hs{-0.03cm}\XNNXX} = {\cal O}(v^2)$ is $p\,$-wave suppressed.\,\,Here we have applied the Feynman rules of fermion-number-violating interactions to derive these cross sections \cite{Denner:1992vza}.\,\,In our study, we will consider the resonant effect for SIMP DM \cite{Choi:2016hid,Ho:2017fte}, where $r^{}_S \simeq 3$, to reduce the values of $\lambda^{}_3$ and $y^{}_N$ to escape from the perturbative bounds.\,\,For the resonant SIMP DM, we have to adopt the Breit-Wigner form for \eqref{XXXNN} with a nonvanishing velocity of DM in the center of mass energy, $(p^{}_1 + p^{}_2 + p^{}_3)^2 \simeq 9{}^{}m_X^2\big( 1+ 2\beta/3\big)$, as \cite{Gondolo:1990dk,Choi:2016hid}
\begin{eqnarray}
(\sigma v^2)^\tf{BW}_{\hs{-0.03cm}\XXXNN}
\Eq
\frac{c^{}_X}{m_X^5}
\frac{\gamma^2_S}{\big(\epsilon^{}_S - 2\beta/3 {}^{} \big)\raisebox{-0.5pt}{$^{\hs{-0.03cm}2}$} + \gamma_S^2}
~,
\quad
c^{}_X
\,=\,
\frac{2\pi {}^{} \lambda_3^2}{y_N^2}
\frac
{r_S^2 \big({}^{}9 - 4{}^{} r_N^2\big)^{\hs{-0.05cm}3/2}}{\big({}^{}r_S^2 - 4{}^{}r_N^2\big)
\raisebox{1pt}{$^{\hs{-0.05cm}3}$}}
~,
\end{eqnarray}
where $\beta \equiv \frac{1}{2}\big(v_1^2+v_2^2+v_3^2 {}^{} \big)$ with $v^{}_i$ the speeds of three initial $X$ particles.\,\,In this expression, the $\epsilon^{}_S$ indicates the level of degeneracy between $m^{}_S$ and $3{}^{}m^{}_X$, and the $\gamma^{}_S$ is the normalized dimensionless width of the resonance, respectively\,:
\begin{eqnarray}
\epsilon^{}_S
&\equiv&
\frac{m_S^2 - 9{}^{}m_X^2}{9{}^{}m_X^2}
\,=\,
\frac{r_S^2}{9}-1
~,\quad
\\[0.1cm]
\gamma^{}_S
&\hs{-0.2cm}\equiv\hs{-0.2cm}&
\frac{m^{}_S \Gamma^{}_S}{9{}^{}m_X^2}
\,=\,
\frac{y_N^2 r_S^2}{144{}^{}\pi}
\bigg(1-\frac{4{}^{}r_N^2}{r_S^2}\bigg)^{\hs{-0.15cm}3/2}
~.
\end{eqnarray}
Here the decay rate of the $S$ particle is given by\footnote{As mentioned in the previous section, the $S$ particle can also decay into three $X$ particles if it is kinematically allowed.\,\,However, since we are interested in the mass region where $m^{}_S \simeq 3{}^{}m^{}_X$, the decay rate of $S \to \bar{X}\hs{-0.03cm}\bar{X}\hs{-0.03cm}\bar{X}$ is then suppressed by phase space even if $\lambda^{}_3 \sim {\cal O}(10)$.\,\,Thus, we ignore this decay mode in our numerical study.}
\begin{eqnarray}
\Gamma^{}_S 
\,=\, 
\Gamma \big(S \to \bar{N}\hs{-0.02cm}\bar{N} {}^{} \big)
\,=\,
\frac{y_N^2 m^{}_S}{16{}^{}\pi}
\bigg(1-\frac{4{}^{}m_N^2}{m_S^2}\bigg)^{\hs{-0.15cm}3/2}
~.
\end{eqnarray}
Employing the formula in Ref.\,\cite{Choi:2017mkk}, the annihilation cross section for the process $\XXXNN$ near the resonance with thermal average is then
\begin{eqnarray}
\langle \sigma v^2 \rangle_{\hs{-0.03cm}\XXXNN}
\,=\,
\frac{x^3}{2} \hs{-0.03cm}
\int_{\hs{-0.03cm}0}^{\infty} \hs{-0.05cm}
\dd \beta \,(\sigma v^2)^\tf{BW}_{\hs{-0.03cm}\XXXNN} \,
\beta^2  \exp\hs{-0.08cm}\big({-}{}^{}x \beta {}^{} \big)
~,
\end{eqnarray}
where $x \equiv m^{}_X/T$ is the dimensionless cosmic time variable with $T$ being the thermal plasma temperature.\,\,For the process $\XXNXN$, we simply take $\langle\sigma v^2\rangle_{\hs{-0.03cm}\XXNXN} \simeq (\sigma v^2)_{\hs{-0.03cm}\XXNXN}$.

%%%%%%%%%%%%%%%%%%%%%%%%%%%%%%%%%%%%%%%%%%%%%%%%%

\begin{figure}[t!]
\hs{0.1cm}
\centering
\includegraphics[width=0.477\textwidth]{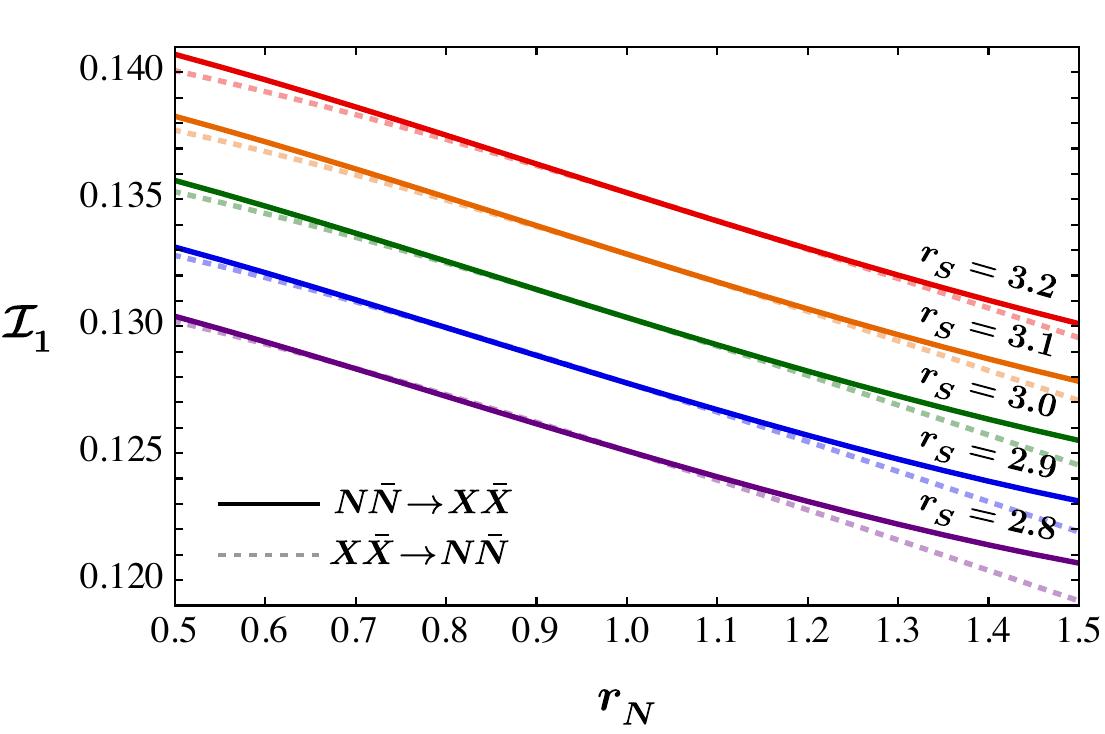}
\hs{0.2cm}
\includegraphics[width=0.47\textwidth]{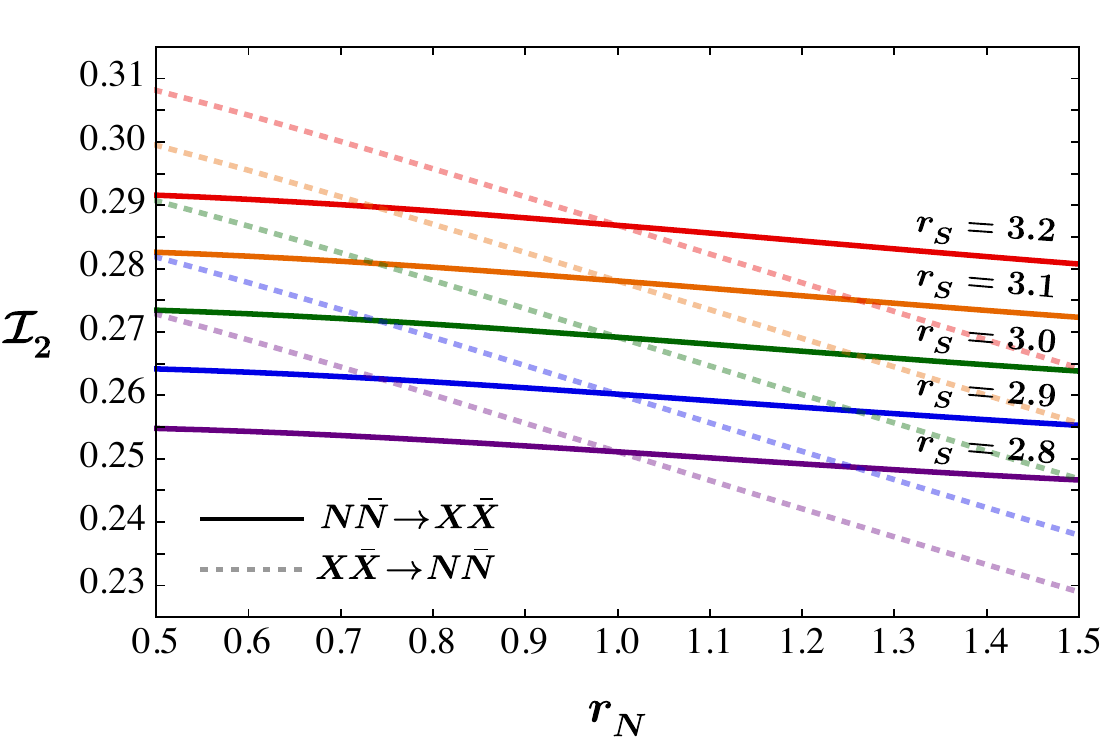}
\vs{-0.3cm}
\caption{The two-loop functions $\,{\cal I}^{}_1$ and $\,{\cal I}^{}_2$ as functions of $r^{}_N$ with different choices of $r^{}_S$ near the resonance.\,\,As indicated, the ${\cal I}^{}_{1,2} \sim {\cal O}(0.1)$ in the mass range of interest.}
\label{fig:I1I2}
\end{figure}

\begin{figure}[t!]
\hs{0.1cm}
\centering
\includegraphics[width=0.41\textwidth]{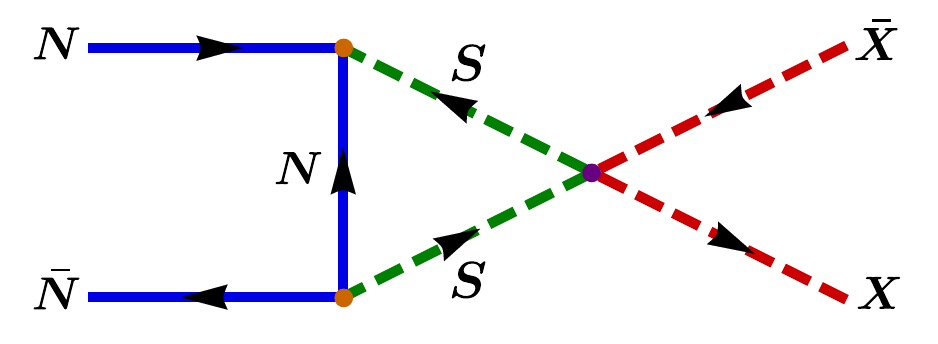}
\hs{0.2cm}
\includegraphics[width=0.35\textwidth]{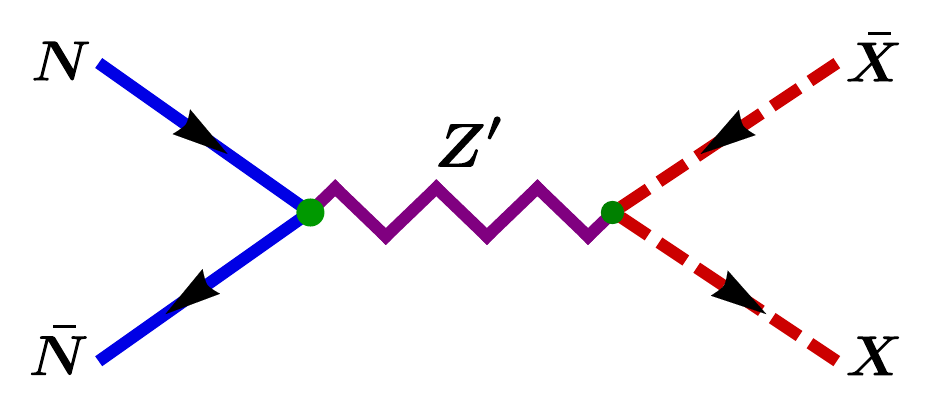}
\caption{The $\NNXX$ process through the one-loop and $Z'$-mediated tree diagrams.}
\label{fig:NNXXZ}
\end{figure}

Next, the thermally-averaged cross sections for the two-loop induced $2 \to 2$ processes are calculated as
\begin{eqnarray}
\hs{-0.8cm}
\langle \sigma v \rangle^{2\tf{-loop}}_{\hs{-0.03cm}\NNXX}
&\hs{-0.2cm}=\hs{-0.2cm}&
\frac
{81{}^{} \lambda_3^4 {}^{}{}^{} y_N^4 \sqrt{r_N^2-1}}
{\pi{}^{}(4\pi)^8{}^{}r_S^4{}^{}{}^{}m_X^2 r^{}_N}
\Bigg[\hs{-0.03cm}
\big(r_N^2-1\big) |{}^{}{}^{}{\cal I}^{}_1|^2 +
\frac
{\big(11-2{}^{}r_N^2\big) |{}^{}{}^{}{\cal I}^{}_1|^2 + 6{}^{}r_N^2|{}^{}{}^{}{\cal I}^{}_2|^2}
{4{}^{}x} 
\Bigg]
\label{cNNXX}
\,,
\\[0.2cm]
\hs{-0.8cm}
\langle \sigma v \rangle^{2\tf{-loop}}_{\hs{-0.05cm}\XXNN}
&\hs{-0.2cm}=\hs{-0.2cm}&
\frac
{81{}^{} \lambda_3^4 {}^{}{}^{} y_N^4 {}^{} r_N^2 \sqrt{1- r_N^2}}
{\pi{}^{}(4\pi)^8{}^{}r_S^4{}^{}{}^{}m_X^2}
\Bigg[\hs{-0.03cm}
\big(1-r_N^2\big) |{}^{}{}^{}{\cal I}^{}_2|^2 +
\frac
{2\big(1+2{}^{}r_N^{-2}\big) |{}^{}{}^{}{\cal I}^{}_1|^2 + 
3\big(5{}^{}r_N^2-2\big)|{}^{}{}^{}{\cal I}^{}_2|^2}
{4{}^{}x}
\Bigg]
\label{cXXNN}
\,,
\end{eqnarray}
where $\,{\cal I}^{}_{1,2} = {\cal I}^{}_{1,2}(r^{}_N,r^{}_S)$ are two-loop functions in the form of quintuple integrals as
\begin{eqnarray}\label{I1I2}
{\cal I}^{}_{1,2}(r^{}_N,r^{}_S) 
\,=\,
\int_{\hs{-0.02cm}0}^1 \hs{-0.05cm} \dd z^{}_1
\int_{\hs{-0.02cm}0}^1 \hs{-0.05cm} \dd z^{}_2
\int_{\hs{-0.02cm}0}^{1-z^{}_2} \hs{-0.05cm} \dd z^{}_3  
\int_{\hs{-0.02cm}0}^{{}^{}z^{}_1(1-z^{}_1)} \hs{-0.05cm} \dd z^{}_4
\int_{\hs{-0.02cm}0}^1 \hs{-0.05cm} \dd z^{}_5  
\,{}^{}{\cal F}^{}_{1,2}(r^{}_N,r^{}_S) 
\end{eqnarray}
with
\begin{eqnarray}
{\cal F}^{}_1(r^{}_N,r^{}_S) 
\Eq
\frac
{r_S^2 {}^{} z_5^2 
\sx{1.1}{\big[} 2P^2 z_5^3 - 
\big(P^2 + 3{}^{}Q^2 \big) z_5^2 + 
\big(2{}^{}Q^2 + 3 \big) z^{}_5 - 2 {}^{} 
\sx{1.1}{\big]}}
{2 \big(P^2 z_5^2 - Q^2 z^{}_5 + 1 \big)\raisebox{1pt}{$^{\hs{-0.05cm}2}$}}
~,
\\[0.15cm]
{\cal F}^{}_2(r^{}_N,r^{}_S) 
\Eq
\frac{r_S^2 (1-z^{}_2-z^{}_3) {}^{} z_5^3 \big(2P^2 z_5^2 - 3{}^{}Q^2 z^{}_5 + 3 \big)}
{2 \big(P^2 z_5^2 - Q^2 z_5 + 1 \big)\raisebox{1pt}{$^{\hs{-0.05cm}2}$}}
~,
\end{eqnarray}
\vs{-0.3cm}
\begin{eqnarray}
P^2
\Eq
\begin{cases}
\,z^{}_4 {}^{} \sx{1.1}{\big[} {}^{} r_N^2 (z^{}_2-z^{}_3+1)(z^{}_2-z^{}_3-1) + 1 \sx{1.1}{\big]}
&\,\,\text{for}  \quad \NNXX
\\[0.3cm]
\,z^{}_4 {}^{} \sx{1.1}{\big[} {}^{} r_N^2 (z^{}_2+z^{}_3 - 1)^2 - (2z^{}_2 - 1)(2z^{}_3 -1) \sx{1.1}{\big]} 
&\,\,\text{for}  \quad \XXNN
\end{cases}
~,\quad
\\[0.2cm]
Q^2
\Eq
\begin{cases}
\,1 + z^{}_4 {}^{} \sx{1.1}{\big[} 2{}^{}r_N^2 (z^{}_2+z^{}_3-1) - r_S^2 (z^{}_2+z^{}_3) + 1\sx{1.1}{\big]}
&\text{for}  \quad \NNXX
\\[0.3cm]
\,1 + z^{}_4 {}^{} \sx{1.1}{\big[} \big(2 - r_S^2\big) (z^{}_2+z^{}_3) -1 \sx{1.1}{\big]}
&\text{for}  \quad \XXNN
\end{cases}
~.
\end{eqnarray}
We present the typical values of $\,{\cal I}^{}_1$ and $\,{\cal I}^{}_2$ for $r^{}_S \simeq 3$ and $3/2 > r^{}_N > 1/2$ in Fig.\,\ref{fig:I1I2}.\,\,Note that the $\langle \sigma v \rangle^{2\tf{-loop}}_{\hs{-0.03cm}\NNXX}$ and $\langle \sigma v \rangle^{2\tf{-loop}}_{\hs{-0.03cm}\XXNN}$ are dominated by the $p\,$-wave contributions if the masses of $N$ and $X$ are degenerate.\,\,It is worth mentioning that the $2 \to 2$ annihilation cross sections in Eqs. \eqref{cNNXX} and \eqref{cXXNN} are in agreement with the ones derived by the effective operator approach, where we introduce $c/(2!\Lambda) X^3 \overline{N\raisebox{0.5pt}{$^\tf{c}$}} \hs{-0.03cm} N$ with $c$ the coupling constant and $\Lambda$ the cutoff scale of the theory \cite{Ho:2021ojb}.\footnote{For instance, in the case of $\NNXX$ with $r^{}_N \simeq 1$ and $\Lambda \sim m^{}_S \simeq 3{}^{}m^{}_X$, the two-loop induced annihilation cross sections in the UV complete model and the effective theory are approximately given by
\begin{eqnarray}
\langle \sigma v \rangle^\tf{UV}_{\hs{-0.03cm}\NNXX}
\approx
\frac
{243{}^{} \lambda_3^4 {}^{}{}^{} y_N^4 \sqrt{r_N^2-1}}
{2\pi{}^{}(4\pi)^8{}^{}x{}^{}m_X^2}
\bigg(\frac{m^{}_X}{m^{}_S}\bigg)^{\hs{-0.13cm}4}
|{}^{}{}^{}{\cal I}^{}_2|^2
~,\quad
\langle \sigma v \rangle^\tf{EFT}_{\hs{-0.03cm}\NNXX}
\approx
\frac
{243{}^{}{}^{}c^4 \sqrt{r_N^2-1}}
{2\pi{}^{}(4\pi)^8{}^{}x{}^{}m_X^2}
\bigg(\frac{m^{}_X}{\Lambda}\bigg)^{\hs{-0.13cm}4}
|{}^{}{}^{}{\cal I}^{}_\Lambda|^2
~,
\end{eqnarray}
where $\,{\cal I}^{}_2 = {\cal I}^{}_2\big(r^{}_N=1,r^{}_S=3\big) \simeq 0.27$ and $\,{\cal I}^{}_\Lambda = {\cal I}^{}_\Lambda\big(r^{}_S = 3\big) \simeq 0.45$ \cite{Ho:2021ojb}.
} In fact, the $X$ and $N$ particles can also annihilate each other via one-loop diagrams with the $\lambda_{X \hs{-0.03cm} S}$ term and $Z'$-mediated diagrams with the dark gauge coupling as shown in Fig.\,\ref{fig:NNXXZ}.\,\,We will discuss their effects in Sec.\,\ref{sec:5} and Sec.\,\ref{sec:6}, respectively.

%%%%%%%%%%%%%%%%%%%%%%%%%%%%%%%%%%%%%%%%%%%%%%%%%

\section{Theoretical \& Experimental constraints}\label{sec:4}

In this section, we take into account various theoretical and experimental restrictions on the masses and couplings of the new particles in the $r$SIMP model.

Theoretically, the quartic, Yukawa, and dark gauge couplings are subject to the conditions of perturbativity.\,\,We impose that \cite{Choi:2021yps,Perez:2021rbo,Allwicher:2021rtd}
\begin{eqnarray}
\lambda^{}_k < 4{}^{}\pi 
~,\quad
y^{}_N < \sqrt{8{}^{}\pi} 
~,\quad
g^{}_\tf{D}
< 4{}^{}\pi 
~,
\end{eqnarray}
where $k = \{h,X,S,\phi, h X, h S, h \phi, X \hs{-0.03cm} S, X \hs{-0.03cm} \phi, S \phi\}$.\,\,Besides, the thermally-averaged annihilation cross sections are bounded from above by partial wave unitarity, which can place bounds on the couplings for given masses.\,\,In the non-relativistic limit, we require that \cite{Namjoo:2018oyn}
\begin{eqnarray}
\langle \sigma v^2 \rangle_{\hs{-0.03cm}\XXXNN}
\leqslant
\frac{192 \sqrt{3} {}^{}{}^{} \pi^2 x^2}{m_X^5}
~,\quad
\langle \sigma v^2 \rangle_{\hs{-0.03cm}\XXNXN}
\leqslant
\frac{16 {}^{}{}^{} \pi^2 x^2}{m_X^5}
\bigg(\hs{-0.05cm} 1 + \frac{2}{r^{}_N}\bigg)^{\hs{-0.15cm}3/2}
~,
\end{eqnarray}
\vs{-0.5cm}
\begin{eqnarray}
\langle \sigma v^2 \rangle_{\hs{-0.03cm}\XNNXX}
\leqslant
\frac{4 {}^{}\pi^2 x^2}{m_X^5}
\bigg(\frac{1}{r_N^2} + \frac{2}{r^{}_N}\bigg)^{\hs{-0.15cm}3/2}
~,
\end{eqnarray}
\vs{-0.5cm}
\begin{eqnarray}
\langle \sigma v \rangle_{\hs{-0.03cm}\NNXX}
\leqslant
\frac{4 {}^{} \sqrt{\pi {}^{} x}}{m_X^2 r_N^{3/2}}
~,\quad
\langle \sigma v \rangle_{\hs{-0.03cm}\XXNN}
\leqslant
\frac{64 {}^{} \sqrt{\pi {}^{} x}}{m_X^2}
~,
\end{eqnarray}
here ${}^{}x{}^{}$ will be set at the freeze-out time of DM.\,\,On the other hand, the quartic couplings must satisfy certain relations to stabilize the vacuum at large scalar field values, where the potential energy ${\cal V}$ is bounded from below.\,\,For simplicity, we only focus on the potential including the $X$ and $S$ fields, and assume that other quartic couplings are negligible but positive.\,\,Under these considerations, we found that \cite{Choi:2016tkj}
\begin{eqnarray}
\hs{-0.2cm}
\lambda^{}_{X,{}^{}{}^{}S} > 0
~,\quad
\lambda^{}_{X\hs{-0.03cm}S} + 2 \sqrt{\lambda^{}_X \lambda^{}_S} {}^{} > 0
~,\quad
|\lambda^{}_3| <
\sqrt{
\frac{
\big(12{}^{}\lambda^{}_X \lambda^{}_S + \lambda^{2}_{X\hs{-0.03cm}S}\big)\raisebox{1pt}{$\hs{-0.05cm}^{3/2}$} +
36{}^{}\lambda^{}_X \lambda^{}_S \lambda^{}_{X\hs{-0.03cm}S} -
\lambdaXS^3}
{54{}^{}\lambda^{}_S}
}
~.
\end{eqnarray}
In particular, the above conditions are reduced to $\lambda^{}_{X,{}^{}{}^{}S} > 0$ and $|\lambda^{}_3| < \big(16{}^{}\lambda^3_X \lambda^{}_S/27 \big)^{\hs{-0.05cm}1/4}$
in the limit $\lambdaXS \to 0$, which turn out to be a stringent constraint in this model.\,\,Notice that these conditions also ensure that $\langle X \rangle = \langle S \rangle = 0$. 

Cosmologically, the light DM would contribute to the effective number of neutrino species, $N^{}_\tf{eff}$.\,\,Assuming the entropy of the universe is conserved and considering the DM particles mainly interact with electrons and positrons, the $N^{}_\tf{eff}{}^{}$ at the CMB temperature is estimated as \cite{Boehm:2013jpa}
\begin{eqnarray}\label{Neff}
\hs{-0.6cm}
N^{}_\tf{eff} {}^{} \big(T^{}_\tf{CMB}\big)
\,=\,
\scalebox{1.2}{\bigg[}
1 + \frac{4}{11} 
\hs{-0.05cm}
\sum_{j{}^{}={}^{}X,N} g^\tf{DM}_{\star s}\big(m^{}_j,T_{\nu\tf{d}}\big)
\hs{-0.05cm}
\scalebox{1.2}{\bigg]}^{\hs{-0.1cm}-{}^{}4/3}
N^\tf{SM}_\tf{eff}\big(T^{}_\tf{CMB}\big)
~,
\end{eqnarray}
where $N^\tf{SM}_\tf{eff}\big(T^{}_\tf{CMB}\big) = 3.044$ in the SM \cite{Bennett:2020zkv,Akita:2020szl}, and $g^\tf{DM}_{\star s}\big(m^{}_j,T_{\nu\tf{d}}\big)$ counts the DM entropy degrees of freedom at neutrino decoupling temperatures, $T_{\nu \tf{d}} \hs{-0.03cm} \simeq 2\,\text{MeV}$\,\,\cite{Escudero:2018mvt}, which has the form as \cite{Lehmann:2020lcv}
\begin{eqnarray}\label{gDM}
g^\tf{DM}_{\star s}\big(m^{}_j,x\big)
\,=\,
\frac{15{}^{}g^{}_j}{4\pi^4}
\mathop{\mathlarger{\int}_{\hs{-0.03cm}r^{}_j{}^{}x}^{{}^{}\infty}} \hs{-0.05cm} dw \, 
\frac{\big(4w^2 - r_j^2{}^{}x^2{}^{}\big)\big(w^2 - r_j^2{}^{}x^2{}^{}\big)^{\hs{-0.05cm}1/2}}{e^w \pm 1}
\end{eqnarray}
with $g^{}_j$ the internal degrees of freedom of particle $j$.\,\,The latest measurement from the Planck satellite gives $N^{}_\tf{eff} = 2.99^{+0.34}_{-0.33}$ (95\% C.L.) \cite{Aghanim:2018eyx}, which can provide lower bounds for DM masses. As we will see in the next section, the masses of DM should be near degenerate in this model. Using Eqs.\,\eqref{Neff} and \eqref{gDM} with this property, we suggest that the $m^{}_{X,N} \gtrsim {\cal O}(10)\,\tx{MeV}$.\,\,Another cosmological constraint is the observed relic abundance of DM. We will discuss it as well in the next section.

For the gauge sector, there is a constraint for the kinetic mixing parameter, depending on the mass of the dark gauge boson.\,\,In this model, the $Z'$ mainly decays into invisible particles, $Z' \to \XX,\NN$, and $S\hs{-0.01cm}\bar{S}$.\,\,Also, we will concentrate on the $Z'$ with a few hundred MeV mass.\,\,In these circumstances, the measurements from the BaBar collaboration cap $\epsilon \lesssim 10^{-3}$ \cite{BaBar:2017tiz,Fabbrichesi:2020wbt}.

%%%%%%%%%%%%%%%%%%%%%%%%%%%%%%%%%%%%%%%%%%%%%%%%%

\section{Relic abundance of DM and reshuffled effect}\label{sec:5}
To estimate the current density of DM in the $r$SIMP model, one has to numerically solve
the coupled Boltzmann equations for the comoving number yields $Y_X$ 
and $Y_N$.\,\,Assuming there is no asymmetry in DM, namely 
$Y_{\white{\bar{\black{X}}}} = Y_{\bar{X}}$ and $Y_{\white{\bar{\black{N}}}} = Y_{\bar{N}}$, 
the Boltzmann equations are given by \cite{Ho:2021ojb}
\begin{eqnarray}
\frac{\dd Y^{}_X}{\dd x}
\Eq
-{}^{}\frac{s(x)^2}{H(x){}^{}x}
\Bigg\{
12 {}^{} \langle \sigma v^2 \rangle_{\hs{-0.03cm}\XXXNN} \hs{-0.01cm}
\sx{1.1}{\bigg[} 
Y^3_X - Y_N^2 \frac{(Y^\tf{eq}_X)^3}{(Y^\tf{eq}_N)^2} 
\sx{1.1}{\bigg]} \hs{-0.05cm}
+ 
2 {}^{} \langle \sigma v^2 \rangle_{\hs{-0.03cm}\XXNXN} {}^{}{}^{}
Y^{\white{q}}_X Y^{\white{q}}_N
\sx{1.1}{\big(} {}^{}
Y^{}_X - Y^\tf{eq}_X
\sx{1.1}{\big)}
\nn
&&\hs{2.0cm}
{-} {}^{}{}^{} \langle \sigma v^2 \rangle_{\hs{-0.03cm}\XNNXX} {}^{}{}^{}
Y^{}_X \hs{-0.05cm}
\sx{1.1}{\bigg[} 
Y_N^2 - Y^{}_X \frac{(Y^\tf{eq}_N)^2}{Y^\tf{eq}_X} 
\sx{1.1}{\bigg]} \hs{-0.05cm}
\Bigg\}
\nn
&&
-{}^{}\frac{s(x)}{H(x){}^{}x}
\Bigg\{
4 {}^{} \langle \sigma v \rangle_{\hs{-0.03cm}\XXNN} \hs{-0.01cm}
\sx{1.1}{\bigg[} 
Y_X^2 - Y_N^2 \frac{(Y^\tf{eq}_X)^2}{(Y^\tf{eq}_N)^2} 
\sx{1.1}{\bigg]} \hs{-0.05cm} 
-
\langle \sigma v \rangle_{\hs{-0.03cm}\NNXX} \hs{-0.01cm}
\sx{1.1}{\bigg[} 
Y_N^2 - Y_X^2 \frac{(Y^\tf{eq}_N)^2}{(Y^\tf{eq}_X)^2} 
\sx{1.1}{\bigg]} \hs{-0.05cm}
\Bigg\}
~,
\label{dYX}
\\[0.2cm]
\frac{\dd Y^{}_N}{\dd x} 
\Eq
-{}^{}\frac{s(x)^2}{H(x){}^{}x}
\Bigg\{
2 {}^{} \langle \sigma v^2 \rangle_{\hs{-0.03cm}\XNNXX} {}^{}{}^{}
Y^{}_X \hs{-0.05cm}
\sx{1.1}{\bigg[} 
Y_N^2 - Y^{}_X \frac{(Y^\tf{eq}_N)^2}{Y^\tf{eq}_X} 
\sx{1.1}{\bigg]} \hs{-0.05cm} 
-
8 {}^{} \langle \sigma v^2 \rangle_{\hs{-0.03cm}\XXXNN} \hs{-0.01cm}
\sx{1.1}{\bigg[} 
Y_X^3 - Y_N^2 \frac{(Y^\tf{eq}_X)^3}{(Y^\tf{eq}_N)^2} 
\sx{1.1}{\bigg]} \hs{-0.05cm}
\Bigg\}
\nn
&&
-{}^{}\frac{s(x)}{H(x){}^{}x}
\Bigg\{
\langle \sigma v \rangle_{\hs{-0.03cm}\NNXX} \hs{-0.01cm}
\sx{1.1}{\bigg[} 
Y_N^2 - Y_X^2 \frac{(Y^\tf{eq}_N)^2}{(Y^\tf{eq}_X)^2} 
\sx{1.1}{\bigg]} \hs{-0.05cm}
- 
4 {}^{} \langle \sigma v \rangle_{\hs{-0.03cm}\XXNN} \hs{-0.01cm}
\sx{1.1}{\bigg[} 
Y_X^2 - Y_N^2 \frac{(Y^\tf{eq}_X)^2}{(Y^\tf{eq}_N)^2} 
\sx{1.1}{\bigg]}  \hs{-0.05cm}
\Bigg\}
~,
\label{dYN}
\end{eqnarray} 
where $Y^\tf{eq}_j$ is the equilibrium comoving number yield of the species $j$ given by
\begin{eqnarray}
Y^\tf{eq}_{j}
\,=\,
\frac{45}{4{}^{}\pi^4}
\frac{g^{}_j}{g^{}_{\star s}(x)} 
\big(r^{}_j {}^{} x\big)^{\hs{-0.05cm}2}
K^{}_2 \hs{-0.03cm} \big(r^{}_j {}^{} x\big)
\,\simeq\,
\frac{45\sqrt{2}}{8{}^{}\pi^{7/2}}
\frac{g^{}_j}{g^{}_{\star s}(x)} {}^{}
(r^{}_j {}^{} x)^{3/2} e^{- r^{}_j {}^{} x}
\end{eqnarray}
with $K^{}_2(x)$ being the modified Bessel function of the second kind.\,\,The $s(x)$ and $H(x)$ are the comoving entropy density and the Hubble parameter, respectively, which are given by
\begin{eqnarray}
s(x) \,=\, 
\frac{2{}^{}\pi^2}{45} g^{}_{\star s}(x) {}^{}{}^{} \frac{m_X^3}{x^3} ~,\quad
H(x) \,=\, 
\sqrt{\frac{\pi^2 g^{}_{\star}(x)}{90}} \frac{m_X^2}{x^2 m^{}_\tf{Pl}} 
\end{eqnarray}
with $g^{}_{\star}\,(g^{}_{\star s})$ being the effective energy (entropy) degrees of freedom of thermal plasma~\cite{Saikawa:2018rcs}, and $m^{}_\tf{Pl} = 2.4 \times 10^{18}\,\tx{GeV}$ the reduced Planck mass.\,\,Now, with an appropriate initial condition $Y^{}_{X,N}(x_\tf{ini.}\hs{-0.03cm}) = Y^\tf{eq}_{X,N}(x_\tf{ini.}\hs{-0.03cm})$, where typically $10 < x_\tf{ini.} \hs{-0.05cm} < 20$, we can obtain the $Y^{}_{X,N}(x)$, and then predict the present density of DM by the relation below \cite{Bhattacharya:2019mmy}
\begin{eqnarray}
\Omega_\tf{DM} \hat{h}^2 
\,=\,
2\big({}^{}
\Omega^{}_X \hat{h}^2 + \Omega^{}_N {}^{} \hat{h}^2 
{}^{}\big) 
\,\simeq\,
5.49 \times 10^5 {}^{}
\sx{0.9}{\bigg(}
\frac{m^{}_X}{\text{MeV}}
\sx{0.9}{\bigg)}
\sx{1.1}{\big(}
Y^0_X + r^{} _N {}^{} Y^0_N
\sx{1.1}{\big)}
~,
\end{eqnarray}
where $Y^0_j = Y^{}_j(x \to \infty)$.\,\,Imposing the observed DM abundance, $\Omega^\tf{obs}_\tf{DM} \hat{h}^2 = 0.12 \pm 0.0012$\,\,\cite{Aghanim:2018eyx}, one can fix the values of $\lambda^{}_3$ and $y^{}_N$ for given masses of $X$, $N$ and $S$.\,\,In the following we will first consider the case without the $2 \to 2$ processes, and then turn it on to see the effects.

\begin{figure}[t!]
\centering
\includegraphics[width=0.48\textwidth]{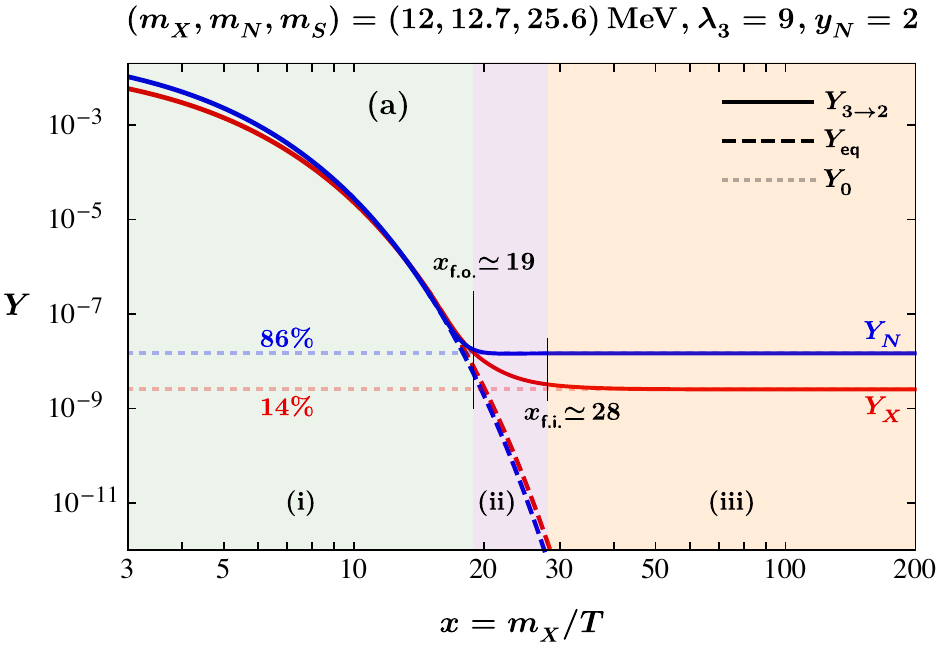}
\hs{0.1cm}
\includegraphics[width=0.48\textwidth]{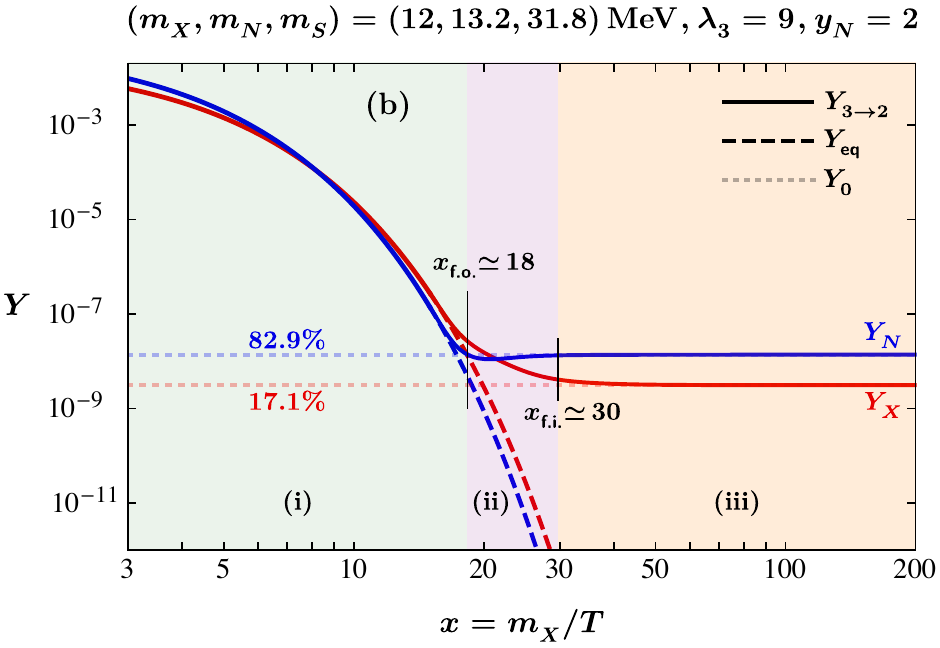}
\\[0.2cm]
\hs{0.02cm}
\includegraphics[width=0.48\textwidth]{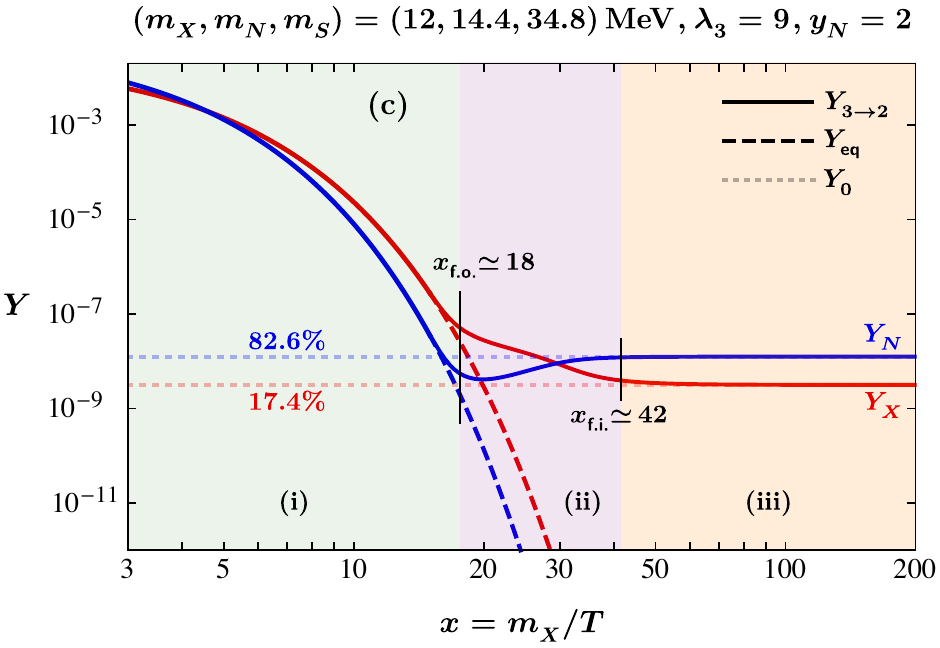}
\hs{0.1cm}
\includegraphics[width=0.48\textwidth]{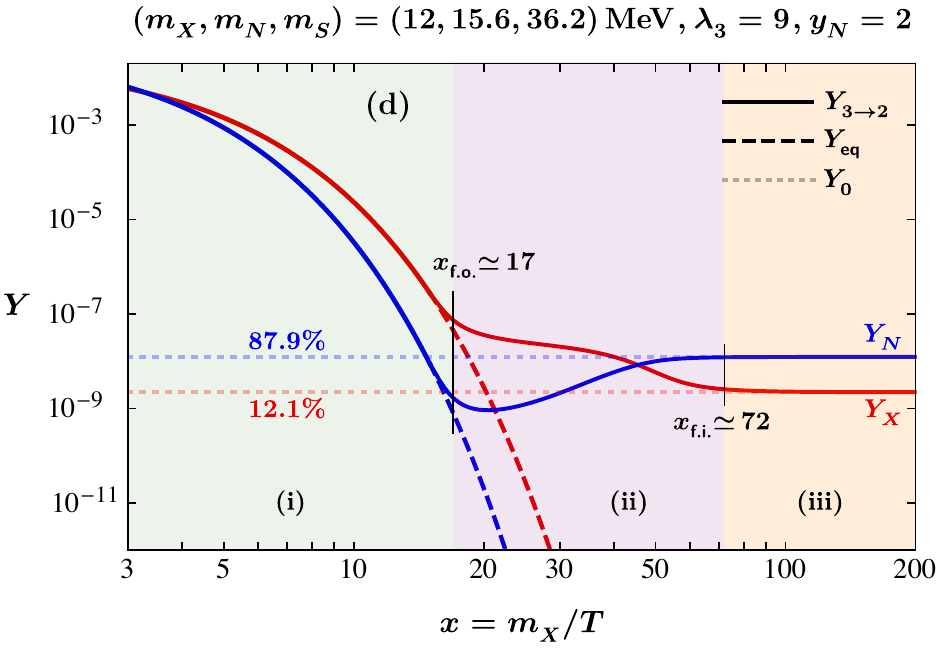}
\vs{-0.3cm}
\caption{The cosmological evolution of the comoving number densities of DM in the absence of the $2 \to 2$ processes for $r^{}_N > 1$ in the $r$SIMP model.\,\,In region (i), the DM particles are in chemical equilibrium via the $3 \to 2$ annihilations.\,\,In region (ii), the number densities of DM are out of the chemical equilibrium and keep changing (increased or decreased) before the freeze-in temperature of DM.\,\,Finally, the DM number densities are frozen until today in region (iii).}
\label{fig:YNX_wo22}
\end{figure}

We present in Fig.\,\ref{fig:YNX_wo22} a few examples of the cosmological evolution of the comoving number densities of DM without the $2 \to 2$ process in the case of $m^{}_N > m^{}_X$, where the color solid lines satisfy the DM relic abundance.\,\,Note that the parameter inputs in these plots may not satisfy other constraints mentioned above.\,\,The plots shown here are merely for demonstration purposes.\,\,As indicated, one can see that both SIMP particles with non-degenerate masses can contribute a sizable amount to the observed DM density.\,\,In particular, there is a phenomenon of the increasing number density of $N$ right after the chemical freeze-out of DM, remarkably in Figs.\,\ref{fig:YNX_wo22}(c) and \ref{fig:YNX_wo22}(d).\footnote{The bouncing effect of DM density after the DM chemical freeze-out was first pointed out in \cite{Katz:2020ywn} and \cite{Shakya:2021pa}.}\,\,To account for this behavior of DM number density, let us first define the freeze-out temperature $\xfo$ and freeze-in temperature $\xfi$ of DM in the following ways\,\,:
\begin{eqnarray}
&&\tx{Freeze-out temp. of $X$}\,:\,
Y^{}_X(x^X_\tf{f.o.}\hs{-0.03cm}) - Y^\tf{eq}_X(x^X_\tf{f.o.}\hs{-0.03cm}) \,\simeq\, Y^\tf{eq}_X(x^X_\tf{f.o.}\hs{-0.03cm}) 
~,
\\[0.1cm]
&&\tx{Freeze-out temp. of $N$}\,:\,
Y^{}_N(x^N_\tf{f.o.}\hs{-0.03cm}) - Y^\tf{eq}_N(x^N_\tf{f.o.}\hs{-0.03cm}) \,\simeq\, Y^\tf{eq}_N(x^N_\tf{f.o.}\hs{-0.03cm}) 
~,
\end{eqnarray}
and we define the freeze-out temperature of DM as a temperature at which both DM particles start to depart from the chemical equilibrium, namely $\xfo \hs{-0.05cm} \equiv \tx{Max}(x^X_\tf{f.o.}, x^N_\tf{f.o.}\hs{-0.03cm})$\,;
\begin{eqnarray}
&&\hs{-1cm}
\tx{Freeze-in temp. of $X$}\,:\,
\tx{Max}\sx{1.2}{\big[}{}^{}
12{}^{}\Gamma_{\hs{-0.03cm}\XXXNN}(x^X_\tf{f.i.}\hs{-0.03cm})\,,
2{}^{}\Gamma_{\hs{-0.03cm}\XXNXN}(x^X_\tf{f.i.}\hs{-0.03cm}) 
\sx{1.2}{\big]}
\,\simeq\,
H(x^X_\tf{f.i.}\hs{-0.03cm}){}^{}{}^{}n^{}_X(x^X_\tf{f.i.}\hs{-0.03cm}) 
~,
\label{fiX}
\\[0.1cm]
&&\hs{-1cm}
\tx{Freeze-in temp. of $N$}\,:\,
8{}^{}\Gamma_{\hs{-0.03cm}\XXXNN}(x^N_\tf{f.i.}\hs{-0.03cm})
\,\simeq\,
H(x^N_\tf{f.i.}\hs{-0.03cm}){}^{}{}^{}n^{}_N(x^N_\tf{f.i.}\hs{-0.03cm}) 
~,
\label{fiN}
\end{eqnarray}
where $\Gamma_{\hs{-0.03cm}\XXXNN}(x) = n_X^3(x) \langle \sigma v^2 \rangle_{\hs{-0.03cm}\XXXNN}$ and $\Gamma_{\hs{-0.03cm}\XXNXN}(x) = n_X^2(x){}^{}n^{}_N(x)\langle \sigma v^2 \rangle_{\hs{-0.03cm}\XXNXN}$ are the $3 \to 2$ annihilation rates per unit volume per unit time with $n_j(x) = s(x){}^{}Y_j(x)$ the number density of DM, and the prefactors are the ones appearing in Eqs.\,\eqref{dYX} and \eqref{dYN}.\,\,Similar to the $\xfo$, we define the freeze-in temperature of DM as a temperature at which both DM number densities begin to be constants, that is $\xfi \hs{-0.05cm} \equiv \tx{Max}(x^X_\tf{f.i.},x^N_\tf{f.i.}\hs{-0.03cm})$.

Now, we take Fig.\,\ref{fig:YNX_wo22}(d) as an example to explain the increasing phenomenon of DM number density after the DM freeze-out temperature.\,\,At high temperatures, the DM number-changing processes, $\XXXNN$, and $\XXNXN$, as well as the conjugate processes (here we ignore the $\XNNXX$ process since it is $p\,$-wave suppressed) maintain the chemical equilibrium of DM such that the actual DM number densities follow the equilibrium DM number densities, $Y^{}_j(x) \simeq Y^\tf{eq}_j(x)$.\,\,Around the $\xfo$, the actual DM number densities are no longer 
tracking the equilibrium DM number densities due to the inefficiency of the chemical equilibrium of DM at lower temperatures.\,\,After the $x^N_\tf{f.o.}$, since the $x^N_\tf{f.i.} > x^N_\tf{f.o.}$ and the process $\XXXNN$ produces two vector-like fermions, the number of $N$ is increased.\,\,Notice that the process $\XXNXN$ does not alter the number of $N$ in total.\,\,The reason this increasing phenomenon is remarkable in Figs.\,\ref{fig:YNX_wo22}(c) and \ref{fig:YNX_wo22}(d) is that the $x^N_\tf{f.i.}$ and $r^{}_N$ are much larger in comparison with Figs.\,\ref{fig:YNX_wo22}(a) and \ref{fig:YNX_wo22}(b).\,\,The former prolongs the time of the increasing number in $N$ and the latter decreases the number of $N$ fastly before the $x^N_\tf{f.o.}$.\,\,On the other hand, the number of $X$ is further decreased after the $x^X_\tf{f.o.}$ because the processes $\XXXNN$ and $\XXNXN$ both annihilate complex scalars until the $x^X_\tf{f.i.}$.\,\,However as we will see immediately, this increasing effect of DM number density would disappear when we switch on the $2 \to 2$ processes.

\begin{figure}[t!]
\centering
\includegraphics[width=0.48\textwidth]{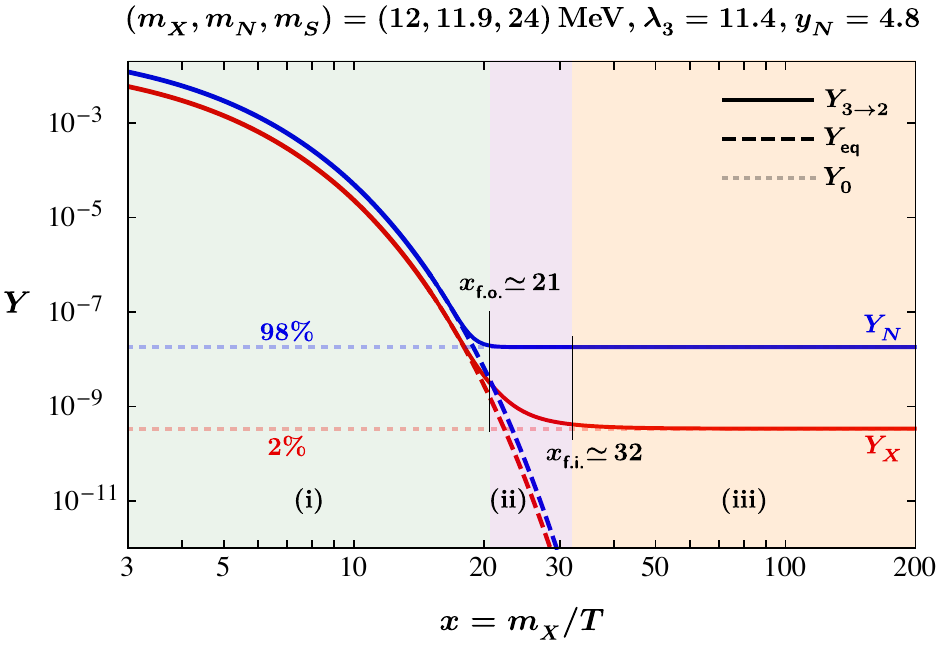}
\hs{0.1cm}
\caption{The cosmological evolution of the comoving number densities of DM 
without including the two-loop induced $2 \to 2$ processes for $r^{}_N < 1$ in the $r$SIMP model.}
\label{fig:YXN_wo22}
\end{figure}

We also show in Fig.\,\ref{fig:YXN_wo22} one example of the cosmological evolution of the comoving number densities of DM without the $2 \to 2$ process in the case of $m^{}_X > m^{}_N$.\,\,We see that in this case there is no increasing phenomenon of DM density after the chemical freeze-out of DM.\,\,This is because $r^{}_N < 1$ and $g^{}_N = 2{}^{}g^{}_X$, meaning the number density of $N$ is always bigger than that of $X$.\,\,Besides, we find that we have to choose large couplings and relatively degenerate masses of DM to satisfy the observed DM density.\,\,Again, the situation would change completely once we turn on the $2 \to 2$ process.

\begin{figure}[t]
\centering
\includegraphics[width=0.48\textwidth]{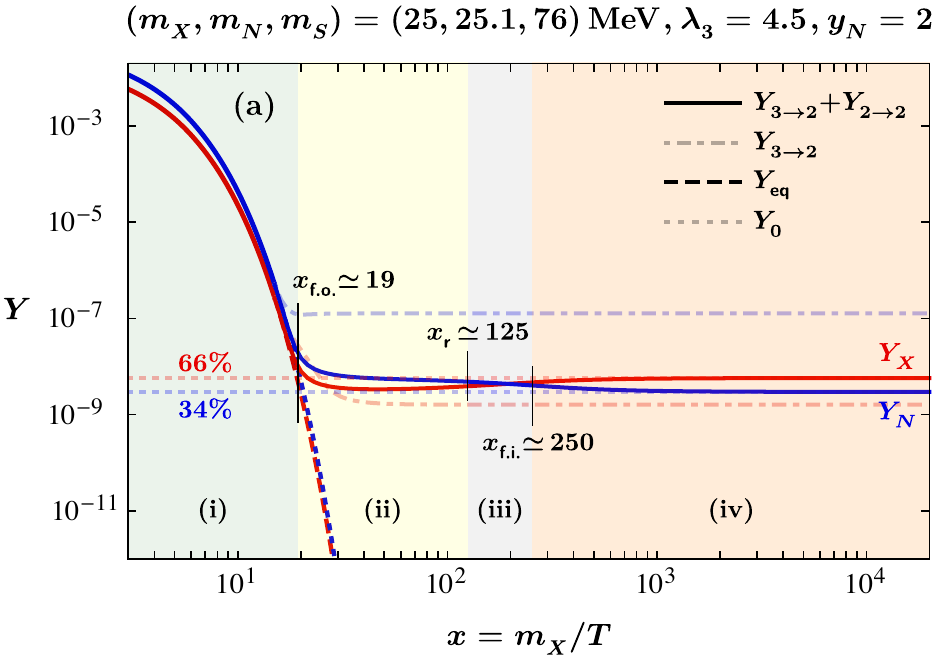}
\hs{0.1cm}
\includegraphics[width=0.48\textwidth]{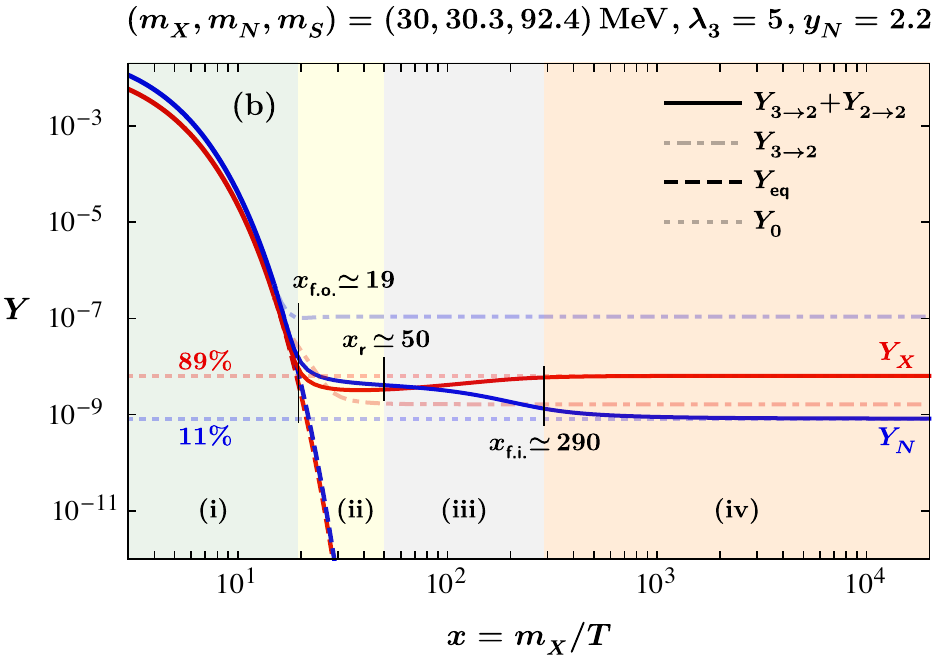}
\\[0.2cm]
\centering
\hs{0.02cm}
\includegraphics[width=0.48\textwidth]{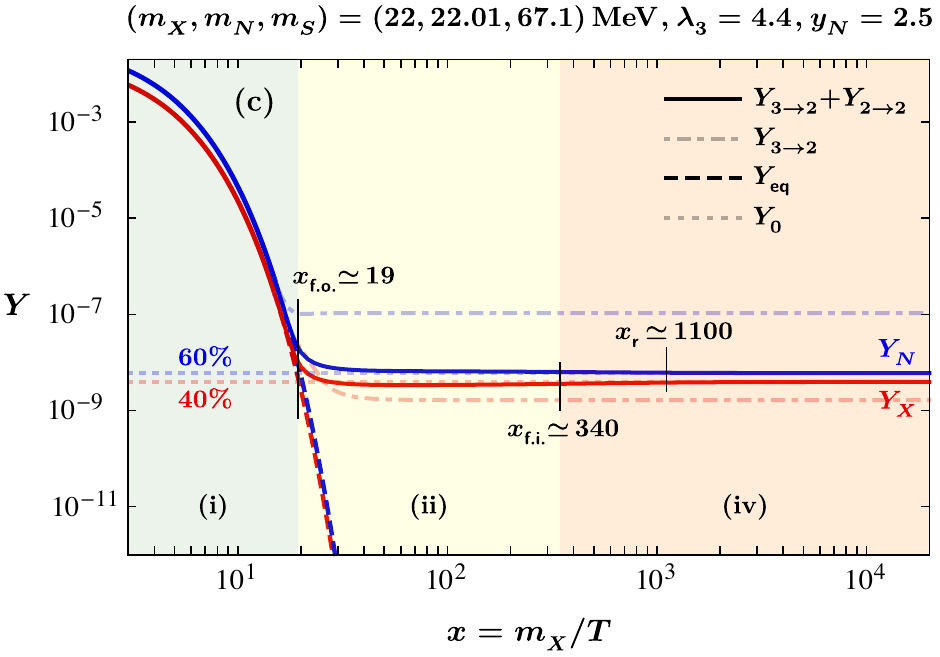}
\hs{0.1cm}
\includegraphics[width=0.48\textwidth]{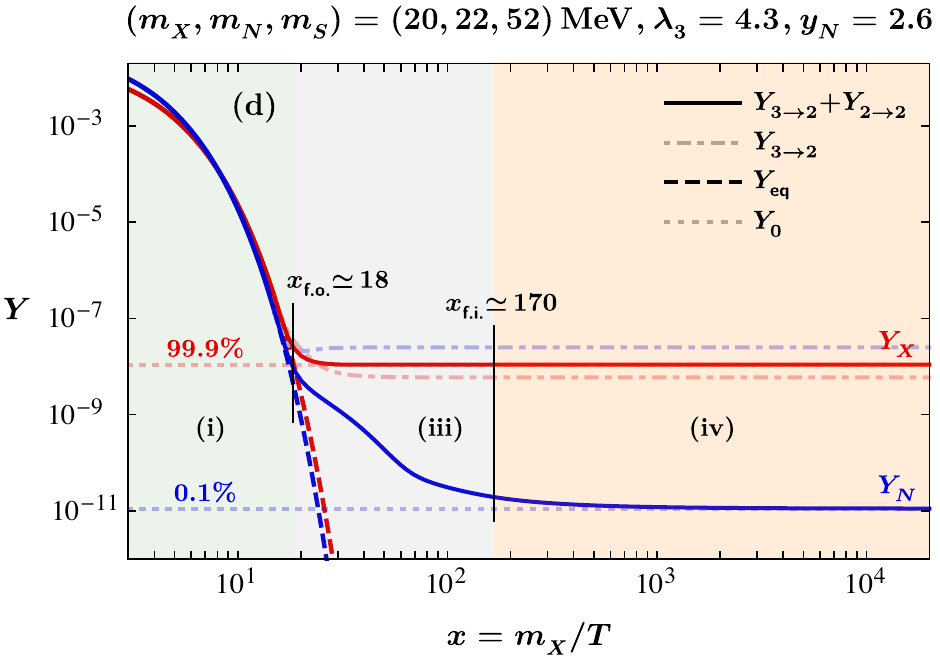}
\vs{-0.3cm}
\caption{Cosmological evolution of the comoving number densities in the presence of the $3 \to 2$ and $2 \to 2$ processes for some benchmark points in the $r$SIMP model for $r^{}_N > 1$.}
\label{fig:YNX_w22}
\end{figure}

We present in Fig.\,\ref{fig:YNX_w22} a few benchmark plots of the cosmological evolution of the comoving number densities of DM with both $3 \to 2$ and $2 \to 2$ processes in the case of $m^{}_X > m^{}_N$.\,\,By comparing Figs.\,\ref{fig:YNX_w22}(a-c) with Fig.\,\ref{fig:YNX_w22}(d), we see that the masses of DM must be nearly degenerate to contribute a non-negligible amount to the total DM relic abundance.\,\,Typically, the evolution of the comoving number density is divided into four stages as shown in color shaded regions of Figs.\,\ref{fig:YNX_w22}(a) and \ref{fig:YNX_w22}(b).\,\,In region (i), the $3 \to 2$ reaction rates are much larger than the Hubble expansion rate, $\Gamma_{3 \to 2} \gg H$, where the $3 \to 2$ processes deplete the DM number densities until the $\xfo \hs{-0.1cm} \simeq 20$.\,\,In region (ii), the DM particles deviate from the chemical equilibrium because of $\Gamma_{3 \to 2} \lesssim H$.\,\,However, the $2 \to 2$ process seems to be inert for a while after the $\xfo\hs{-0.1cm}$ even if the reaction rate of the $2 \to 2$ process governs over that of the $3 \to 2$ process.\,\,This is because the reaction rate of the forward $2 \to 2$ process $\NNXX$ is partially cancelled by that of the backward $2 \to 2$ process $\XXNN$, attributing to the degeneracy of DM masses in the $r$SIMP scenario.\,\,To understand this more clearly, one can look at the last term of Eq.\,\eqref{dYX}, where
\begin{eqnarray}
\langle \sigma v \rangle_{\hs{-0.03cm}\NNXX} \hs{-0.01cm}
\sx{1.1}{\bigg[} 
Y_N^2 - Y_X^2 \frac{(Y^\tf{eq}_N)^2}{(Y^\tf{eq}_X)^2} 
\sx{1.1}{\bigg]}
\,=\,
\langle \sigma v \rangle_{\hs{-0.03cm}\NNXX} \hs{-0.01cm}
\sx{1.1}{\Big[} 
Y_N^2 - 4{}^{}Y_X^2 r^3_N{}^{}e^{-2(r^{}_N-1){}^{}x}
\sx{1.1}{\Big]}
\end{eqnarray}
with the first (second) term in the square bracket the reaction rate of the forward (backward) $2 \to 2$ process.\,\,At high temperatures with $r^{}_N \sim 1$, we have $ r^3_N{}^{}e^{-2(r^{}_N-1){}^{}x} \sim 1$ and $Y^{}_N \sim 2{}^{}Y^{}_X$ right after the $\xfo$.\,\,As a consequence,
this term vanishes and gives no physical effect until the reshuffled temperature, $x^{}_\tf{r} \equiv 1/(2|r^{}_N-1|)$, after which the backward reaction is exponentially-suppressed.\,\,That is to say,
the $X$ particles do not have enough kinetic energy to overcome the mass gap, $m^{}_N - m^{}_X$, to annihilate back into the $N$ particles.\,\,In region (iii), the forward $2 \to 2$ reaction becomes active, the $N$ particles annihilate into the $X$ particles during this stage.\,\,Note that since the $2 \to 2$ process preserves the total number of DM, it would only redistribute the number densities of DM until the $\xfi$, which now is defined as
\begin{eqnarray}
&&\tx{Freeze-in temp. of $X$}\,:\,
\Gamma_{\hs{-0.03cm}\NNXX}(x^X_\tf{f.i.}\hs{-0.03cm})
\,\simeq\,
H(x^X_\tf{f.i.}\hs{-0.03cm}){}^{}{}^{}n^{}_X(x^X_\tf{f.i.}\hs{-0.03cm}) 
~,
\\[0.1cm]
&&\tx{Freeze-in temp. of $N$}\,:\,
\Gamma_{\hs{-0.03cm}\NNXX}(x^N_\tf{f.i.}\hs{-0.03cm})
\,\simeq\,
H(x^N_\tf{f.i.}\hs{-0.03cm}){}^{}{}^{}n^{}_N(x^N_\tf{f.i.}\hs{-0.03cm}) 
~,
\end{eqnarray}
where $\Gamma_{\hs{-0.03cm}\NNXX}(x) = n_N^2(x) \langle \sigma v \rangle_{\hs{-0.03cm}\NNXX}$.\,\,In region (iv), the number densities of DM are frozen until the present day.\,\,In Fig.\,\ref{fig:YNX_w22}(c), there is no reshuffled period because the masses of DM are so degenerate ($r^{}_N = 1.00045$) that the $x^{}_\tf{r} > \xfi$.\,\,Finally, we see that in Fig.\,\ref{fig:YNX_w22}(d) the increasing phenomenon of $N$ is washed out by the $2 \to 2$ process after the $\xfo$, and the non-degenerate masses of DM lead to almost no abundance of $N$.

\begin{figure}[t!]
\centering
\includegraphics[width=0.48\textwidth]{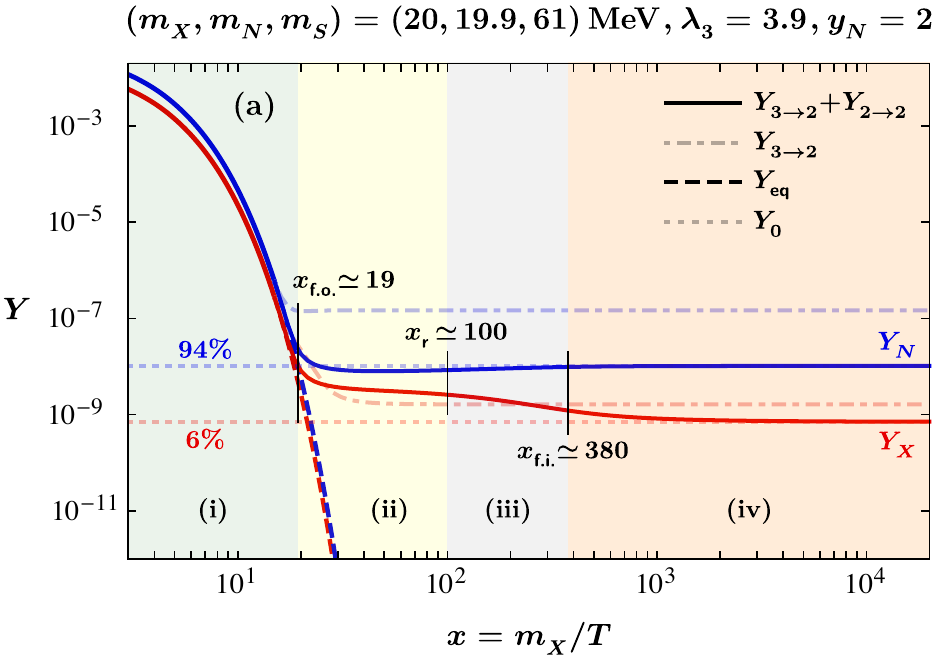}
\hs{0.1cm}
\includegraphics[width=0.48\textwidth]{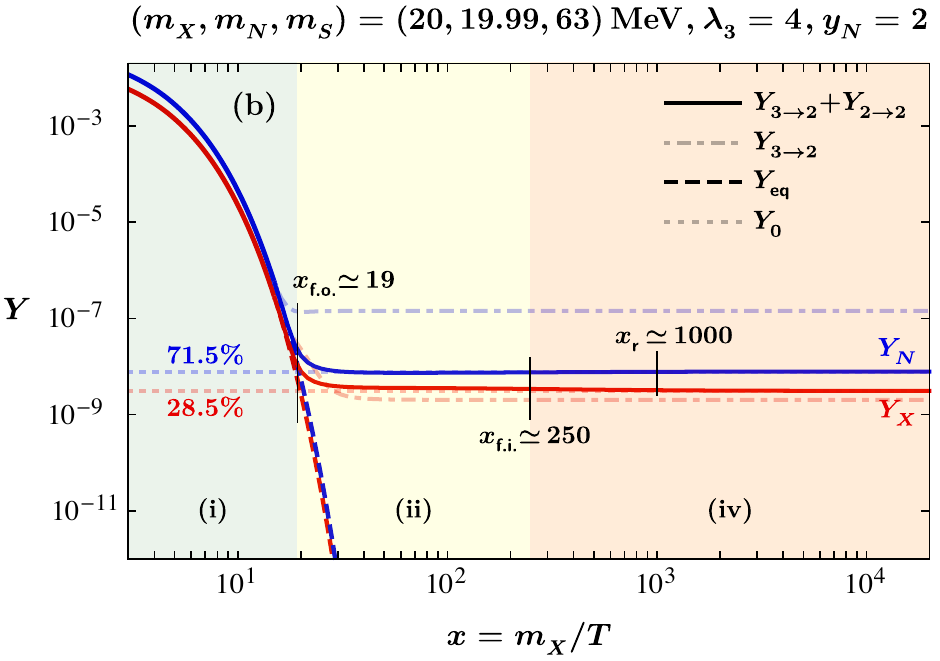}
\vs{-0.3cm}
\caption{Cosmological evolution of the comoving number densities in the presence of the $3 \to 2$ and $2 \to 2$ processes for some benchmark points in the $r$SIMP model for $r^{}_N < 1$.}
\label{fig:YXN_w22}
\end{figure}

Likewise, we show in Fig.\,\ref{fig:YXN_w22} two typical plots of the cosmological evolution of the comoving number densities of DM with both $3 \to 2$ and $2 \to 2$ processes in the case of $m^{}_X > m^{}_N$.\,\,By comparing Fig.\,\ref{fig:YXN_w22}(a) with Fig.\,\ref{fig:YNX_wo22}, we find that we can choose relatively small couplings to satisfy the relic abundance of DM.\,\,Similar to Fig.\,\ref{fig:YNX_w22}(d), we show again there is no reshuffled effect if the masses of DM are extremely degenerate ($r^{}_N =0.9995$).

We briefly summarize the importance of the $2 \to 2$ processes for the cosmological evolution of the comoving DM number densities in the $r$SIMP model.\,\,First, the two-loop induced $2 \to 2$ processes are closely related to the tree-level $3 \to 2$ processes and their reaction rates cannot be omitted.\,\,Second, involving these $2 \to 2$ processes to the $3 \to 2$ processes can alter not only the fractions of DM particles but also total DM number densities.\,\,It is clear to compare the solid lines ($Y_{3 \to 2} + Y_{2 \to 2}$) and dashed lines ($Y_{3 \to 2}$ only) in Figs.\,\ref{fig:YNX_w22} and \ref{fig:YXN_w22} for displaying the differences, where the DM density is overproduced without the $2 \to 2$ processes.\,\,This is easy to understand since the $2 \to 2$ processes strengthen the chemical equilibrium of DM around the DM freeze temperature.\,\,It is crucial to include the two-loop induced $2 \to 2$ annihilations in order to get the correct thermal relic abundance of multi-component SIMP DM.

Before closing this section, let us discuss the effect of non-zero $\lambdaXS$.\,\,As shown in Fig.\,\ref{fig:NNXXZ}, the dominant contributions for the $2 \to 2$ processes may come from the one-loop diagrams.\,\,Thus, with large values of $\lambdaXS$, we can expect that the reshuffled effect is even stronger than that induced by the two-loop diagrams.\,\,However, since $\lambdaXS$ is nothing to do with the $3 \to 2$ processes, we can naively turn it off to keep our model belonging to the two-component SIMP DM scenario.\,\,Of course, one can choose a special $\lambda_{X \hs{-0.03cm} S}$ value (which can be positive or negative) such that there is a destructive interference between one-loop and two-loop diagrams to avoid the reshuffled effect.\,\,But we do not consider this fine-tuning case in our analysis, since that would not be the generic situation.

%%%%%%%%%%%%%%%%%%%%%%%%%%%%%%%%%%%%%%%%%%%%%%%%

\section{SIMP conditions : Thermalization \& Annihilation}\label{sec:6}
As in the typical SIMP paradigm, the DM particles should maintain the kinetic equilibrium with SM particles until the freeze-out temperature of DM.\,\,Hence, the interactions between the dark and SM sectors are required in the $r$SIMP model.\,\,Since the U$(1)^{}_\tf{D}$ symmetry introduced in the model is gauged, then it is natural to have a vector-portal coupling connecting these two sectors.\,\,On the other hand, as we have shown in the previous section, the preferred mass scale of DM in the $r$SIMP scenario is around ${\cal O}(20)\,\tx{MeV}$.\,\,It follows that the freeze-out temperature of DM is $T_f \simeq m^{}_X/20 \simeq {\cal O}(1)\,\tx{MeV}$, thereby the relativistic degrees of freedom in the thermal plasma the DM particles mainly interact with are electron and positron.\footnote{The neutrinos and photon are also relativistic particles in the thermal plasma, however, they can only interact with the DM particles via one-loop diagrams or the kinetic mixing which are much suppressed in this model.}\,\,Accordingly, we then consider the following Lagrangian based on Eq.\,\eqref{Drho} for the thermalization of the DM and $e^\pm$ as
\begin{eqnarray}\label{Zprime}
{\cal L}^{}_{Z'}
\,=\,
-
\Big[{}^{}
i g^{}_\tf{D} {\cal Q}^{}_X 
\big( X^\ast \partial^\rho X - X \partial^\rho X^\ast \big)
+ g^{}_\tf{D} {\cal Q}^{}_N \overline{N} \gamma^\rho N
+ g^{}_e {}^{}{}^{} c^{}_\tf{W} {}^{} \epsilon \, \overline{e} \gamma^\rho e {}^{} 
\Big]
Z'_\rho
~,
\end{eqnarray}
where ${\cal Q}^{}_X$ and ${\cal Q}^{}_N$ are dark charges of the $X$ and $N$ particles, respectively.

To determine how large the gauge coupling is sufficient for an efficient kinetic equilibrium, one has to compute the energy transfer rate of the DM and SM particles and then impose the thermalization condition.\,\,Assuming electron and positron are massless at the $T_f$, the energy transfer rate between the DM particles and $e^\pm$ is given by \cite{Gondolo:2012vh}
\begin{eqnarray}\label{Gammae}
\gamma^{}_{e}(T)
\Eq 
\sum_{j{}^{}={}^{}X,{}^{}N}
\frac{1}{192 {}^{} \pi^3 m^3_j {}^{} T} \hs{-0.05cm}
\mathop{\mathlarger{\int}_{\hs{-0.03cm}0}^{{}^{}\infty}} \hs{-0.05cm}
\dd E^{}_e
\,\frac{e^{E^{}_e/T}}{\big(e^{E^{}_e/T} + 1\big)\raisebox{0.05pt}{$\hs{-0.03cm}^2$}}
\mathop{\mathlarger{\int}^{{}^{}0}_{\hs{-0.05cm}-4 E_e^2}} \hs{-0.05cm} \dd^{} t_j
\,(-{}^{}{}^{}t^{}_j)\,\overline{\big|{\cal M}_{j e \to je}(t^{}_j, E^{}_e)\big|^{\hs{-0.03cm}2}} ~,
\end{eqnarray}
where $E^{}_e$ is the energy of $e^\pm$, $t^{}_{j} = \big({}^{}{}^{}p^{}_j -p_j'\big)\raisebox{1pt}{$\hs{-0.05cm}^2$}$, and $\overline{|{\cal M}_{j e \to je}|^2}$ here is the squared scattering amplitude and an overline represents the usual sum (average) over final (initial) spins.\,\,Using Eq.\,\eqref{Zprime}, the squared amplitudes of the DM particles scattering off the $e^\pm$ in the $m^{}_e = 0$ limit are calculated as 
\begin{eqnarray}
\overline{\big|{\cal M}_{X\hs{-0.03cm}e \to X\hs{-0.03cm}e}(t^{}_X, E^{}_e)\big|^{\hs{-0.03cm}2}}
\Eq
4\bigg(\frac{c^{}_{X\hs{-0.03cm}e}}{t^{} _X - m^2_{Z'}}\bigg)^{\hs{-0.13cm}2}
\Big[{}^{}
s^2_{X\hs{-0.03cm}e} +
\big({}^{}t^{}_X-2{}^{}m_X^2\big) s^{}_{X\hs{-0.03cm}e} +
m_X^4
\Big] 
~,
\label{MXeXe}
\\[0.1cm]
\overline{\big|{\cal M}_{N\hs{-0.03cm}e\to N\hs{-0.03cm}e}(t^{}_N, E^{}_e)\big|^{\hs{-0.03cm}2}}
\Eq
4
\bigg(\frac{c^{}_{N\hs{-0.03cm}e}}{t^{}_N - m^2_{Z'}} \bigg)^{\hs{-0.13cm}2}
\Big[{}^{}
s_{N\hs{-0.03cm}e}^2 + 
\big(t^{}_N - 2{}^{}m_N^2\big) s^{}_{N\hs{-0.03cm}e} + \tfrac{1}{2}{}^{}t_N^2 +
m_N^4
\Big] 
~,
\label{MNeNe}
\end{eqnarray}
where $c^{}_{je} \equiv g^{}_\tf{D}{}^{}g^{}_\tf{e}{}^{}c^{}_\tf{W}{}^{}\epsilon{}^{}{\cal Q}^{}_j{}^{}$, and $s^{}_{je} = \big({}^{}{}^{}p^{}_j + p^{}_e{}^{}\big)\raisebox{1pt}{$\hs{-0.05cm}^2$}$.\,\,Since the SIMP DM are non-relativistic and the $e^\pm$ are relativistic particles at the $T_f$, $E_j \simeq m_j \gg T_f \simeq E^{}_e$, thus $s_j \simeq \big(m_j + E^{}_e{}^{}\big)\raisebox{1pt}{$\hs{-0.05cm}^2$}$ in the center of mass (CM) frame of $j$ and $e^\pm$.\,\,Plugging Eqs.\,\eqref{MXeXe} and \eqref{MNeNe} with this approximate form of $s^{}_j$ into Eq.\,\eqref{Gammae} and taking the leading order in $E^{}_e$ for the integrations, for $r^{}_N \sim 1$ we arrive at
\begin{eqnarray}\label{gammae}
\gamma^{}_{e}(T)
\,=\,
\frac{31{}^{}\pi^3}{189{}^{}x^6}
\frac{m_X^5}{m_{Z'}^4}
\sx{1.1}{\big(}
c_{X\hs{-0.03cm}e}^2 +
c_{N\hs{-0.03cm}e}^2 {}^{}{}^{}
\sx{1.1}{\big)}
~.
\end{eqnarray}
Imposing the thermalization condition of the DM and $e^\pm$, $\gamma^{}_e(x) \gtrsim H(x){}^{}x^2$\,\cite{Choi:2019zeb}, at the time of freeze-out, we then obtain the lower bound of the gauge coupling as
\begin{eqnarray}\label{gDlower}
g^{}_\tf{D}
\,\gtrsim\,
\frac{0.2}{\sqrt{{\cal Q}_X^2 + {\cal Q}_N^2}}
\bigg(\frac{\epsilon}{10^{-3}}\bigg)^{\hs{-0.17cm}-1}
\bigg(\frac{m^{}_{Z'}}{250\,\text{MeV}}\bigg)^{\hs{-0.15cm}2}
\bigg(\frac{m^{}_X}{20\,\text{MeV}}\bigg)^{\hs{-0.17cm}-3/2} ~.
\end{eqnarray}
Here we have set $\xfo \simeq 20$ and $g^{}_{\star}(\xfo\hs{-0.03cm}) \simeq 10.75$.\,\,Employing Eq.\,\eqref{gammae}, we can also determine the highest kinetic decoupling temperature $x^{}_\tf{k.d.}$ of the DM particles from the thermal plasma by the conditions, $\gamma^{}_e(x^{}_\tf{k.d.}\hs{-0.03cm}) \simeq 2 H(x^{}_\tf{k.d.}\hs{-0.03cm})$ \cite{Gondolo:2012vh} and $\gamma^{}_e(\xfo\hs{-0.03cm}) \simeq H(\xfo\hs{-0.03cm}){}^{}x_\tf{f.o.}^2\hs{-0.03cm}$.\,\,Solving these equations, we find that $x^{}_\tf{k.d.} \hs{-0.03cm} \simeq x_\tf{f.o.}^{3/2}/\sqrt[4]{2} \,\simeq 75 < x^{}_\tf{f.i.}$, which implies that $\Gamma^{}_\tf{el} < \Gamma^{2\tf{-loop}}_{2 \to 2}$.\footnote{We have checked numerically that by using the general formula of $\gamma^{}_e(T)$ in Ref.\,\cite{Gondolo:2012vh} and the squared scattering amplitudes with $m^{}_e \neq 0$, the $x^{}_\tf{k.d.} \simeq 120-140$ for $m^{}_X \simeq 20-30\,\tx{MeV}$, which is still less than the $x^{}_\tf{f.i.}$.}\,\,Notice that since the total number and entropy of the DM particles are conserved after the chemical freeze-out and their masses are near degenerate, the DM temperatures after the kinetic decoupling are $T^{}_{X,{}^{}N} \propto R^{-2}$ with $R = R(x)$ the cosmic scale factor, just like usual DM in WIMP or SIMP scenarios.

To achieve the SIMP mechanism, one also needs to suppress the $2 \to 2$ annihilations for the WIMP scenario.\,\,In the $r$SIMP model, such $2 \to 2$ processes are $\XX \to e^+e^-$ and $\NN \to e^+e^-$ through the $Z'$ exchange diagrams.\,\,Applying the crossing symmetry to Eqs.\,\eqref{MXeXe} and \eqref{MNeNe}, we can easily get the squared annihilation amplitudes of these processes as \cite{Lehmann:2020lcv}
\begin{eqnarray}
\overline{\big|{\cal M}_{\XX \to e^+e^-} \hs{-0.03cm}\big|^{\hs{-0.03cm}2}}
\Eq
-{}^{}8\bigg(\frac{c^{}_{X\hs{-0.03cm}e}}{s^{} _X - m^2_{Z'}}\bigg)^{\hs{-0.13cm}2}
\Big[{}^{}
t^2_{X\hs{-0.03cm}e} +
\big(s^{}_X-2{}^{}m_X^2\big) t^{}_{X\hs{-0.03cm}e}  +
m_X^4
\Big] ~,
\label{MXXee}
\\[0.1cm]
\overline{\big|{\cal M}_{\NN \to e^+e^-} \hs{-0.03cm}\big|^{\hs{-0.03cm}2}}
\Eq
4
\bigg(\frac{c^{}_{N\hs{-0.03cm}e}}{s^{}_N - m^2_{Z'}} \bigg)^{\hs{-0.13cm}2}
\Big[{}^{}
t_{N\hs{-0.03cm}e}^2 + 
\big(s^{}_N - 2{}^{}m_N^2\big) t^{}_{N\hs{-0.03cm}e} + \tfrac{1}{2}{}^{}s_N^2 +
m_N^4
\Big] 
~,
\label{MNNee}
\end{eqnarray}
where $s^{}_j = \big({}^{}{}^{}p^{}_{\white{\bar{\black{j}}}}+ p^{}_{\bar{j}}{}^{}\big)\raisebox{1pt}{$\hs{-0.05cm}^2$}$ and $t^{}_{je} = \big({}^{}{}^{}p^{}_j - p^{}_e\big)\raisebox{1pt}{$\hs{-0.05cm}^2$}$.\,\,The resultant thermally-averaged annihilation cross sections are calculated as \cite{Cheung:2012gi}
\begin{eqnarray}
\langle \sigma v \rangle_{\hs{-0.03cm}\XX \to e^+e^-}
\,=\,
\frac{c_{X\hs{-0.03cm}e}^2}{\pi {}^{} x}\frac{m_X^2}{m_{Z'}^4}
~,\quad
\langle \sigma v \rangle_{\hs{-0.03cm}\NN \to e^+e^-}
\,=\,
\frac{c_{N\hs{-0.03cm}e}^2}{\pi}\frac{m_N^2}{m_{Z'}^4} 
~,
\end{eqnarray}
where we have used the fact that $s^{}_j \simeq 4{}^{}m_j^2 \ll m_{Z'}^2$ in the CM frame of the DM pair.\,\,Since the $\langle \sigma v \rangle_{\hs{-0.03cm}\XX \to e^+e^-}$ is dominated by $p\,$-wave contribution, the reaction of the $2 \to 2$ annihilation for the WIMP scenario is then approximated as
\begin{eqnarray}
\Gamma_\tf{ann}(x)
\,=\,
\sum_{j{}^{}={}^{}X,{}^{}N} 
n^{}_j(x) \langle \sigma v \rangle_{\hs{-0.03cm}j{}^{}{}^{}\bar{j} \to e^+e^-}
\approx\,
n^{}_N(x) \langle \sigma v \rangle_{\hs{-0.03cm}\NN \to e^+e^-}
~,
\end{eqnarray}
where $n^{}_j(x) = g^{}_j r_j^3 m_X^3 e^{-x} /(2{}^{}\pi x)^{3/2}$.\,\,Now, to make this reaction is subdominant in the $r$SIMP scenario, we demand that $\Gamma_\tf{ann}(\xfo)
\ll H(\xfo) \simeq \Gamma^{}_{3 \to 2}$ during the freeze-out temperature.\,\,With this requirement and $r^{}_N \sim 1$, we yield the upper bound of the gauge coupling as
\begin{eqnarray}\label{gDupper}
g^{}_\tf{D} 
\,\ll\,
\frac{3}{|{\cal Q}^{}_N|}
\bigg(\frac{\epsilon}{10^{-3}}\bigg)^{\hs{-0.17cm}-1}
\bigg(\frac{m^{}_{Z'}}{250\,\text{MeV}}\bigg)^{\hs{-0.15cm}2}
\bigg(\frac{m^{}_X}{20\,\text{MeV}}\bigg)^{\hs{-0.17cm}-3/2} ~.
\end{eqnarray} 
Here again, we have chosen $g^{}_{\star}(\xfo\hs{-0.03cm}) \simeq 10.75$ with $\xfo \hs{-0.1cm} \simeq 20$.\,\,Therefore, saturating the marginal values of $g^{}_\tf{D}$ given in Eq.\,\eqref{gDlower}, we can have a successful $r$SIMP scenario.

\begin{figure}[t!]
\centering
\includegraphics[width=0.48\textwidth]{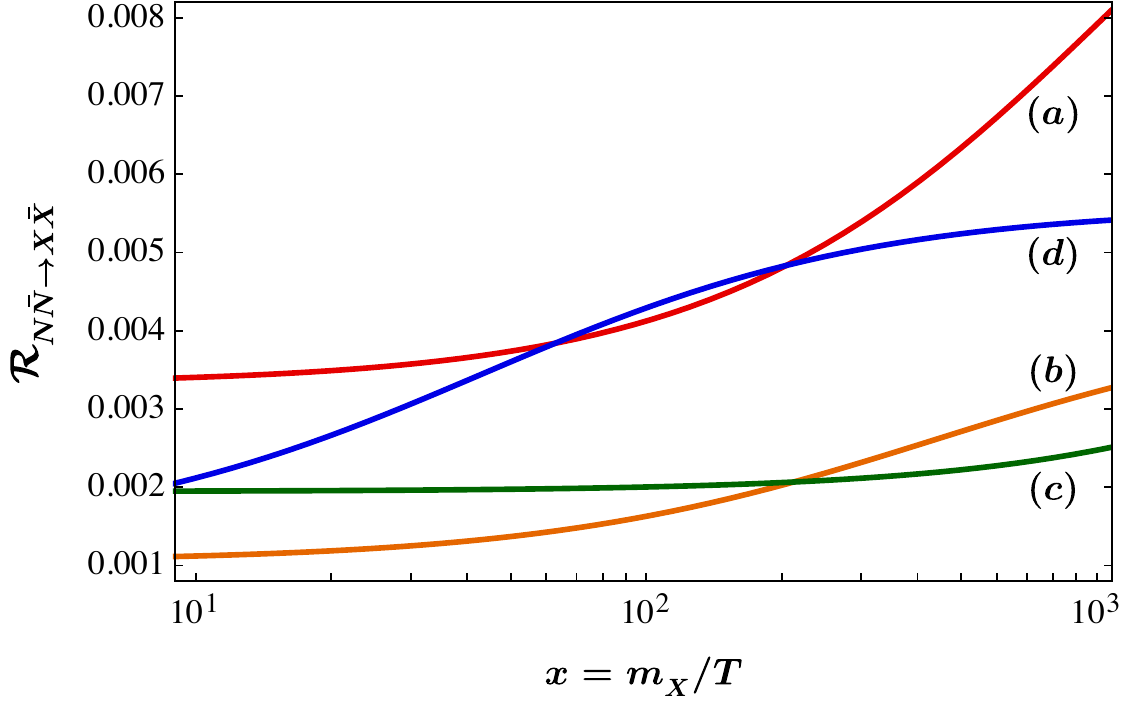}
\hs{0.2cm}
\includegraphics[width=0.48\textwidth]{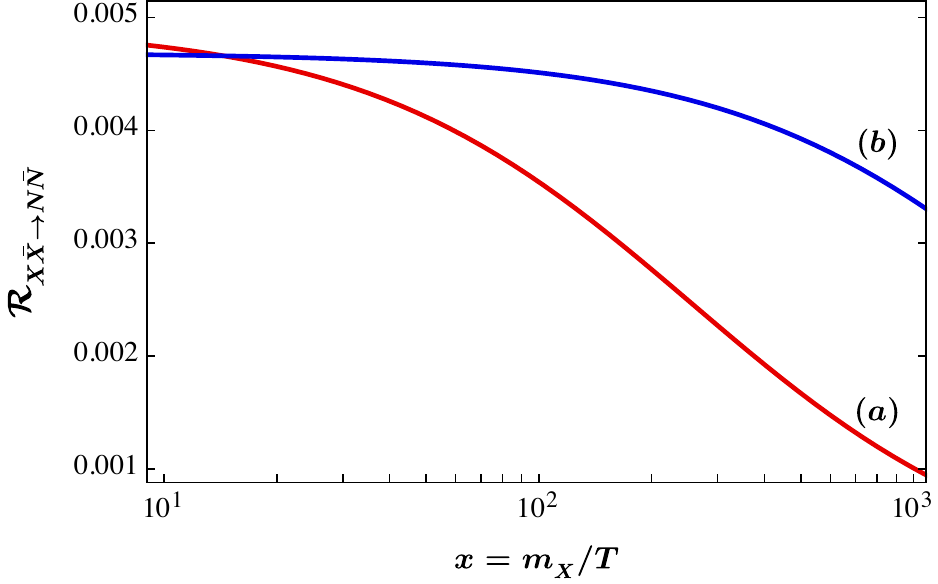}
\vs{-0.3cm}
\caption{The ${\cal R}_{\NNXX}$ and ${\cal R}_{\XXNN}$ as functions of $x$ with the parameter inputs given in Figs.\,\ref{fig:YNX_w22} and \ref{fig:YXN_w22}.\,\,Here we have fixed the $g^{}_\tf{D} $ to the minimal value of Eq.\,\eqref{gDlower} with $\epsilon = 10^{-3}$ and $m^{}_{Z'} = 250\,\tx{MeV}$, and $3 {\cal Q}^{}_X = 2 {\cal Q}^{}_N$ for making these plots.}
\label{fig:RZ}
\end{figure}

As we already mentioned in Sec.\,\ref{sec:3}, there are also tree-level $Z'$-mediated diagrams for the $2 \to 2$ processes in addition to the two-loop diagrams.\,\,Using Eq.\,\eqref{Zprime} again, the corresponding thermally-averaged cross sections are calculated as
\begin{eqnarray}
\langle \sigma v \rangle^{Z'}_{\hs{-0.03cm}\NNXX}
\Eq
\frac
{c_{X\hs{-0.03cm}N}^2 m_X^2}
{4{}^{}\pi{}^{}m_{Z'}^4}
\frac{ \sqrt{r_N^2-1}}{r^{}_N}
\Bigg(\hs{-0.05cm}
r_N^2-1+
\frac
{11-2{}^{}r_N^2}
{4{}^{}x} 
\hs{-0.05cm}\Bigg)
\label{cNNXXZ}
~,
\\[0.15cm]
\langle \sigma v \rangle^{Z'}_{\hs{-0.05cm}\XXNN}
\Eq
\frac
{c_{X\hs{-0.03cm}N}^2 m_X^2}{8{}^{}\pi{}^{}x{}^{}m_{Z'}^4}
\sqrt{1- r_N^2} 
\sx{1.1}{\big(} 2 + r_N^2 \sx{1.1}{\big)}
\label{cXXNNZ}
~,
\end{eqnarray}
where $c^{}_{X\hs{-0.03cm}N} \equiv g_\tf{D}^2 {\cal Q}^{}_X {\cal Q}^{}_N$.\,\,In Fig.\,\ref{fig:RZ}, we show the ratios of the cross sections induced by the $Z'$-mediated diagrams to the ones induced by the two-loop diagrams with the parameter inputs referring to Figs.\,\ref{fig:YNX_w22} and \ref{fig:YXN_w22}, where
\begin{eqnarray}
{\cal R}^{}_{\NNXX}  \,\equiv\,
\frac{\langle \sigma v \rangle^{Z'}_{\hs{-0.03cm}\NNXX}}
{\langle \sigma v \rangle^{2\tf{-loop}}_{\hs{-0.03cm}\NNXX}}
~,\quad
{\cal R}^{}_{\XXNN} \,\equiv\,
\frac{\langle \sigma v \rangle^{Z'}_{\hs{-0.03cm}\XXNN}}
{\langle \sigma v \rangle^{2\tf{-loop}}_{\hs{-0.03cm}\XXNN}}
~.
\end{eqnarray}
As indicated, the contribution of the $Z'$-mediated diagram for the $2 \to 2$ process is subdominant to that of the two-loop diagram.\,\,Notice that, unlike the $\lambdaXS$, we cannot switch $g^{}_\tf{D}$ off to evade the reshuffled mechanism.\,\,As we have discussed in this section, a sufficiently large dark gauge coupling is required to maintain the kinetic equilibrium between the DM and SM particles.\,\,There are a couple of factors that make the ${\cal R}_{\NNXX}$ and ${\cal R}_{\XXNN}$ much smaller than the unity although the $\langle \sigma v \rangle^{2\tf{-loop}}_{\hs{-0.03cm}\NNXX}$ and $\langle \sigma v \rangle^{2\tf{-loop}}_{\hs{-0.03cm}\XXNN}$ are suppressed by the two-loop factor $(4\pi)^8$.\,\,Firstly, we have to choose strong couplings, $\lambda^{}_3{}^{}{}^{}y^{}_N \sim {\cal O}(10)$, to satisfy the relic abundance of DM.\,\,Secondly, the SIMP conditions suggest that $c^{}_{X\hs{-0.03cm}N} \simeq 0.2 {\cal Q}^{}_X {\cal Q}^{}_N / ({\cal Q}_X^2 + {\cal Q}_N^2) \sim 0.02$ with $3 {\cal Q}^{}_X = 2 {\cal Q}^{}_N$.\,\,Thirdly, the mass of $Z'$ in the tree-level graphs is heavier than that of $S$ in the two-loop diagrams, where $m^{}_{Z'} \sim 4{}^{}m^{}_S$.\,\,As a result, the ${\cal R}^{}_{\NNXX}$, for instance, is roughly equal to $(4\pi)^8(c_{X\hs{-0.03cm}N}^2/\lambda_3^4{}^{}y_N^4)(m_S^4 / m_{Z'}^4 \hs{-0.03cm}) \ll 1$.

%%%%%%%%%%%%%%%%%%%%%%%%%%%%%%%%%%%%%%%%%%%%%%%%

\section{Observational signature : DM self-interacting cross section}\label{sec:7}

\begin{figure}[t!]
\hs{0.1cm}
\centering
\includegraphics[width=0.4\textwidth]{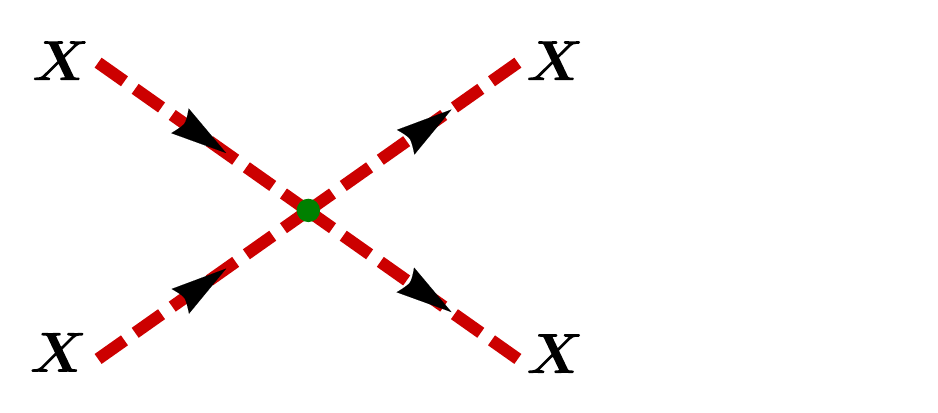}
\hs{-2cm}
\includegraphics[width=0.4\textwidth]{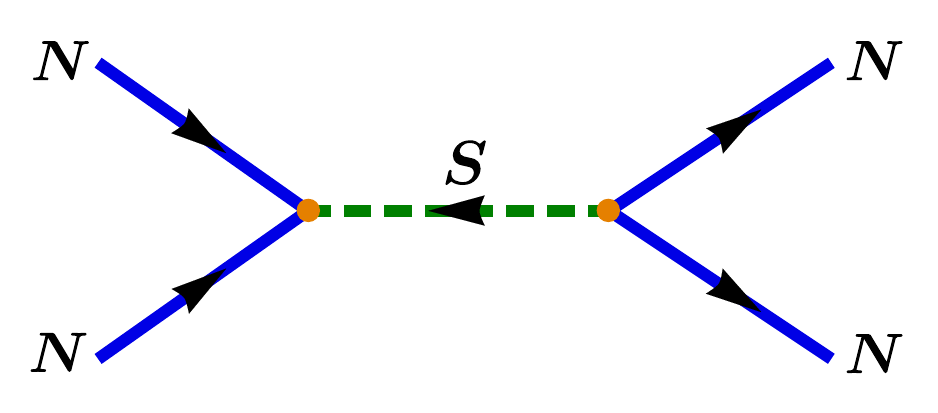}
\vs{-0.3cm}
\caption{The dominant Feynman diagrams of DM self-interacting processes for $X$ and $N$, where the other processes can be obtained by rotating these diagrams.}
\label{fig:self}
\vs{0.3cm}
\end{figure}

In this model, both $X$ and $N$ particles can have self-interactions via the contact coupling in Eq.\,\eqref{potential} and the Yukawa coupling in Eq.\,\eqref{Yukawa}, respectively, as displayed in Fig.\,\ref{fig:self}.\,\,There are also self-interactions of DM through the $Z'$-mediated diagrams akin to Fig.\,\ref{fig:NNXXZ}.\,\,However, these contributions are subleading due to small dark gauge coupling and heavy $Z'$ mass.\,\,In general, there is no well-defined effective self-interacting cross section for two-component DM scenarios.\,\,With the degeneracy of DM masses, we fairly define the self-interacting cross section as follows
\begin{eqnarray}\label{selfcs}
\frac{\sigma^{}_\tf{self}}{m^{}_\tf{DM}}
\,=\,
{\cal R}_X^2
\frac{\sigma^{}_X}{m^{}_X} +
{\cal R}_N^2
\frac{\sigma^{}_N}{m^{}_N}
~,
\end{eqnarray}
where ${\cal R}^{}_X$ and ${\cal R}^{}_N$ are the fractions of DM particles given by
\begin{eqnarray}
{\cal R}^{}_X \,=\, \frac{\Omega^{}_X}{\Omega^{}_X + \Omega^{}_N}  ~,\quad
{\cal R}^{}_N \,=\, \frac{\Omega^{}_N}{\Omega^{}_X + \Omega^{}_N}  ~,
\end{eqnarray}
and the self-interacting cross sections of $X$ and $N$ are computed as
\begin{eqnarray}
\sigma^{}_X
\Eq
\tfrac{1}{4}
\big(
\sigma^{}_{X\hs{-0.03cm}X \to X\hs{-0.03cm}X} + 
\sigma^{}_{X\hs{-0.03cm}\bar{X} \to X\hs{-0.03cm}\bar{X}} +
\sigma^{}_{\bar{X}\hs{-0.03cm}\bar{X} \to \bar{X}\hs{-0.03cm}\bar{X}} 
\big)
\,=\,
\frac{\lambda_X^2}{8{}^{}\pi{}^{}m_X^2}
~,
\\[0.1cm]
\sigma^{}_N
\Eq
\tfrac{1}{4}
\big(
\sigma^{}_{N\hs{-0.03cm}N \to N\hs{-0.03cm}N} + 
\sigma^{}_{N\hs{-0.03cm}\bar{N} \to N\hs{-0.03cm}\bar{N}} +
\sigma^{}_{\bar{N}\hs{-0.03cm}\bar{N} \to \bar{N}\hs{-0.03cm}\bar{N}} 
\big)
\,=\,
\frac{y_N^4}{16{}^{}\pi{}^{}m_X^2}\frac{r_N^2}{r_S^4}
~.
\end{eqnarray}
Note that the $\sigma^{}_{N\hs{-0.03cm}N \to N\hs{-0.03cm}N}$ and $\sigma^{}_{\bar{N}\hs{-0.03cm}\bar{N} \to \bar{N}\hs{-0.03cm}\bar{N}}$ are velocity-suppressed.\,\,When the ${\cal R}_N$ goes to 0, Eq.\,\eqref{selfcs} reduces to the usual definition of the self-interacting cross section for complex scalar DM.

\begin{table}[hbpt!]
\begin{center}
\def\arraystretch{1.2}
\begin{tabular}{|c|c|c|c|c|c|c|c|}
\hline
~$\lambda^{}_X$~ & ~$\lambda^{}_S$~ & ~$\lambda^{}_3$~ & ~$y^{}_N$~ & ~$\big(m^{}_X,m^{}_N,m^{}_S\big)/\tx{MeV}$~ & ~${\cal R}^{}_X$~
& ~${\cal R}^{}_N$~ & ~$\sigma^{}_\tf{self}/m^{}_\tf{DM}\,(\tx{cm}^2/\tx{g})$~ 
\\[0.05cm]
\hline 
~$4.4$~ & ~$10.0$~ & ~$4.7$~ & ~$3.0$~ & ~$(20,20.02,59.6)$~ & ~$0.56$~ & ~$0.44$~ & ~$6.70$~ 
\\\hline 
~$4.2$~ & ~$9.0$~ & ~$4.4$~ & ~$2.5$~ & ~$(22,22.01,67)$~ & ~$0.40$~ & ~$0.60$~ & ~$2.34$~ 
\\\hline 
~$4.5$~ & ~$8.0$~ & ~$4.5$~ & ~$2.0$~ & ~$(25,25.1,76)$~ & ~$0.66$~ & ~$0.34$~ & ~$4.92$~ 
\\\hline 
~$4.0$~ & ~$10.0$~ & ~$4.3$~ & ~$2.5$~ & ~$(25,25.2,77)$~ & ~$0.86$~ & ~$0.14$~ & ~$6.66$~ 
\\\hline 
~$5.0$~ & ~$9.0$~ & ~$5.0$~ & ~$2.2$~ & ~$(30,30.3,92.4)$~ & ~$0.89$~ & ~$0.11$~ & ~$6.31$~ 
\\\hline  
\end{tabular}
\caption{The benchmark points in the $r$SIMP model for $r^{}_N > 1$.}
\label{tab:2}
\end{center}
\vs{-0.5cm}
\end{table}
\begin{table}[hbpt!]
\begin{center}
\def\arraystretch{1.2}
\begin{tabular}{|c|c|c|c|c|c|c|c|}
\hline
~$\lambda^{}_X$~ & ~$\lambda^{}_S$~ & ~$\lambda^{}_3$~ & ~$y^{}_N$~ & ~$\big(m^{}_X,m^{}_N,m^{}_S\big)/\tx{MeV}$~ & ~${\cal R}^{}_X$~
& ~${\cal R}^{}_N$~ & ~$\sigma^{}_\tf{self}/m^{}_\tf{DM}\,(\tx{cm}^2/\tx{g})$~ 
\\[0.05cm]
\hline 
~$5.9$~ & ~$6.2$~ & ~$5.2$~ & ~$2.6$~ & ~$(15,14.9,43.5)$~ & ~$0.01$~ & ~$0.99$~ & ~$0.82$~ 
\\\hline 
~$4.0$~ & ~$8.0$~ & ~$4.0$~ & ~$2.0$~ & ~$(20,19.99,63)$~ & ~$0.28$~ & ~$0.72$~ & ~$1.45$~ 
\\\hline 
~$5.0$~ & ~$4.0$~ & ~$3.9$~ & ~$2.0$~ & ~$(20,19.9,61)$~ & ~$0.06$~ & ~$0.94$~ & ~$0.20$~ 
\\\hline 
~$7.5$~ & ~$4.0$~ & ~$5.4$~ & ~$1.8$~ & ~$(25,24.9,76)$~ & ~$0.07$~ & ~$0.93$~ & ~$0.18$~ 
\\\hline 
~$6.5$~ & ~$6.5$~ & ~$5.6$~ & ~$1.3$~ & ~$(28,27.9,85.4)$~ & ~$0.14$~ & ~$0.86$~ & ~$0.32$~ 
\\\hline  
\end{tabular}
\caption{The benchmark points in the $r$SIMP model for $r^{}_N < 1$.}
\label{tab:3}
\end{center}
\vs{-0.5cm}
\end{table}

To alleviate the discrepancy between simulations and observations, several analyses have set the bounds on the self-interacting cross section of DM.\,\,For instance, there are constraints of $0.1\,\tx{cm}^2/\tx{g} < \sigma^{}_\tf{self}/m^{}_\tf{DM} < 1\,\tx{cm}^2/\tx{g}$ from Milky Way and cluster scales \cite{Tulin:2013teo}.\,\,The Bullet cluster also imposes a similar upper bound, $\sigma^{}_\tf{self}/m^{}_\tf{DM} < 1\,\tx{cm}^2/\tx{g}$ \cite{Markevitch:2003at,Clowe:2003tk}.\,\,Nevertheless, it has been studied in Ref.\,\cite{Kamada:2016euw} that the self-interacting DM with baryons can explain the diverse rotation curves of spiral galaxies if $\sigma^{}_\tf{self}/m^{}_\tf{DM} = 3\,\tx{cm}^2/\tx{g}$.\,\,Therefore, to cover all of these observations, we then consider an optimistic bound, $0.1\,\tx{cm}^2/\tx{g} < \sigma^{}_\tf{self}/m^{}_\tf{DM} < 10\,\tx{cm}^2/\tx{g}$ \cite{Chu:2018fzy,Tulin:2013teo} in our study before the consensus for the value of the DM self-interacting cross section.

We list in Tabs.\,\ref{tab:2} and \ref{tab:3} a few benchmark points satisfying all the constraints mentioned above with the predictions of the DM self-interacting cross section,\footnote{The unitarity of S-matrix sets a conservative bound for the amplitude of self-interacting scattering, where $|{\cal M}^{}_\tf{self}| < 16{}^{}\pi$~\cite{Biswas:2021dan,Namjoo:2018oyn}, by which the quartic couplings $\lambda^{}_{X,{}^{}S} < 4\pi$.} in the cases of $m^{}_N > m^{}_X$ and $m^{}_X > m^{}_N$, respectively.\,\,As can be seen in Tab.\,\ref{tab:2}, the prediction of $\sigma^{}_\tf{self}/m^{}_\tf{DM}$ is typically larger than $1\,\tx{cm}^2/\tx{g}$ but still  well within the bound, $10\,\tx{cm}^2/\tx{g}$.\,\,This is easy to understand since the density of DM is dominated by the $X$ due to the reshuffled effect and we have to choose a sufficiently large $\lambda^{}_X$ to make the vacuum stable.\,\,In principle, one may consider heavier DM masses to suppress the $\sigma^{}_\tf{self}/m^{}_\tf{DM} \propto1/m_X^3$.\,\,However, we have to enhance the $\lambda^{}_3$ and $\lambda^{}_X$ at the same time to fulfill the DM relic abundance and the vacuum stability, respectively.\,\,The small values of $\sigma^{}_\tf{self}/m^{}_\tf{DM}$ can only be obtained if the DM masses are highly degenerate, with which the density of DM is dominated by the $N$ (no reshuffling in this case) as displayed in the third row of Tab.\,\ref{tab:2}.\,\,Hence, there is a tension among the constraints in the case of $m^{}_N > m^{}_X$.\,\,On the other hand, the size of $\sigma^{}_\tf{self}/m^{}_\tf{DM}$ can be smaller than or comparable with $1\,\tx{cm}^2/\tx{g}$ in the case of $m^{}_X > m^{}_N$ as indicated in Tab.\,\ref{tab:3}.\,\,There are two reasons for this occurrence.\,\,Firstly, the reshuffled effect reduces the number of the $X$ particle.\,\,Secondly, the self-interacting cross section of $N$ is suppressed by the mass of the mediator $S$.\,\,Therefore, it is much easier for the latter case to adjust the parameters to satisfy the DM self-interacting cross section and other constraints.\,\,Future observations and simulations may pin down the value of $\sigma^{}_\tf{self}/m^{}_\tf{DM}$ which can be used to test the reshuffled effect in this model.

%%%%%%%%%%%%%%%%%%%%%%%%%%%%%%%%%%%%%%%%%%%%%%%%

\section{Discussions \& Conclusion}\label{sec:8}

We discuss some future investigations for the $r$SIMP model.\,\,Since the DM masses are about $20$\,MeV, the DM-$e^-$ scattering experiments can be used to test the allowed parameter space in this model~\cite{Hochberg:2021pkt,Blanco:2021hlm,Griffin:2021znd,Liang:2021zkg}.\,\,According to Ref.\,\cite{Hochberg:2021pkt}, the lower limit of the DM-$e^-$ scattering cross section can reach $\sigma_e\simeq 8.4 \times 10^{-41}$\,cm$^2$ for $m^{}_{X,N} \sim 20$\,MeV.\,\,It can be transferred to $g^2_\tf{D} \epsilon^2 \big(4.95 {}^{} {\cal R}_X {\cal Q}^2_X + 6.50 {}^{} {\cal R}_N {\cal Q}^2_N\big) \lesssim 10^{-6}$ for $m^{}_{Z'} \sim 250$\,MeV in our $r$SIMP model.\,\,On the other hand, the dark boson $Z'$ is about hundreds MeV and mainly decays to $\XX,\NN$, and $S\hs{-0.01cm}\bar{S}$.\,\,The Belle II~\cite{Belle-II:2018jsg}, KLEVER~\cite{KLEVERProject:2019aks}, LDMX@SLAC~\cite{LDMX:2018cma} and LDMX@CERN~\cite{LDMX:2018cma,Raubenheimer:2018mwt} experiments can be applied to the invisible searches of the $Z'$\,\cite{Fabbrichesi:2020wbt}.\,\,In particular, the LDMX@CERN experiment can constrain $3.0\times 10^{-6}\leq \epsilon\leq 1.4\times 10^{-4}$ for $0.1\,\tx{GeV} \leq m^{}_{Z'}\leq 1\,\tx{GeV}$.

Except for the SIMP scenario, the WIMP scenario can also be realized in this model.\footnote{Fermion and scalar two-component DM with the discrete $\Zf$ symmetry in the WIMP scenario has recently been studied in Ref.~\cite{Yaguna:2021rds}. However, compared with~\cite{Yaguna:2021rds}, the residual $\Zf$ symmetry in our model is an accidental symmetry after the gauged U$(1)^{}_\tf{D}$ symmetry breaking, and the phenomenology in our model can be quite distinct from theirs.}\,\,Akin to the vector portal~\cite{Holdom:1985ag,Okun:1982xi} and Higgs portal~\cite{Patt:2006fw,Lebedev:2021xey} DM models, the typical DM annihilation channels are $\NN \to Z' \to f\bar{f}$, $\XX \to Z' \to f\bar{f}$, $\XX \to \phi, h \to f\bar{f},VV, \phi\phi, hh$ and four-points interaction $\XX \to \phi \phi, hh$.\,\,Besides, the secluded WIMP DM scenario~\cite{Pospelov:2007mp} for processes $\NN, \XX \to Z'Z'$ can also be achieved when $m^{}_{N,X} > m^{}_{Z'}$.\,\,Also, instead of assuming tiny mass splitting between $S_\tx{R}$ and $S_\tx{I}$ in Eq.\,\eqref{Scalar_mass}, we can set $m^{}_{S_\tx{I}} \sim m^{}_{X,N} \sim m^{}_{S_\tx{R}}/3$ such that $S_\tx{I}$ can be the DM candidate as well and our model becomes three-component DM.\,\,Not only the typical scalar DM annihilation channels in the Higgs portal but also the new DM semi-annihilation channel $N S_\tx{I} \to \bar{N} Z'$ and DM self-interaction channel $S_\tx{I} X \to X\hs{-0.03cm}X$ can occur.\,\,Furthermore, in the SIMP scenario, $S_\tx{I}$ can also be annihilated via $S_\tx{I} \NN \to \bar{N}\hs{-0.03cm}\bar{N}$, $S_\tx{I} N\hs{-0.03cm}N \to \XX, S_\tx{I} \XX \to \bar{N}\hs{-0.03cm}\bar{N}$, and their conjugate processes.\,\,These details are beyond the scope of this work and we would like to study them in the future. 

In summary, we propose a novel scalar and fermion two-component SIMP DM model with a $\Zf$ symmetry.\,\,This residual $\Zf$ symmetry is an accidental symmetry after the gauged U$(1)^{}_\tf{D}$ symmetry breaking instead of a subgroup via the Krauss-Wilczek mechanism.\footnote{For multi-component DM models with the Krauss-Wilczek manner, see \cite{Choi:2021yps} for U$(1)^{}_\tf{D} \to \mathbb{Z}^{}_2 \times \mathbb{Z}^{}_3$\,; and see \cite{Ho:2016aye} for U$(1)^{}_{\tf{B}-\tf{L}} \to \Zf$.}\,\,With the help of an extra complex scalar $S$ as a mediator between the SIMP particles $X$ and $N$, we can have $3 \to 2$ number-changing processes as shown in Fig.\,\ref{fig:ann} which determine the DM relic density in this model.\,\,Note this complex scalar $S$ also has $\Zf$ symmetry compared with other mediators in SIMP models.\,\,Moreover, the SIMP DM particles can maintain kinetic equilibrium with the thermal bath until the freeze-out temperature of DM via the vector portal $Z'$ interactions with SM particles.\,\,To satisfy the thermalization condition and suppress the annihilation rate for the WIMP scenario, the lower and upper bounds of the  U$(1)^{}_\tf{D}$ gauge coupling $g^{}_\tf{D}$ are estimated in Eq.\,\eqref{gDlower} and \eqref{gDupper} which can be tested in future experiments. 

An appealing feature of the multi-component SIMP DM model is that an unavoidable two-loop induced $2 \to 2$ process tightly connects to the $3 \to 2$ process.\,\,This process would reshuffle the SIMP DM number densities after the chemical freeze-out of DM.\,\,We underline that the $2 \to 2$ process in this kind of model is important and cannot be neglected.\,\,Including $2 \to 2$ processes with $3 \to 2$ processes in a multi-component SIMP model will not only change the fractions of DM particles but also the total DM number yields.\,\,As a result, model parameters to explain the correct relic density can be dramatically changed compared with only involving the $3 \to 2$ processes.\,\,Finally, the size of DM self-interacting cross section is also a feature in this model.\,\,Usually, the SIMP models predict inevitably large  $\sigma^{}_\tf{self}/m^{}_\tf{DM}$.\,\,However, thanks to the redistribution behavior of SIMP DM number densities, the predictions of $\sigma^{}_\tf{self}/m^{}_\tf{DM} < 1\,\tx{cm}^2/\tx{g}$ are still possible in our model.\,\,Therefore, future observations and simulations of DM self-interactions can help to distinguish the $r$SIMP model from the usual SIMP models. 

\acknowledgments

We would like to thank Fagner C. Correia and Chao-Jung Lee for useful discussions.\,\,This work is supported by KIAS Individual Grants under Grant No.\,PG081201 (SYH), No.\,PG075301 (CTL), and No.\,PG021403 (PK), and also in part by National Research Foundation of Korea (NRF) Grant No.\,NRF2019R1A2C3005009 (PK).

\newpage
\appendix
\section{The derivation of the $3 \to 2$ annihilation cross sections}

Using Eq.\,\eqref{Yukawa} with the Feynman rule given in \cite{Denner:1992vza}, the annihilation amplitudes of the $3 \to 2$ processes $X(p^{}_1)X(p^{}_2)X(p^{}_3) \to \bar{N}(q_1',s_1') \bar{N}(q_2',s_2'), X(p^{}_1)X(p^{}_2)N(q^{}_1,s^{}_1) \to \bar{X}(p'_3) \bar{N}(q_2',s_2')$ and $X(p^{}_1)N(q^{}_1,s^{}_1)N(q^{}_2,s^{}_2) \to \bar{X}(p'_2) \bar{X}(p'_3)$ are written as
\begin{eqnarray}
{\cal M}_{\hs{-0.03cm}\XXXNN} 
\Eq
\frac{\lambda^{}_3 {}^{} y^{}_N}{(q'_1+q'_2)^2 - m_S^2 + i m^{}_S \Gamma^{}_S} {}^{} \overline{u(q'_2,s'_2)} {}^{}{}^{} v(q'_1,s'_1)
~,\quad
\\[0.1cm]
{\cal M}_{\hs{-0.03cm}\XXNXN} 
\Eq
\frac{\lambda^{}_3 {}^{} y^{}_N}
{(q^{}_1-q'_2)^2 - m_S^2} {}^{} 
\overline{u(q'_2,s'_2)} {}^{}{}^{} u(q^{}_1,s^{}_1)
~,\quad
\\[0.1cm]
{\cal M}_{\hs{-0.03cm}\XNNXX} 
\Eq
\frac{\lambda^{}_3 {}^{} y^{}_N}{(q^{}_1+q^{}_2)^2 - m_S^2} {}^{} 
\overline{v(q^{}_1,s^{}_1)} {}^{}{}^{} u(q^{}_2,s^{}_2)
~.
\end{eqnarray}
Here we have omitted the sign and $i$ for simplicity.\,\,In particular, we add the decay width in the propagator of $\XXXNN$ process in order to see the resonance effect in a correct way.\,\,The squared amplitudes appearing in the Boltzmann equations are
\begin{eqnarray}
\hs{-0.3cm}
\overline{\big|{\cal M}_{\hs{-0.03cm}\XXXNN}\big|\raisebox{1pt}{$^{\hs{-0.03cm}2}$}} 
&\hs{-0.2cm}=\hs{-0.2cm}&
\frac{\lambda^2_3 {}^{} y^2_N}{12}
\frac{{}^{}{}^{}q'_1 \cdot q'_2 - m_N^2}
{\big[(q'_1+q'_2)^2 - m_S^2\big]\raisebox{1pt}{$^{\hs{-0.03cm}2}$} + m_S^2 \Gamma_S^2}
\,=\,
\frac{\lambda^2_3 {}^{} y^2_N}{24}
\frac{9{}^{}m_X^2 - 4{}^{} m_N^2}
{\big(9{}^{}m_X^2 - m_S^2 {}^{}\big)\raisebox{1pt}{$^{\hs{-0.03cm}2}$} + m_S^2 \Gamma_S^2}
~,\quad
\\[0.1cm]
\hs{-0.3cm}
\overline{\big|{\cal M}_{\hs{-0.03cm}\XXNXN}\big|\raisebox{1pt}{$^{\hs{-0.03cm}2}$}} 
&\hs{-0.2cm}=\hs{-0.2cm}&
\frac{\lambda^2_3 {}^{} y^2_N}{2}
\frac{q^{}_1 \cdot q'_2 + m_N^2}
{\big[(q^{}_1 - q'_2)^2 - m_S^2\big]\raisebox{1pt}{$^{\hs{-0.03cm}2}$}}
\,=\,
\frac{\lambda^2_3 {}^{} y^2_N}{2}
\frac{m^{}_N \big(m^{}_X + m^{}_N\big) 
\big[\big(m^{}_X + m^{}_N\big)\raisebox{1pt}{$^{\hs{-0.03cm}2}$} + m_N^2 \big]}
{\big[m_S^2\big(m^{}_X + m^{}_N\big) + 2 {}^{} m_X^2 m^{}_N \big]\raisebox{1pt}{$^{\hs{-0.03cm}2}$}}
\,,\quad
\\[0.1cm]
\hs{-0.3cm}
\overline{\big|{\cal M}_{\hs{-0.03cm}\XNNXX}\big|\raisebox{1pt}{$^{\hs{-0.03cm}2}$}} 
&\hs{-0.2cm}=\hs{-0.2cm}&
\frac{\lambda^2_3 {}^{} y^2_N}{4}
\frac{q^{}_1 \cdot q^{}_2 - m_N^2}
{\big[(q^{}_1+q^{}_2)^2 - m_S^2 \big]\raisebox{1pt}{$^{\hs{-0.03cm}2}$}}
\,=\,{\cal O}(x^{-1})
~,
\end{eqnarray}
after taking the average over initial and final spins, and including the symmetry factors for identical particles in the initial or final states, where for the second equalities we have used the Mandelstam variables for the $3 \to 2$ process, $p^{}_1 + p^{}_2 + p^{}_3 \to q^{}_4 + q^{}_5$, in the CM frame of the initial particles at the nonrelativistic limit as \cite{Ho:2021pqw}
\begin{eqnarray}
s^{}_{jk} 
\Eq 
\big({}^{}{}^{}p^{}_j + p^{}_k\big)\raisebox{1pt}{$^{\hs{-0.03cm}2}$} 
\,\approx\,
\big(m^{}_j + m^{}_k {}^{} \big)\raisebox{1pt}{$^{\hs{-0.03cm}2}$}
~,
\\[0.2cm]
s^{}_{45}
\Eq
\big(q^{}_4 + q^{}_5 \big)\raisebox{1pt}{$^{\hs{-0.03cm}2}$} 
\,\approx\,
\big(m^{}_1 + m^{}_2 + m^{}_3 {}^{} \big)\raisebox{1pt}{$^{\hs{-0.03cm}2}$}
~,
\\[0.1cm]
t^{}_{k\ell} 
\Eq 
\big({}^{}{}^{}p^{}_k - q^{}_\ell {}^{}\big)\raisebox{1pt}{$^{\hs{-0.03cm}2}$}
\,\approx\,
\big(m^{}_k - m^{}_\ell {}^{} \big)\raisebox{1pt}{$^{\hs{-0.03cm}2}$}
- \frac{2{}^{}m^{}_k{}^{}\mu^{}_{45} \Delta m}{m^{}_\ell}
\end{eqnarray}
with $j,k = \{1,2,3\},\,\ell=\{4,5\}$, $\mu^{}_{45} = m^{}_4 m^{}_5/(m^{}_4 + m^{}_5)$, and $\Delta m = m^{}_1 + m^{}_2 + m^{}_3 - m^{}_4 - m^{}_5$. Notice that these Mandelstam variables satisfy the following relation
\begin{eqnarray}
\hs{-0.5cm}
s^{}_{12} + s^{}_{13} + s^{}_{23} + s^{}_{45}
+ t^{}_{14} + t^{}_{24} + t^{}_{34} + t^{}_{15} + t^{}_{25} + t^{}_{35}
\,=\, 3\big(m_1^2 + m_2^2 + m_3^2 + m_4^2 + m_5^2 {}^{}\big)
~.
\end{eqnarray}
Using the nonrelativistic cross section formula of the $3 \to 2$ process \cite{Ho:2021pqw}
\begin{eqnarray}
(\sigma v^2)^{}_{\hs{-0.03cm}123 \to 45}
\Eq
\frac{\overline{\big|{\cal M}^{}_{123 \to 45}\big|\raisebox{0.5pt}{$^{\hs{-0.01cm}2}$}}}
{64{}^{} \pi {}^{} m^{}_1 m^{}_2 m^{}_3 \big(m^{}_1 + m^{}_2 + m^{}_3\big)\raisebox{0.5pt}{$^{\hs{-0.03cm}2}$}} 
\sqrt{{\cal K}\hs{-0.03cm}\sx{1.2}{\big[}\big(m^{}_1 + m^{}_2 + m^{}_3\big)\raisebox{0.5pt}{$^{\hs{-0.03cm}2}$}, m_4^2, m_5^2 \sx{1.2}{\big]}}
~,
\end{eqnarray}
where ${\cal K}\big(a,b,c\big) = a^2+b^2+c^2-2\big(a{}^{}b+b{}^{}c+a{}^{}c\big)$, one can readily derive Eqs.\,\eqref{XXXNN} and \eqref{XXNXN}.

\section{The derivation of the two-loop induced $2 \to 2$ annihilation cross sections}

\begin{figure}[htpb!]
\centering
\hs{0.5cm}
\includegraphics[width=0.6\textwidth]{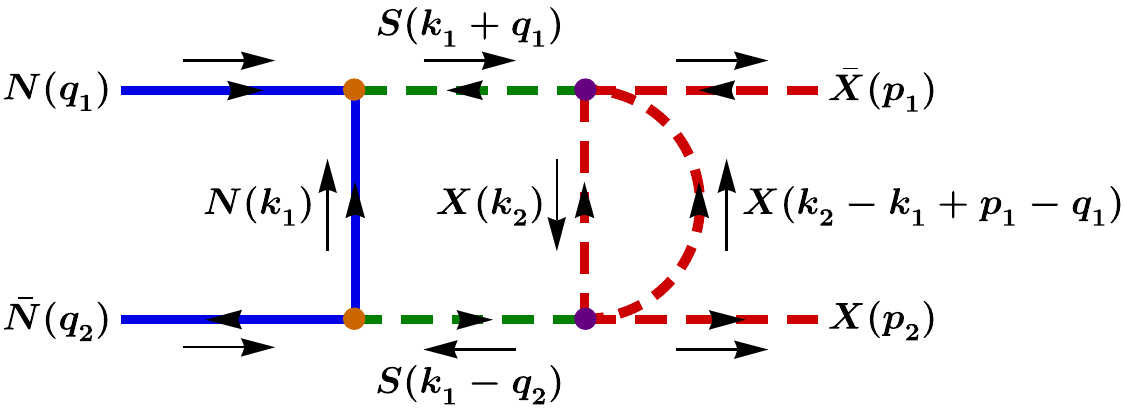}
\caption{The 4-momentum flows for the two-loop diagram of the $2 \to 2$ process $\NNXX$.}
\label{fig:pflow}
\end{figure}

Before writing down the annihilation amplitudes for the two-loop induced $2 \to 2$ processes, let us first assign the 4-momentum flows for the two-loop diagrams.\,\,We start with the process $N(q^{}_1,s^{}_1)\bar{N}(q^{}_2,s^{}_2) \to X(p^{}_1) \bar{X}(p^{}_2)$, in which the 4-momentum flows is shown in Fig.\,\ref{fig:pflow}.
\newline
There are two technical parts of computing this process.\,\,First, there is a UV divergence coming from the scalar loop (the red semi-circle) in this diagram.\,\,We can absorb this UV divergence by introducing a counterterm $\delta \lambdaXS |X|^2|S|^2$ into Eq.\,\eqref{Yukawa}.\,\,Second, we will encounter loop integrals with a logarithmic function that needs a special trick to proceed with the calculations.

Now, we compute the amplitude of the one-loop process $\SSXX$ in the two-loop diagram with the counterterm as
\begin{eqnarray}
{\cal M}^{}_{\SSXX}\big({}^{}{}^{}p^2\big)
\,=\,
\delta \lambdaXS
+ i
\frac{\lambda_3^2}{2}
\mathop{\mathlarger{\int}} \hs{-0.15cm} 
\frac{\dd^d k^{}_2}{(2\pi)^d} {}^{}
\frac{1}{(k_2^2-m_X^2+ i{}^{}\varepsilon)\big[(k^{}_2 - p)^2-m_X^2 + i{}^{}\varepsilon\big]}
~,
\end{eqnarray}
where $\varepsilon = 0^+$ and $p \,=\, k^{}_1 - p^{}_1 + q^{}_1$.\,\,Using the Feynman parametrization and performing the loop integration, one can obtain
\begin{eqnarray}
{\cal M}^{}_{\SSXX}\big({}^{}{}^{}p^2\big)
\,=\,
\delta \lambdaXS
-
\frac{\lambda_3^2}{2(4\pi)^2} \hs{-0.1cm}
\mathop{\mathlarger{\int}_{\hs{-0.02cm}0}^1} \hs{-0.1cm} \dd z^{}_1 
\Big\{
\tf{D} - \ln \hs{-0.08cm} \sx{1.2}{\big[} m_X^2 - z^{}_1(1-z^{}_1) {}^{}{}^{} p^2 - i{}^{}\varepsilon\sx{1.2}{\big]} 
\Big\}
~,
\end{eqnarray}
where $\tf{D} = 2/(4-d) - \gamma^{}_\tf{E} + \ln(4\pi)$ with $d \to 4$ and $\gamma^{}_\tf{E}$ the Euler's constant.\,\,Imposing the renormalizable condition at zero momentum limit,
\begin{eqnarray}\label{rc}
{\cal M}^{}_{\SSXX}\big({}^{}{}^{}p^2 = 0\big) \,=\, \lambdaXS ~,
\end{eqnarray}
we can fix the counterterm $\delta \lambdaXS$ and remove the divergence.\,\,Notice that the renormalization condition \eqref{rc} can be determined numerically only with experimental input from  $\SSXX$ scattering as usual.\,\,Then, the finite $\SSXX$ amplitude at the one-loop is given by
\begin{eqnarray}\label{MSSXX}
{\cal M}^{}_{\SSXX}\big({}^{}{}^{}p^2\big)
\,=\,
\lambdaXS
+
\frac{\lambda_3^2}{2(4\pi)^2}
\mathop{\mathlarger{\int}_{\hs{-0.02cm}0}^1} \hs{-0.1cm} \dd z^{}_1 
\ln \hs{-0.08cm} 
\sx{1.1}{\bigg[}\frac{m_X^2 - z^{}_1(1-z^{}_1) {}^{}{}^{} p^2 - i{}^{}\varepsilon}{m_X^2}\sx{1.1}{\bigg]}
~.
\end{eqnarray}
In the following calculation, we will set the $\lambdaXS$ equal to 0 as explained in Sec.\ref{sec:5}.\footnote{Note that the nonzero $\lambdaXS$ in the two-loop diagram produces a finite result without an additional divergence, and so our arguments in the following paragraphs shall not be spoiled at all.}

With Eq.\,\eqref{MSSXX}, the amplitude of the two-loop induced process $\NNXX$ is written by
\begin{eqnarray}\label{MNNXX}
\hs{-1cm}
&&
{\cal M}^{2\tf{-loop}}_{\hs{-0.03cm}\NNXX}
\nn[0.1cm]
\hs{-1cm}
\Eq
\frac{\lambda_3^2{}^{}{}^{}y_N^2}{2(4\pi)^2} 
\mathop{\mathlarger{\int}_{\hs{-0.02cm}0}^1} \hs{-0.1cm} \dd z^{}_1
\mathop{\mathlarger{\int}} \hs{-0.15cm} 
\frac{\dd^4 k^{}_1}{(2\pi)^4}
\frac{
\overline{v(q^{}_2,s^{}_2)} \big(\slashed{k^{}_1} + m^{}_N\big) u(q^{}_1,s^{}_1) \,
V \sx{1.1}{\big[}(k^{}_1 - p^{}_1 + q^{}_1)^2 \sx{1.1}{\big]}}
{\big(k_1^2-m_N^2 + i{}^{}\varepsilon\big)
\big[(k^{}_1 + q^{}_1)^2-m_S^2 + i{}^{}\varepsilon\big]
\big[(k^{}_1 - q^{}_2)^2-m_S^2 + i{}^{}\varepsilon\big]}
~,
\end{eqnarray}
where for convenience we define
\begin{eqnarray}
V\big({}^{}{}^{}p^2\big)
\,\equiv\,
\ln \hs{-0.08cm} 
\sx{1.1}{\bigg[}
\frac{m_X^2 - z^{}_1(1-z^{}_1) {}^{}{}^{} p^2 - i{}^{}\varepsilon}{m_X^2}
\sx{1.1}{\bigg]}
~.
\end{eqnarray}
Note that we do not have to introduce a counterterm for computing \eqref{MNNXX} which is definitely UV finite since the effective interaction $\NN\XX$ corresponds to a dimension five operator.\,\,Applying the Feynman parametrization again and using the equation of motions of the spinors, $\slashed{p} {}^{}{}^{} u(p) \,=\, m {}^{} u(p),\,\overline{v(p)} {}^{}{}^{} \slashed{p} \,=\, -{}^{} m {}^{}{}^{}\overline{v(p)}$, we get
\begin{eqnarray} 
\hs{-0.5cm}
{\cal M}^{2\tf{-loop}}_{\hs{-0.03cm}\NNXX}
=
\frac{\lambda_3^2 {}^{} y_N^2}{(4\pi)^2}\,\overline{v(q^{}_2,s^{}_2)}
\mathop{\mathlarger{\int}_{\hs{-0.02cm}0}^1} \hs{-0.05cm} \dd z^{}_1
\mathop{\mathlarger{\int}_{\hs{-0.02cm}0}^1} \hs{-0.05cm} \dd z^{}_2
\mathop{\mathlarger{\int}_{\hs{-0.02cm}0}^{1-z^{}_2}} \hs{-0.05cm} \dd z^{}_3 {}^{}
\Big[
m^{}_N ({}^{}1-z^{}_2-z^{}_3) {}^{}
{\cal J}^{}_1
+
\slashed{P} {\cal J}^{}_2
\Big] u(q^{}_1,s^{}_1)
\,,
\end{eqnarray}
where $P \,=\, p^{}_1 - (1-z^{}_2) {}^{} q^{}_1 - z^{}_3 {}^{} q^{}_2\,$, and
\begin{eqnarray}
{\cal J}^{}_1
\Eq
\mathop{\mathlarger{\int}} \hs{-0.15cm} 
\frac{\dd^4 \ell}{(2\pi)^4} 
\frac{1}
{\big(\ell^2 - \Delta + i{}^{}\varepsilon \big)
\raisebox{1pt}{$^{\hs{-0.03cm}3}$}} {}^{}
V\sx{1.1}{\big[}(\ell - P)^2\sx{1.1}{\big]}
~,\quad
\\[0.1cm]
{\cal J}^{}_2
\Eq
\frac{1}{P^2}
\mathop{\mathlarger{\int}} \hs{-0.15cm} 
\frac{\dd^4 \ell}{(2\pi)^4}
\frac{\ell \cdot P}
{\big(\ell^2 - \Delta + i{}^{}\varepsilon \big)
\raisebox{1pt}{$^{\hs{-0.03cm}3}$}} {}^{}
V\sx{1.1}{\big[}(\ell - P)^2\sx{1.1}{\big]}
\end{eqnarray}
with $\Delta = m_N^2 ({}^{}1-z^{}_2-z^{}_3)^2 + m_S^2(z^{}_2+z^{}_3) - (q^{}_1+q^{}_2)^2z^{}_2 z^{}_3{}^{}$, here the 4-momentum $k^{}_1$ has been shifted as $\ell = k^{}_1 + (z^{}_2 {}^{} q^{}_1 - z^{}_3 {}^{} q^{}_2{}^{})$.\,\,To proceed the loop integrations with a logarithmic function, we will consider the following parametrization,
\begin{eqnarray}
\ln \hs{-0.08cm} 
\sx{1.1}{\bigg[}
\frac{m_X^2 - z^{}_1(1-z^{}_1) {}^{}{}^{} p^2 - i{}^{}\varepsilon}{m_X^2}
\sx{1.1}{\bigg]}
\,=\,
\mathop{\mathlarger{\int}_{\hs{-0.02cm}0}^{{}^{}z^{}_1(1-z^{}_1)}} \hs{-0.05cm} \dd z^{}_4\,
\frac{p^2}{p^2 z^{}_4  - m_X^2 + i{}^{}\varepsilon}
~.
\end{eqnarray}
Let us now use it to compute the ${\cal J}^{}_1$ integral, which becomes
\begin{eqnarray}
{\cal J}^{}_1
\,=\,
\mathop{\mathlarger{\int}_{\hs{-0.02cm}0}^{{}^{}z^{}_1(1-z^{}_1)}} \hs{-0.05cm} 
\frac{\dd z^{}_4}{z^{}_4}
\mathop{\mathlarger{\int}} \hs{-0.15cm} 
\frac{\dd^4 \ell}{(2\pi)^4} {}^{}
\frac{\ell^2}
{\big[(\ell + P)^2 - \Delta + i{}^{}\varepsilon \big]
\raisebox{1pt}{$^{\hs{-0.03cm}3}$}\big(\ell^2 - m_X^2/z^{}_4 + i{}^{}\varepsilon\big)}
~.
\end{eqnarray}
Here we have shifted the loop momentum $\ell \to \ell + P$.\,\,Next, utilizing the Feynman parameter $z^{}_5{}^{}$,
\begin{eqnarray}
\frac{1}{A^3 B}
\,=\,
\mathop{\mathlarger{\int}_{\hs{-0.02cm}0}^1} \hs{-0.05cm} \dd z^{}_5 {}^{}
\frac{3 {}^{} z_5^2}{\big[z^{}_5 A + (1-z^{}_5) B \big]\raisebox{1pt}{$^{\hs{-0.03cm}4}$}}
~,
\end{eqnarray}
one can arrive at
\begin{eqnarray}
\hs{-0.8cm}
{\cal J}^{}_1
\Eq
3
\mathop{\mathlarger{\int}_{\hs{-0.02cm}0}^{{}^{}z^{}_1(1-z^{}_1)}} \hs{-0.05cm} 
\frac{\dd z^{}_4}{z^{}_4}
\mathop{\mathlarger{\int}_{\hs{-0.02cm}0}^1} \hs{-0.05cm} \dd z^{}_5 \, z_5^2
\mathop{\mathlarger{\int}} \hs{-0.15cm} 
\frac{\dd^4 \kappa}{(2\pi)^4} {}^{}
\frac{\kappa^2 + P^2 z_5^2}
{( \kappa^2 - \Box + i{}^{}\varepsilon \big)\raisebox{1pt}{$^{\hs{-0.03cm}4}$}}
~,
\end{eqnarray}
where ${}^{}\Box {}^{}= P^2 z_5^2 - \big(Q^2 + m_X^2/z^{}_4 \big) z^{}_5 + m_X^2/z^{}_4$ with $Q^2 = P^2 - \Delta$, and the 4-momentum ${}^{}\ell{}^{}$ has been shifted as $\kappa = \ell + Pz^{}_5$.\,\,Employing the formulas of the loop integrations, it yields
\begin{eqnarray}
\hs{-0.5cm}
{\cal J}^{}_1
\Eq
-{}^{}\frac{i}{(4\pi)^2}
\mathop{\mathlarger{\int}_{\hs{-0.02cm}0}^{{}^{}z^{}_1(1-z^{}_1)}} \hs{-0.05cm} \dd z^{}_4
\mathop{\mathlarger{\int}_{\hs{-0.02cm}0}^1} \hs{-0.05cm} \dd z^{}_5 \,
\frac{z_5^2 \big[ z^{}_4 P^2 z_5^2 - 2\big(z^{}_4 Q^2 + m_X^2 \big) z^{}_5 + 2{}^{}m_X^2 \big]}
{2\big[z^{}_4 P^2 z_5^2 - \big(z^{}_4 Q^2 + m_X^2\big) z^{}_5 + m_X^2 \big]\raisebox{1pt}{$^{\hs{-0.03cm}2}$}}
~.
\end{eqnarray}
Using the same techniques, we can obtain the result of the ${\cal J}^{}_2$ integration as 
\begin{eqnarray}
{\cal J}^{}_2 \,=\, 
{\cal J}^{}_1 
+
\frac{i}{(4\pi)^2}
\mathop{\mathlarger{\int}_{\hs{-0.02cm}0}^{{}^{}z^{}_1(1-z^{}_1)}} \hs{-0.05cm} \dd z^{}_4
\mathop{\mathlarger{\int}_{\hs{-0.02cm}0}^1} \hs{-0.05cm} \dd z^{}_5 \,
\frac{z_5^3 \big[ 2z^{}_4 P^2 z_5^2 - 3\big(z^{}_4 Q^2 + m_X^2 \big) z^{}_5 + 3{}^{}m_X^2 \big]}
{2\big[z^{}_4 P^2 z_5^2 - \big(z^{}_4 Q^2 + m_X^2\big) z^{}_5 + m_X^2 \big]\raisebox{1pt}{$^{\hs{-0.03cm}2}$}}
~.
\end{eqnarray}
Taking the nonrelativistic limit, the $P^2$ and $Q^2$ in the ${\cal J}^{}_1$ and ${\cal J}^{}_2$ reduce to
\begin{eqnarray}
P^2 
&\approx &
m_ X^2 + m_N^2 (z^{}_2 - z^{}_3 + 1)(z^{}_2 - z^{}_3-1)  
~,\quad
\\[0.15cm]
Q^2 
&\approx &
m_X^2 + 2{}^{}m_N^2 (z^{}_2 + z^{}_3-1) - m_S^2 (z^{}_2 + z^{}_3)  ~,
\end{eqnarray}
here we have used the Mandelstam variables $s = (q^{}_1+q^{}_2)^2 \approx 4{}^{}m_N^2,\,t=({}^{}{}^{}p^{}_1 -q^{}_1)^2 \approx m_X^2 -m_N^2$, and $u = ({}^{}{}^{}p^{}_1 
- q^{}_2)^2 \approx m_X^2 -m_N^2$ in the CM frame of the initial particles.

With these integration results, Eq.\,\eqref{MNNXX} becomes
\begin{eqnarray} 
{\cal M}^{2\tf{-loop}}_{\hs{-0.03cm}\NNXX}
\,=\,
\overline{v(q^{}_2,s^{}_2)} \,
\big( C^{}_1 {}^{} \slashed{p^{}_1} + C^{}_2 {}^{} m^{}_N \big) 
u(q^{}_1,s^{}_1)
~,
\end{eqnarray}
here we have used the equation of motions of the spinors once more, and
\begin{eqnarray}
C^{}_1
\Eq
\frac{\lambda_3^2{}^{}{}^{}y_N^2}{(4\pi)^2}
\mathop{\mathlarger{\int}_{\hs{-0.02cm}0}^1} \hs{-0.05cm} \dd z^{}_1
\mathop{\mathlarger{\int}_{\hs{-0.02cm}0}^1} \hs{-0.05cm} \dd z^{}_2
\mathop{\mathlarger{\int}_{\hs{-0.02cm}0}^{1-z^{}_2}} \hs{-0.05cm} \dd z^{}_3 \,
{\cal J}^{}_2
~,\quad
\\[0.15cm]
C^{}_2 
\Eq
\frac{\lambda_3^2{}^{}{}^{}y_N^2}{(4\pi)^2}
\mathop{\mathlarger{\int}_{\hs{-0.02cm}0}^1} \hs{-0.05cm} \dd z^{}_1
\mathop{\mathlarger{\int}_{\hs{-0.02cm}0}^1} \hs{-0.05cm} \dd z^{}_2
\mathop{\mathlarger{\int}_{\hs{-0.02cm}0}^{1-z^{}_2}} \hs{-0.05cm} \dd z^{}_3 \,
({}^{}1-z^{}_2-z^{}_3) 
\big( {\cal J}^{}_1 - {\cal J}^{}_2 \big)
~.
\end{eqnarray}
The matrix element squared appearing in the Boltzmann equations is
\begin{eqnarray}
\overline{\big|{\cal M}^{2\tf{-loop}}_{\hs{-0.03cm}\NNXX}\big|\raisebox{1pt}{$^{\hs{-0.03cm}2}$}} 
\Eq
\frac{1}{2}
\Big[ 
m_N^2 s + t{}^{}u - \big(m_N^2+ m_X^2 \hs{-0.03cm}\big)\raisebox{1.2pt}{$^{\hs{-0.03cm}2}$} 
\Big] |C^{}_1|^2  
+
\frac{1}{2} m_N^2 \big(s - 4{}^{}m_N^2\big) |C^{}_2|^2  
\nn[0.1cm]
&&+\,
m_N^2 (t-u) {}^{} \tx{Re}\big(C_1^\ast C^{}_2\big)
~,
\end{eqnarray}
after taking the average over initial and final spins and including the appropriate symmetry factors for identical particles in the initial or final states.\,\,Using the partial wave expansion, the resultant $2 \to 2$ cross section up to the $p\,$-wave is given by
\begin{eqnarray}\label{cNNXXapp}
(\sigma v)^{2\tf{-loop}}_{\hs{-0.03cm}\NNXX}
\,=\,
\frac{m_X^2}{16{}^{}\pi}
\frac{\sqrt{r_N^2-1}}{r^{}_N}
\Bigg[
\big(r_N^2-1\big) |C^{}_1|^2 +
\frac{\big(11-2{}^{}r_N^2\big) |C^{}_1|^2 + 6{}^{}r_N^2|C^{}_2|^2}{24} {}^{} v^2
\Bigg]
~.
\end{eqnarray}
Finally with the redefinitions, $P^2 \to P^2/z^{}_4$ and $Q^2 \to Q^2/z^{}_4 - m_X^2$, one can define the two-loop functions given in Eq.\,\eqref{I1I2} associated with the $C^{}_1$ and $C{}_2$ after some algebra, and then derive Eq.\,\eqref{cNNXX} by taking a thermal average of the above result.

Lastly, the procedure for computing the two-loop induced process $\XXNN$ is almost the same as the one we have demonstrated so far.\,\,On the other hand, the calculation with nonzero $\lambdaXS$ is straightforward.\,\,Thus, we do not show these details here.


\begin{thebibliography}{200}

%\cite{Baer:2014eja}
\bibitem{Baer:2014eja}
H.~Baer, K.~Y.~Choi, J.~E.~Kim and L.~Roszkowski,
%``Dark matter production in the early Universe: beyond the thermal WIMP paradigm,''
Phys. Rept. \textbf{555}, 1-60 (2015)
doi:10.1016/j.physrep.2014.10.002
[arXiv:1407.0017 [hep-ph]].
%290 citations counted in INSPIRE as of 22 Dec 2021

%\cite{Hochberg:2014kqa}
\bibitem{Hochberg:2014kqa}
Y.~Hochberg, E.~Kuflik, H.~Murayama, T.~Volansky and J.~G.~Wacker,
%``Model for Thermal Relic Dark Matter of Strongly Interacting Massive Particles,''
Phys. Rev. Lett. \textbf{115}, no.2, 021301 (2015)
%doi:10.1103/PhysRevLett.115.021301
[arXiv:1411.3727 [hep-ph]].
%223 citations counted in INSPIRE as of 30 May 2021

%\cite{Katz:2020ywn}
\bibitem{Katz:2020ywn}
A.~Katz, E.~Salvioni and B.~Shakya,
%``Split SIMPs with Decays,''
JHEP \textbf{10}, 049 (2020)
%doi:10.1007/JHEP10(2020)049
[arXiv:2006.15148 [hep-ph]].
%5 citations counted in INSPIRE as of 10 Jun 2021

%\cite{Choi:2021yps}
\bibitem{Choi:2021yps}
S.~M.~Choi, J.~Kim, P.~Ko and J.~Li,
%``A multi-component SIMP model with $U(1)_X \rightarrow Z_2 \times Z_3$,''
[arXiv:2103.05956 [hep-ph]].
%0 citations counted in INSPIRE as of 31 May 2021

%\cite{Baek:2013dwa}
\bibitem{Baek:2013dwa}
S.~Baek, P.~Ko and W.~I.~Park,
%``Hidden sector monopole, vector dark matter and dark radiation with Higgs portal,''
JCAP \textbf{10}, 067 (2014)
%doi:10.1088/1475-7516/2014/10/067
[arXiv:1311.1035 [hep-ph]].
%77 citations counted in INSPIRE as of 21 Jun 2021

%\cite{Aoki:2016glu}
\bibitem{Aoki:2016glu}
M.~Aoki and T.~Toma,
%``Implications of Two-component Dark Matter Induced by Forbidden Channels and Thermal Freeze-out,''
JCAP \textbf{01}, 042 (2017)
%doi:10.1088/1475-7516/2017/01/042
[arXiv:1611.06746 [hep-ph]].
%9 citations counted in INSPIRE as of 21 Jun 2021

%\cite{Daido:2019tbm}
\bibitem{Daido:2019tbm}
R.~Daido, S.~Y.~Ho and F.~Takahashi,
%``Hidden monopole dark matter via axion portal and its implications for direct detection searches, beam-dump experiments, and the H$_{0}$ tension,''
JHEP \textbf{01}, 185 (2020)
%doi:10.1007/JHEP01(2020)185
[arXiv:1909.03627 [hep-ph]].
%8 citations counted in INSPIRE as of 21 Jun 2021

%\cite{Herms:2019mnu}
\bibitem{Herms:2019mnu}
J.~Herms and A.~Ibarra,
%``Probing multicomponent FIMP scenarios with gamma-ray telescopes,''
JCAP \textbf{03}, 026 (2020)
%doi:10.1088/1475-7516/2020/03/026
[arXiv:1912.09458 [hep-ph]].
%4 citations counted in INSPIRE as of 03 Jul 2021

%\cite{Yaguna:2021rds}
\bibitem{Yaguna:2021rds}
C.~E.~Yaguna and \'O.~Zapata,
%``Fermion and scalar two-component dark matter from a $Z_4$ symmetry,''
[arXiv:2112.07020 [hep-ph]].
%0 citations counted in INSPIRE as of 05 Jan 2022

%\cite{DiazSaez:2021pmg}
\bibitem{DiazSaez:2021pmg}
B.~D\'\i{}az S\'aez, P.~Escalona, S.~Norero and A.~R.~Zerwekh,
%``Fermion singlet dark matter in a pseudoscalar dark matter portal,''
JHEP \textbf{10}, 233 (2021)
doi:10.1007/JHEP10(2021)233
[arXiv:2105.04255 [hep-ph]].
%2 citations counted in INSPIRE as of 21 Jan 2022

%\cite{Lee:1977ua}
\bibitem{Lee:1977ua}
B.~W.~Lee and S.~Weinberg,
%``Cosmological Lower Bound on Heavy Neutrino Masses,''
Phys. Rev. Lett. \textbf{39}, 165-168 (1977)
%doi:10.1103/PhysRevLett.39.165
%1351 citations counted in INSPIRE as of 28 May 2021

%\cite{Pospelov:2007mp}
\bibitem{Pospelov:2007mp}
M.~Pospelov, A.~Ritz and M.~B.~Voloshin,
%``Secluded WIMP Dark Matter,''
Phys. Lett. B \textbf{662}, 53-61 (2008)
doi:10.1016/j.physletb.2008.02.052
[arXiv:0711.4866 [hep-ph]].
%900 citations counted in INSPIRE as of 04 Jan 202

%\cite{Pospelov:2008jd}
\bibitem{Pospelov:2008jd}
M.~Pospelov and A.~Ritz,
%``Astrophysical Signatures of Secluded Dark Matter,''
Phys. Lett. B \textbf{671}, 391-397 (2009)
doi:10.1016/j.physletb.2008.12.012
[arXiv:0810.1502 [hep-ph]].
%529 citations counted in INSPIRE as of 17 Jan 2022

%\cite{Hochberg:2014dra}
\bibitem{Hochberg:2014dra}
Y.~Hochberg, E.~Kuflik, T.~Volansky and J.~G.~Wacker,
%``Mechanism for Thermal Relic Dark Matter of Strongly Interacting Massive Particles,''
Phys. Rev. Lett. \textbf{113}, 171301 (2014)
%doi:10.1103/PhysRevLett.113.171301
[arXiv:1402.5143 [hep-ph]].

%\cite{Tulin:2017ara}
\bibitem{Tulin:2017ara}
S.~Tulin and H.~B.~Yu,
%``Dark Matter Self-interactions and Small Scale Structure,''
Phys. Rept. \textbf{730}, 1-57 (2018)
doi:10.1016/j.physrep.2017.11.004
[arXiv:1705.02358 [hep-ph]].
%527 citations counted in INSPIRE as of 22 Dec 2021

%\cite{Brooks:2012vi}
\bibitem{Brooks:2012vi}
A.~M.~Brooks and A.~Zolotov,
%``Why Baryons Matter: The Kinematics of Dwarf Spheroidal Satellites,''
Astrophys. J. \textbf{786}, 87 (2014)
doi:10.1088/0004-637X/786/2/87
[arXiv:1207.2468 [astro-ph.CO]].
%218 citations counted in INSPIRE as of 22 Dec 2021

%\cite{Ho:2021ojb}
\bibitem{Ho:2021ojb}
S.~Y.~Ho, P.~Ko and C.~T.~Lu,
%``Reshuffled SIMP Dark Matter,''
[arXiv:2107.04375 [hep-ph]].
%1 citations counted in INSPIRE as of 02 Nov 2021

%\cite{Krauss:1988zc}
\bibitem{Krauss:1988zc}
L.~M.~Krauss and F.~Wilczek,
%``Discrete Gauge Symmetry in Continuum Theories,''
Phys. Rev. Lett. \textbf{62}, 1221 (1989)
doi:10.1103/PhysRevLett.62.1221
%639 citations counted in INSPIRE as of 05 Jan 2022

%\cite{Hochberg:2015vrg}
\bibitem{Hochberg:2015vrg}
Y.~Hochberg, E.~Kuflik and H.~Murayama,
%``SIMP Spectroscopy,''
JHEP \textbf{05}, 090 (2016)
doi:10.1007/JHEP05(2016)090
[arXiv:1512.07917 [hep-ph]].
%107 citations counted in INSPIRE as of 26 Dec 2021

%\cite{Hochberg:2018rjs}
\bibitem{Hochberg:2018rjs}
Y.~Hochberg, E.~Kuflik, R.~Mcgehee, H.~Murayama and K.~Schutz,
%``Strongly interacting massive particles through the axion portal,''
Phys. Rev. D \textbf{98}, no.11, 115031 (2018)
doi:10.1103/PhysRevD.98.115031
[arXiv:1806.10139 [hep-ph]].
%56 citations counted in INSPIRE as of 21 Jan 2022

%\cite{Markevitch:2003at}
\bibitem{Markevitch:2003at}
M.~Markevitch, A.~H.~Gonzalez, D.~Clowe, A.~Vikhlinin, L.~David, W.~Forman, C.~Jones, S.~Murray and W.~Tucker,
%``Direct constraints on the dark matter self-interaction cross-section from the merging galaxy cluster 1E0657-56,''
Astrophys. J. \textbf{606}, 819-824 (2004)
doi:10.1086/383178
[arXiv:astro-ph/0309303 [astro-ph]].
%719 citations counted in INSPIRE as of 13 Dec 2021

%\cite{Clowe:2003tk}
\bibitem{Clowe:2003tk}
D.~Clowe, A.~Gonzalez and M.~Markevitch,
%``Weak lensing mass reconstruction of the interacting cluster 1E0657-558: Direct evidence for the existence of dark matter,''
Astrophys. J. \textbf{604}, 596-603 (2004)
doi:10.1086/381970
[arXiv:astro-ph/0312273 [astro-ph]].
%540 citations counted in INSPIRE as of 13 Dec 2021

%\cite{Massey:2015dkw}
\bibitem{Massey:2015dkw}
R.~Massey, L.~Williams, R.~Smit, M.~Swinbank, T.~D.~Kitching, D.~Harvey, M.~Jauzac, H.~Israel, D.~Clowe and A.~Edge, \textit{et al.}
%``The behaviour of dark matter associated with four bright cluster galaxies in the 10 kpc core of Abell 3827,''
Mon. Not. Roy. Astron. Soc. \textbf{449}, no.4, 3393-3406 (2015)
doi:10.1093/mnras/stv467
[arXiv:1504.03388 [astro-ph.CO]].
%137 citations counted in INSPIRE as of 05 Jan 2022

%\cite{Kahlhoefer:2015vua}
\bibitem{Kahlhoefer:2015vua}
F.~Kahlhoefer, K.~Schmidt-Hoberg, J.~Kummer and S.~Sarkar,
%``On the interpretation of dark matter self-interactions in Abell 3827,''
Mon. Not. Roy. Astron. Soc. \textbf{452}, no.1, L54-L58 (2015)
doi:10.1093/mnrasl/slv088
[arXiv:1504.06576 [astro-ph.CO]].
%114 citations counted in INSPIRE as of 05 Jan 2022

%\cite{Denner:1992vza}
\bibitem{Denner:1992vza}
A.~Denner, H.~Eck, O.~Hahn and J.~Kublbeck,
%``Feynman rules for fermion number violating interactions,''
Nucl. Phys. B \textbf{387}, 467-481 (1992)

%\cite{Choi:2016hid}
\bibitem{Choi:2016hid}
S.~M.~Choi and H.~M.~Lee,
%``Resonant SIMP dark matter,''
Phys. Lett. B \textbf{758}, 47-53 (2016)
%doi:10.1016/j.physletb.2016.04.055
[arXiv:1601.03566 [hep-ph]].

%\cite{Ho:2017fte}
\bibitem{Ho:2017fte}
S.~Y.~Ho, T.~Toma and K.~Tsumura,
%``A Radiative Neutrino Mass Model with SIMP Dark Matter,''
JHEP \textbf{07}, 101 (2017)
%doi:10.1007/JHEP07(2017)101
[arXiv:1705.00592 [hep-ph]].

%\cite{Gondolo:1990dk}
\bibitem{Gondolo:1990dk} 
  P.~Gondolo and G.~Gelmini,
  %``Cosmic abundances of stable particles: Improved analysis,''
  Nucl.\ Phys.\ B {\bf 360}, 145 (1991).
  %doi:10.1016/0550-3213(91)90438-4
  %%CITATION = doi:10.1016/0550-3213(91)90438-4;%%

%\cite{Choi:2017mkk}
\bibitem{Choi:2017mkk}
S.~M.~Choi, H.~M.~Lee and M.~S.~Seo,
%``Cosmic abundances of SIMP dark matter,''
JHEP \textbf{04}, 154 (2017)
%doi:10.1007/JHEP04(2017)154
[arXiv:1702.07860 [hep-ph]].

%\cite{Perez:2021rbo}
\bibitem{Perez:2021rbo}
P.~F.~Perez and A.~D.~Plascencia,
%``Theory of Dirac Dark Matter: Higgs Decays and EDMs,''
[arXiv:2112.02103 [hep-ph]].
%0 citations counted in INSPIRE as of 07 Dec 2021

%\cite{Allwicher:2021rtd}
\bibitem{Allwicher:2021rtd}
L.~Allwicher, P.~Arnan, D.~Barducci and M.~Nardecchia,
%``Perturbative unitarity constraints on generic Yukawa interactions,''
JHEP \textbf{10}, 129 (2021)
doi:10.1007/JHEP10(2021)129
[arXiv:2108.00013 [hep-ph]].
%6 citations counted in INSPIRE as of 16 Nov 2021

%\cite{Namjoo:2018oyn}
\bibitem{Namjoo:2018oyn}
M.~H.~Namjoo, T.~R.~Slatyer and C.~L.~Wu,
%``Enhanced n-body annihilation of dark matter and its indirect signatures,''
JHEP \textbf{03}, 077 (2019)
%doi:10.1007/JHEP03(2019)077
[arXiv:1810.09455 [astro-ph.CO]].
%7 citations counted in INSPIRE as of 15 Jun 2021

%\cite{Choi:2016tkj}
\bibitem{Choi:2016tkj}
S.~M.~Choi, Y.~J.~Kang and H.~M.~Lee,
%``On thermal production of self-interacting dark matter,''
JHEP \textbf{12}, 099 (2016)
doi:10.1007/JHEP12(2016)099
[arXiv:1610.04748 [hep-ph]].
%41 citations counted in INSPIRE as of 17 Nov 2021

%\cite{Boehm:2013jpa}
\bibitem{Boehm:2013jpa}
C.~Boehm, M.~J.~Dolan and C.~McCabe,
%``A Lower Bound on the Mass of Cold Thermal Dark Matter from Planck,''
JCAP 08, 041 (2013)
%doi:10.1088/1475-7516/2013/08/041
[arXiv:1303.6270 [hep-ph]].
%184 citations counted in INSPIRE as of 17 May 2021

%\cite{Bennett:2020zkv}
\bibitem{Bennett:2020zkv}
J.~J.~Bennett, G.~Buldgen, P.~F.~De Salas, M.~Drewes, S.~Gariazzo, S.~Pastor and Y.~Y.~Y.~Wong,
%``Towards a precision calculation of $N_{\rm eff}$ in the Standard Model II: Neutrino decoupling in the presence of flavour oscillations and finite-temperature QED,''
JCAP 04, 073 (2021)
%doi:10.1088/1475-7516/2021/04/073
[arXiv:2012.02726 [hep-ph]].
%11 citations counted in INSPIRE as of 15 May 2021

%\cite{Akita:2020szl}
\bibitem{Akita:2020szl}
K.~Akita and M.~Yamaguchi,
%``A precision calculation of relic neutrino decoupling,''
JCAP \textbf{08}, 012 (2020)
doi:10.1088/1475-7516/2020/08/012
[arXiv:2005.07047 [hep-ph]].
%63 citations counted in INSPIRE as of 20 Nov 2021

%\cite{Escudero:2018mvt}
\bibitem{Escudero:2018mvt}
M.~Escudero,
%``Neutrino decoupling beyond the Standard Model: CMB constraints on the Dark Matter mass with a fast and precise $N_{\rm eff}$ evaluation,''
JCAP 02, 007 (2019)
doi:10.1088/1475-7516/2019/02/007
[arXiv:1812.05605 [hep-ph]].
%60 citations counted in INSPIRE as of 15 May 2021

%\cite{Lehmann:2020lcv}
\bibitem{Lehmann:2020lcv}
B.~V.~Lehmann and S.~Profumo,
%``Cosmology and prospects for sub-MeV dark matter in electron recoil experiments,''
Phys. Rev. D 102, no.2, 023038 (2020)
%doi:10.1103/PhysRevD.102.023038
[arXiv:2002.07809 [hep-ph]].
%4 citations counted in INSPIRE as of 15 May 2021

%\cite{Aghanim:2018eyx}
\bibitem{Aghanim:2018eyx}
N.~Aghanim \textit{et al.} [Planck],
%``Planck 2018 results. VI. Cosmological parameters,''
Astron. Astrophys. 641, A6 (2020).
%doi:10.1051/0004-6361/201833910
[arXiv:1807.06209 [astro-ph.CO]].

%\cite{BaBar:2017tiz}
\bibitem{BaBar:2017tiz}
J.~P.~Lees \textit{et al.} [BaBar],
%``Search for Invisible Decays of a Dark Photon Produced in ${e}^{+}{e}^{-}$ Collisions at BaBar,''
Phys. Rev. Lett. \textbf{119}, no.13, 131804 (2017)
doi:10.1103/PhysRevLett.119.131804
[arXiv:1702.03327 [hep-ex]].
%270 citations counted in INSPIRE as of 27 Nov 2021

%\cite{Fabbrichesi:2020wbt}
\bibitem{Fabbrichesi:2020wbt}
M.~Fabbrichesi, E.~Gabrielli and G.~Lanfranchi,
%``The Dark Photon,''
doi:10.1007/978-3-030-62519-1
[arXiv:2005.01515 [hep-ph]].
%101 citations counted in INSPIRE as of 27 Nov 2021

%\cite{Saikawa:2018rcs}
\bibitem{Saikawa:2018rcs}
K.~Saikawa and S.~Shirai,
%``Primordial gravitational waves, precisely: The role of thermodynamics in the Standard Model,''
JCAP \textbf{05}, 035 (2018)
%doi:10.1088/1475-7516/2018/05/035
[arXiv:1803.01038 [hep-ph]].

%\cite{Bhattacharya:2019mmy}
\bibitem{Bhattacharya:2019mmy}
S.~Bhattacharya, P.~Ghosh and S.~Verma,
%``SIMPler realisation of Scalar Dark Matter,''
JCAP 01, 040 (2020)
%doi:10.1088/1475-7516/2020/01/040
[arXiv:1904.07562 [hep-ph]].

%\cite{Shakya:2021pa}
\bibitem{Shakya:2021pa}
B. Shakya, E. Salvioni and J.T. Ruderman, Depleting or producing dark matter, in The 26th International Symposium on Particle Physics, String Theory, and Cosmology (PASCOS 2021), Online Conference, Korea, 14-18 June 2021.

%\cite{Gondolo:2012vh}
\bibitem{Gondolo:2012vh}
P.~Gondolo, J.~Hisano and K.~Kadota,
%``The Effect of quark interactions on dark matter kinetic decoupling and the mass of the smallest dark halos,''
Phys. Rev. D \textbf{86}, 083523 (2012)
%doi:10.1103/PhysRevD.86.083523
[arXiv:1205.1914 [hep-ph]].
%54 citations counted in INSPIRE as of 13 Jun 2021

%\cite{Choi:2019zeb}
\bibitem{Choi:2019zeb}
S.~M.~Choi, H.~M.~Lee, Y.~Mambrini and M.~Pierre,
%``Vector SIMP dark matter with approximate custodial symmetry,''
JHEP \textbf{07}, 049 (2019)
%doi:10.1007/JHEP07(2019)049
[arXiv:1904.04109 [hep-ph]].
%19 citations counted in INSPIRE as of 13 Jun 2021

%\cite{Cheung:2012gi}
\bibitem{Cheung:2012gi}
K.~Cheung, P.~Y.~Tseng, Y.~L.~S.~Tsai and T.~C.~Yuan,
%``Global Constraints on Effective Dark Matter Interactions: Relic Density, Direct Detection, Indirect Detection, and Collider,''
JCAP \textbf{05}, 001 (2012)
doi:10.1088/1475-7516/2012/05/001
[arXiv:1201.3402 [hep-ph]].
%122 citations counted in INSPIRE as of 15 Nov 2021

%\cite{Tulin:2013teo}
\bibitem{Tulin:2013teo}
S.~Tulin, H.~B.~Yu and K.~M.~Zurek,
%``Beyond Collisionless Dark Matter: Particle Physics Dynamics for Dark Matter Halo Structure,''
Phys. Rev. D \textbf{87}, no.11, 115007 (2013)
doi:10.1103/PhysRevD.87.115007
[arXiv:1302.3898 [hep-ph]].
%388 citations counted in INSPIRE as of 13 Dec 2021

%\cite{Kamada:2016euw}
\bibitem{Kamada:2016euw}
A.~Kamada, M.~Kaplinghat, A.~B.~Pace and H.~B.~Yu,
%``How the Self-Interacting Dark Matter Model Explains the Diverse Galactic Rotation Curves,''
Phys. Rev. Lett. \textbf{119}, no.11, 111102 (2017)
doi:10.1103/PhysRevLett.119.111102
[arXiv:1611.02716 [astro-ph.GA]].
%164 citations counted in INSPIRE as of 13 Dec 2021

%\cite{Chu:2018fzy}
\bibitem{Chu:2018fzy}
X.~Chu, C.~Garcia-Cely and H.~Murayama,
%``Velocity Dependence from Resonant Self-Interacting Dark Matter,''
Phys. Rev. Lett. \textbf{122}, no.7, 071103 (2019)
doi:10.1103/PhysRevLett.122.071103
[arXiv:1810.04709 [hep-ph]].
%33 citations counted in INSPIRE as of 13 Dec 2021

%\cite{Biswas:2021dan}
\bibitem{Biswas:2021dan}
A.~Biswas and S.~Khan,
%``$(g-2)_{e,\,\mu}$ and strongly interacting dark matter with collider implications,''
[arXiv:2112.08393 [hep-ph]].
%0 citations counted in INSPIRE as of 19 Dec 2021

%\cite{Hochberg:2021pkt}
\bibitem{Hochberg:2021pkt}
Y.~Hochberg, Y.~Kahn, N.~Kurinsky, B.~V.~Lehmann, T.~C.~Yu and K.~K.~Berggren,
%``Determining Dark-Matter\textendash{}Electron Scattering Rates from the Dielectric Function,''
Phys. Rev. Lett. \textbf{127}, no.15, 151802 (2021)
doi:10.1103/PhysRevLett.127.151802
[arXiv:2101.08263 [hep-ph]].
%16 citations counted in INSPIRE as of 04 Jan 2022

%\cite{Blanco:2021hlm}
\bibitem{Blanco:2021hlm}
C.~Blanco, Y.~Kahn, B.~Lillard and S.~D.~McDermott,
%``Dark Matter Daily Modulation With Anisotropic Organic Crystals,''
Phys. Rev. D \textbf{104}, 036011 (2021)
doi:10.1103/PhysRevD.104.036011
[arXiv:2103.08601 [hep-ph]].
%9 citations counted in INSPIRE as of 04 Jan 2022 

%\cite{Griffin:2021znd}
\bibitem{Griffin:2021znd}
S.~M.~Griffin, K.~Inzani, T.~Trickle, Z.~Zhang and K.~M.~Zurek,
%``Extended calculation of dark matter-electron scattering in crystal targets,''
Phys. Rev. D \textbf{104}, no.9, 095015 (2021)
doi:10.1103/PhysRevD.104.095015
[arXiv:2105.05253 [hep-ph]].
%11 citations counted in INSPIRE as of 04 Jan 2022

%\cite{Liang:2021zkg}
\bibitem{Liang:2021zkg}
Z.~L.~Liang, C.~Mo and P.~Zhang,
%``In-medium screening effects for the Galactic halo and solar-reflected dark matter detection in semiconductor targets,''
Phys. Rev. D \textbf{104}, no.9, 096001 (2021)
doi:10.1103/PhysRevD.104.096001
[arXiv:2107.01209 [hep-ph]].
%3 citations counted in INSPIRE as of 04 Jan 2022

%\cite{Belle-II:2018jsg}
\bibitem{Belle-II:2018jsg}
E.~Kou \textit{et al.} [Belle-II],
%``The Belle II Physics Book,''
PTEP \textbf{2019}, no.12, 123C01 (2019)
[erratum: PTEP \textbf{2020}, no.2, 029201 (2020)]
doi:10.1093/ptep/ptz106
[arXiv:1808.10567 [hep-ex]].
%824 citations counted in INSPIRE as of 04 Jan 2022

%\cite{KLEVERProject:2019aks}
\bibitem{KLEVERProject:2019aks}
F.~Ambrosino \textit{et al.} [KLEVER Project],
%``KLEVER: An experiment to measure BR($K_L\to\pi^0\nu\bar{\nu}$) at the CERN SPS,''
[arXiv:1901.03099 [hep-ex]].
%34 citations counted in INSPIRE as of 04 Jan 2022

%\cite{LDMX:2018cma}
\bibitem{LDMX:2018cma}
T.~\r{A}kesson \textit{et al.} [LDMX],
%``Light Dark Matter eXperiment (LDMX),''
[arXiv:1808.05219 [hep-ex]].
%119 citations counted in INSPIRE as of 04 Jan 2022 

%\cite{Raubenheimer:2018mwt}
\bibitem{Raubenheimer:2018mwt}
T.~Raubenheimer, A.~Beukers, A.~Fry, C.~Hast, T.~Markiewicz, Y.~Nosochkov, N.~Phinney, P.~Schuster and N.~Toro,
%``DASEL: Dark Sector Experiments at LCLS-II,''
[arXiv:1801.07867 [physics.acc-ph]].
%18 citations counted in INSPIRE as of 04 Jan 2022

%\cite{Holdom:1985ag}
\bibitem{Holdom:1985ag}
B.~Holdom,
%``Two U(1)'s and Epsilon Charge Shifts,''
Phys. Lett. B \textbf{166}, 196-198 (1986)
doi:10.1016/0370-2693(86)91377-8
%1950 citations counted in INSPIRE as of 04 Jan 2022

%\cite{Okun:1982xi}
\bibitem{Okun:1982xi}
L.~B.~Okun,
%``LIMITS OF ELECTRODYNAMICS: PARAPHOTONS?,''
Sov. Phys. JETP \textbf{56}, 502 (1982)
ITEP-48-1982.
%408 citations counted in INSPIRE as of 04 Jan 2022

%\cite{Patt:2006fw}
\bibitem{Patt:2006fw}
B.~Patt and F.~Wilczek,
%``Higgs-field portal into hidden sectors,''
[arXiv:hep-ph/0605188 [hep-ph]].
%702 citations counted in INSPIRE as of 04 Jan 2022

%\cite{Lebedev:2021xey}
\bibitem{Lebedev:2021xey}
O.~Lebedev,
%``The Higgs portal to cosmology,''
Prog. Part. Nucl. Phys. \textbf{120}, 103881 (2021)
doi:10.1016/j.ppnp.2021.103881
[arXiv:2104.03342 [hep-ph]].
%16 citations counted in INSPIRE as of 04 Jan 2022

%\cite{Ho:2016aye}
\bibitem{Ho:2016aye}
S.~Y.~Ho, T.~Toma and K.~Tsumura,
%``Systematic $U(1)_{B–L}$ extensions of loop-induced neutrino mass models with dark matter,''
Phys. Rev. D \textbf{94}, no.3, 033007 (2016)
doi:10.1103/PhysRevD.94.033007
[arXiv:1604.07894 [hep-ph]].
%23 citations counted in INSPIRE as of 25 Dec 2021

%\cite{Ho:2021pqw}
\bibitem{Ho:2021pqw}
S.~Y.~Ho and C.~T.~Lu,
%``Comment on ''New Freezeout Mechanism for Strongly Interacting Dark Matter'',''
[arXiv:2108.06471 [hep-ph]].
%0 citations counted in INSPIRE as of 15 Dec 2021

\end{thebibliography}
\end{document}